\documentclass[letterpaper,hyper]{JHEP3}
\usepackage[centertags]{amsmath}
\usepackage{array,multirow}
\usepackage{amssymb}
\usepackage{graphicx}
\usepackage{color}

\usepackage{epsfig}
\def\cn{{\mathcal N}}
\def\tr{{\rm Tr}}

\def\l{\lambda}
\def\lb{\bar{\lambda}}

\def\mb{\bar{\mu}}
\def\wb{\bar{w}}
\def\e{\eta}
\def\a{\alpha}
\def\ad{\dot{\alpha}}
\def\ab{\bar{a}}

\def\Qb{\bar{Q}}
\def\o[#1]{{\rm O}\left({#1}\right)}
\def\dotl[#1,#2]{\left\langle #1, #2 \right\rangle}
\def\dotlb[#1,#2]{[ #1, #2 ]}
\def\dotp[#1,#2]{(#1) \cdot (#2)}
\def\>{\rangle}
\def\<{\langle}
\date{}
\title {The Next-to-Simplest Quantum Field Theories}
\author{Shailesh Lal$^{a}$ and 
Suvrat Raju$^{a,b}$ \\
$^a$ Harish-Chandra Research Institute, Jhunsi, Allahabad 211019. \\
$^{b}$ Tata Institute of Fundamental Research, Homi Bhabha Road, Mumbai 400005.
}
\abstract{We describe new on-shell recursion relations for tree-amplitudes in 
${\cn=1}$ and ${\cn=2}$ gauge theories and use these to show that the structure of the 
S-matrix in pure ${\cn=1}$ and ${\cn=2}$ gauge theories resembles that of pure Yang-Mills.
We proceed to study gluon scattering in gauge theories coupled to 
matter in arbitrary representations. The contribution of matter to individual
bubble and triangle coefficients can depend on the fourth and sixth order Indices of the matter representation respectively.
So, the condition
that one-loop amplitudes be free of bubbles and triangles can be written as 
a set of 
linear Diophantine 
equations involving these higher-order Indices. 
These equations simplify for supersymmetric theories. We present new examples of supersymmetric theories that have only boxes (and no triangles or bubbles at one-loop) and non-supersymmetric theories that are free of bubbles. In particular,
our results indicate that one-loop scattering amplitudes in the ${\cn=2}$, SU(N) theory
with a symmetric tensor hypermultiplet and an anti-symmetric tensor hypermultiplet are simple like those in the ${\cn=4}$ theory. }
\preprint{HRI/ST/0911\\
TIFR/TH/09-08}
\begin{document}
\section{Introduction}
S-matrix elements in gauge theories often have beautiful properties that do not extend to correlation functions. As a result these properties are invisible in the Feynman diagram expansion. The past few years have seen the development of new on-shell techniques that not only make some of these properties manifest but also provide a computationally efficient approach to perturbation theory.

The first example of these remarkable properties was discovered in 1986, when Parke and Taylor wrote down a formula for Maximally-Helicity-Violating (MHV) amplitudes \cite{Parke:1986gb}. The Parke-Taylor formula provides a very compact expression for the 
tree-level scattering of two negative-helicity gluons with 
an arbitrary number of positive-helicity gluons.
Moreover, it is very hard to derive this formula from the 
Yang-Mills (YM) Lagrangian.

Then, in  2004, Britto et al. \cite{Britto:2004ap,Britto:2005fq} discovered that tree-level scattering amplitudes in gauge theories
obey recursion relations (now called the BCFW recursion relations) that relate higher 
point amplitudes to products of lower point amplitudes. These recursion relations allow us to 
construct any tree amplitude in a gauge theory  starting with just the on-shell three point amplitude; more strikingly, they never make any reference to off-shell quantities! They also make the Parke-Taylor formula manifest.

Thus, contrary to what one might expect from the Lagrangian, gauge scattering amplitudes are nicer than scattering amplitudes of scalars or fermions which do not see these simplifications.

In \cite{Brandhuber:2008pf,ArkaniHamed:2008gz} the BCFW recursion relations were extended to $\cn=4$ Super Yang Mills (SYM). Roughly speaking this works by utilizing supersymmetry to relate scattering amplitudes of the fermions and scalars in the theory to those of gauge bosons. 
More precisely, since all the particles in the $\cn = 4$ SYM theory lie in a single irreducible representation of the supersymmetry algebra, we can use supersymmetry to convert {\em any} two
particles into {\em negative} helicity gauge bosons. This makes $\cn=4$ tree-scattering amplitudes even nicer than pure gauge scattering amplitudes.  

At one-loop, on-shell methods 
have been used since the early nineties by Bern et al.  \cite{Bern:1991aq,Bern:1993mq,Bern:1994zx,Bern:1994cg,Bern:1995db} as a computationally efficient method of calculating phenomenologically relevant amplitudes.
In addition to this, just as they do at tree-level,  on-shell methods reveal new structures in the one-loop S-matrix.  In any quantum field theory, the one-loop S-matrix can be expanded in a basis of scalar box, triangle and bubble integrals  apart from a possible rational remainder. The no-triangle property \cite{Bern:1994zx,Bern:2005bb,BjerrumBohr:2006yw} states that any one-loop amplitude in the $\cn=4$ SYM theory can be written as a linear combination of scalar box integrals only.

This structure at one-loop is related to the properties of the tree level $\cn=4$ S-matrix.  Arkani-Hamed et al. \cite{ArkaniHamed:2008gz} showed that the no-triangle property follows immediately from the nice behavior of tree level $\cn = 4$ SYM amplitudes under the BCFW deformation.

So as we have seen above, on-shell methods provide a new perspective on S-matrix elements in perturbative theories. The properties of the $\cn=4$ theory alluded to above (and similar simplifications in the $\cn=8$ supergravity S-matrix \cite{BjerrumBohr:2008ji}), led  Arkani-Hamed et al. to suggest that $\cn=4$ SYM and $\cn=8$ SUGRA  might be the `simplest' quantum field theories despite having very complicated Lagrangians.

The question we ask in this paper is whether there are any other gauge theories
with a simple one-loop S-matrix. Is it possible to alter the matter
content of the ${\cn=4}$ theory but still ensure that at least for 
gluon scattering, the S-matrix is free of triangles and bubbles?  The authors of \cite{Badger:2008rn} showed that multi-photon amplitudes in QED, with a sufficient number of external photons,  have a no-triangle 
property that is similar to ${\cn=4}$ SYM. Our focus, in this paper,
is on non-Abelian gauge theories.

It turns out that demanding that the gluon S-matrix be free 
of triangles and bubbles is equivalent to imposing some
group-theoretic constraints on the matter representations in the theory. We
tabulate these constraints and solve them. This gives us new examples
of theories that are free of bubbles and triangles. We also find 
some new theories that have boxes and triangles but no bubbles. 

In particular, one of the theories we find this way is the
${\cn=2}$, SU(N) theory with a hypermultiplet transforming as the symmetric
tensor and another hypermultiplet transforming as the anti-symmetric tensor of SU(N). We find that gluon scattering amplitudes in this theory are free of 
triangles and bubbles just like the ${\cn=4}$ theory. This is true at
finite $N$, without any need to take the planar limit. 
This theory is superconformal and, in fact, has a gravity dual 
that is an orientifold of AdS$_5 \times S^5$ \cite{Ennes:2000fu}. So, several
properties of the {\em planar} ${\cn=4}$ S-matrix, such as dual superconformal invariance, descend to this theory \cite{Bershadsky:1998cb}. Our results go beyond this parent-daughter equivalence and show that, 
even at finite $N$ when non-planar contributions are taken into account, 
one-loop gluon scattering amplitudes in this theory are simple as in the ${\cn=4}$ theory.

As a prelude to studying loop amplitudes, we start by studying tree amplitudes in $\cn=1$ and $\cn=2$ theories. 
So,  we first write down a new set of BCFW-like recursion relations that can be applied to $\cn=1$ and $\cn=2$ gauge theories. Structurally,
these recursion relations are very similar to the recursion relations of 
pure YM theory. In particular, the BCFW extension that we need to perform is dependent on the helicities of the particles that we choose to extend. Nevertheless, in pure ${\cn=1}$ and ${\cn=2}$ theories, these recursion relations
are enough to calculate any tree-amplitude.  In theories with matter, just like
in non-supersymmetric gauge theories with matter, the recursion relations
are useful as long as a sufficient number of external particles belong
to the vector multiplet.

We then use these recursion relations to study the one-loop S-matrix of pure $\cn=1$ and $\cn =2$ theories. Somewhat expectedly, we find that these theories have both bubbles and triangles in their one-loop expansion. Hence, from a structural viewpoint, at tree-level and at one-loop, the S-matrix of pure $\cn=1$ and $\cn=2$ theories is closer to that of non-supersymmetric Yang Mills than to that of the $\cn=4$ theory. Clearly, these theories are too plain; so we add
some matter.

In section \ref{sec:matter}, we consider the effect of the addition of matter, in arbitrary representations, on one-loop gluon amplitudes. By going to the so-called $q$-lightcone gauge, we are able to isolate the representation dependent
color-factors that appear in individual bubble and triangle coefficients. We now find some surprising results. 

The contribution of matter, in a given representation, to the one-loop $\beta$ function is captured by the (quadratic)  Index of the representation. 
We find that the contribution of matter to an individual
bubble coefficient can depend not only on the quadratic  Index 
but also on the 4$^{\rm th}$ order Indices of the representation! In fact, since the $\beta$ function depends on the bubble coefficients,
at first sight this might seem to be a contradiction. However, the $\beta$ function receives contributions
from several bubble diagrams. While the 4$^{\rm th}$ order Indices contribute
to the S-matrix through bubbles, when we consider the net UV divergence in the 
theory the dependence on the 4$^{\rm th}$ order Indices cancels!

Similarly, the contribution of matter to triangles depends on the second,
fourth, fifth and sixth order Indices of the matter representation. There is no 
analogous result for boxes which are sensitive to the entire character of the
representation and not just a few of its invariants. 

For supersymmetric theories, these statements are modified a bit. Due
to cancellations between fermions and bosons, a ${\cn=1}$ chiral multiplet 
contributes to bubbles only through its quadratic Index and to triangles
through its quadratic, fourth and fifth order Indices. When the chiral
multiplets are in self-conjugate representations (this happens 
automatically if they are part of ${\cn=2}$ hypermultiplets), the fifth order Indices vanish
leaving behind only the second and fourth order Indices.

The fact that triangles and bubbles are sensitive to only a
few invariants of the representation has an interesting consequence. It implies
that triangles and bubbles will vanish in a theory that has matter in 
a representation with the same higher order Indices as the matter representation (multiple copies of the adjoint) that occurs in the ${\cn=4}$ theory! 

In a non-supersymmetric theory, we need to mimic all
the Indices of the ${\cn=4}$ theory up to the fourth order to get rid of bubbles 
and up to the sixth order to get rid of both triangles and bubbles. These constraints simplify for supersymmetric theories. 
In supersymmetric theories, the vanishing of the one-loop $\beta$ function
is enough to ensure the absence of bubbles. To make both bubbles and triangles
vanish, we need to mimic the Indices of the ${\cn=4}$ theory up to the fifth order.

We present some explicit examples of theories that have only boxes
and no triangles or bubbles in section \ref{sec:nexttosimplest}. These 
theories are all supersymmetric. However, we do provide examples
of non-supersymmetric theories that have triangles but no bubbles. 

In our description above, we explained that on-shell methods have revealed several new properties of scattering amplitudes that are invisible in the Lagrangian. However, it also happens that some properties of the S-matrix are hard 
to see in the on-shell approach and this has been the subject of some recent investigations \cite{Hodges:2009hk,Nguyen:2009jk,ArkaniHamed:2009dn}. A by-product of our analysis 
is to provide some new examples of this at one-loop. For example, the $\beta$ function of a theory  is proportional to the ratio of a sum of several bubble coefficients and the tree amplitude. From the S-matrix approach, it is very difficult to see why this sum of disparate bubble coefficients should miraculously simplify to give something that is numerically proportional to the tree-amplitude.

In fact, each individual bubble coefficient leads to UV-divergences that would 
appear to spoil the renormalizability of the theory. The tree-level analogue of this is that individual terms in the BCFW sum contain poles that cannot appear in a local quantum field theory  \cite{Hodges:2009hk}. 
When all bubble coefficients are added, these `spurious' terms cancel (just 
like `spurious' poles cancel at tree-level). It would be interesting to understand this property directly from the S-matrix approach.

A brief overview of this paper is as follows. In section \ref{sec:review}, we establish our notation and present a lightning review of the BCFW technique. In section \ref{sec:recrelation}, we present new recursion relations for tree amplitudes in $\cn=1,2$ theories. In section \ref{sec:oneloop}, we examine the one-loop S-matrix of pure $\cn=1,2$ theories and show that both triangle and bubble diagrams occur here. In section \ref{sec:matter}, we analyze the effect of the addition of matter on one-loop gluon amplitudes.
We show that contribution of matter to bubbles and triangles depends
on only a few higher order Indices of the matter representation. 
Imposing the vanishing of bubbles and triangles leads to linear Diophantine equations in these higher-order Indices.
We analyze these constraints and present some explicit examples of theories with a simple S-matrix in section \ref{sec:nexttosimplest}. 
 We present our conclusions in section \ref{sec:conclusions}.  Section \ref{sec:examples} contains several explicit calculations while the appendices
contain some additional details.

\section{Review}
\label{sec:review}
Here we establish some notation and briefly review spinor helicity variables and the standard BCFW extension. 

Given an on-shell momentum for a massless particle, we can decompose it into spinors using
\begin{equation}
p_{\alpha \dot{\alpha}} = p_{\mu} \sigma^{\mu}_{\alpha \dot{\alpha}} = \l_{\alpha} \lb_{\dot{\alpha}}.
\end{equation}
Our $\sigma$ matrix conventions are the same as \cite{Wess:1992cp}. We can take dot products of two momenta using
\begin{equation}
2 p_1 \cdot p_2 = \< \l_1, \l_2 \> \left[ \lb_1, \lb_2 \right],
\end{equation}
where
\begin{equation}
\dotl[\l_1, \l_2] = \epsilon^{\alpha \beta} (\l_1)_{\alpha} (\l_2)_{\beta}=(\l_1)_{\alpha} \l_{2}^{\alpha}, \qquad  \dotlb[\lb_1, \lb_2] = \epsilon^{\dot{\alpha} \dot{\beta}} (\lb_1)_{\dot{\alpha}} (\lb_2)_{\dot{\beta}} = (\lb_1)_{\dot{\alpha}} \lb_{2}^{\dot{\alpha}},
\end{equation}
and $\epsilon^{1 2} = 1$.
In terms of these spinors, gauge boson polarization vectors can be chosen to be
\begin{equation}
\epsilon^+_{\alpha \dot{\alpha}} = \sqrt{2} {\mu_{\alpha} \lb_{\dot{\alpha}} \over \dotl[ \l, \mu ]},  \qquad 
\epsilon^-_{\alpha \dot{\alpha}} = \sqrt{2} {\l_{\alpha} \mb_{\dot{\alpha}} \over \dotlb[\lb, \mb]},
\end{equation}
where $\mu,\mb$ are arbitrary spinors.  Thus, tree-amplitudes become rational functions of these spinors (see \cite{Dixon:1996wi} and references there for applications of spinors to amplitudes).

\subsection{On-Shell Methods at Tree-Level}
Consider an arbitrary gauge boson amplitude with $n$ particles. We denote the helicity of the first gauge boson by $\sigma_1$  and that of the $n^{\rm th}$ gauge boson by $\sigma_n$. Now, deform the momenta and polarization vectors of these particles according to
\begin{equation}
\label{bcfwextension}
\begin{split}
\l_1(z) &= \l_1, \:  \lb_1(z) = \lb_1 + z \lb_n, \:  \l_n(z)=\l_n-z\l_1, \:  \lb_n(z)=\lb_n,  \quad  {\rm if~} (\sigma_1, \sigma_n) = (-1,1), \\
\l_1(z) &= \l_1 + \l_n z, \:  \lb_1(z) = \lb_1, \: \l_n(z)=\l_n, \:  \lb_n(z) = \lb_n - z \lb_1, \quad  {\rm otherwise.} 
\end{split}
\end{equation}
Here $z$ is an arbitrary complex number. Note that while $z$ can become large which makes individual components of the momenta associated with particle $1$ and $n$ large, each momentum stays on shell. This is called the BCFW extension. It was shown in \cite{Britto:2004ap,Britto:2005fq} that, under this extension, the tree amplitude
\begin{equation}
\label{largezbehaviour}
{\mathcal A}^{\mathrm t}\left(\{\sigma_1, \l_1(z),\lb_1(z)\} \ldots \{\sigma_n, \l_n(z),\lb_n(z)\}\right)  \xrightarrow[z \rightarrow \infty]{} {\rm O}\left({1 \over z}\right).
\end{equation}
This surprising result allows us to write down recursion relations for tree amplitudes. Tree amplitudes develop simple poles in $z$ whenever an intermediate propagator goes on shell and the residue at each pole is just the product of two smaller tree amplitudes. Since the amplitude dies off at large $z$, we can completely reconstruct it from these residues i.e. from lower point tree amplitudes. This leads to the BCFW recursion relations.

The BCFW recursion relations are valid for any gauge theory but take on a very simple form for $U(N)$ gauge theories. In a $U(N)$ gauge theory, it is useful to define the color-ordered amplitudes,
\begin{equation}
\label{colorordered}
{\mathcal A}^{\mathrm t} = 2^{n \over 2} \sum_{\pi \in S_n/Z_n} A^{\mathrm t}\left(\{\sigma_{\pi(1)}, \l_{\pi(1)}, \lb_{\pi(1)}\}, \ldots \{\sigma_{\pi(n)}, \l_{\pi(n)}, \lb_{\pi(n)}\}\right)  \tr_F\left(T^{a_{\pi(1)}} \ldots T^{a_{\pi(n)}}\right),
\end{equation}
where $a_i$ indexes the color of particle $i$ and the trace of the product of generators is  taken in the fundamental representation. The sum is over the set of all permutations modulo cyclic permutations. The coefficients $A^{\mathrm t}$ are called color-ordered amplitudes and are obtained by summing over all double line graphs with the same cyclic ordering of external particles.

For color-ordered amplitudes, the BCFW recursion relations are
\begin{equation}
\label{bcfwrecursion}
\begin{split}
&A^{\mathrm t}\left(\{\sigma_1, \l_1(z),\lb_1(z)\} \ldots \{\sigma_n, \l_n(z),\lb_n(z)\}\right)\\ &=\sum_{\substack{j=2 \\ \sigma_{\rm int} = \pm 1}}^{n-2} {1 \over \left(p_1(z) + \sum_{i=2}^j p_i\right)^2} \left[A^{\mathrm t}\left(\{\sigma_1,\l_1(z_p^j), \lb_1(z_p^j)\} \ldots \{\sigma_j,\l_j, \lb_j\}, \{\sigma_{\rm int},\l_{\rm int}, \lb_{\rm int}\}\right)\right. \\
&\times \left. A^{\mathrm t}\left(\{-\sigma_{\rm int},-\l_{\rm int}, \lb_{\rm int}\},\{\sigma_{j+1},\l_{j+1}, \lb_{j+1}\} \ldots \{\sigma_{n},\l_{n}(z_p^j), \lb_{n}(z_p^j)\}\right) \right],
\end{split}
\end{equation}
where $z_p^j,p_{\rm int}$ are defined by
\begin{equation}
\left(p_1(z_p^j) + \sum_{i=2}^j p_i\right)^2 = 0,  \qquad  p_{\rm int} = p_1(z_p^j) + \sum_{i=2}^j p_i,
\end{equation}
and the sum over $\sigma_{\rm int}$ runs over all possible intermediate helicities.

A very similar set of recursion relations can be derived for tree amplitudes in a gauge theory in any number of dimensions \cite{ArkaniHamed:2008yf} (see also \cite{Cheung:2009dc} and \cite{Boels:2009bv} for extensions of spinor helicity technology to higher dimensions).

\section{Tree-Level Recursion in $\cn=1$ and $\cn=2$ Theories}
\label{sec:recrelation}

We now describe how the BCFW recursion relations can be generalized to theories with $\cn=1,2$ supersymmetry. First, we introduce some book-keeping notation --- a generalization of Nair's on-shell superspace \cite{Nair:1988bq} (see also \cite{Mandelstam:1982cb})  that allows us to efficiently keep track of on-shell states in the $\cn=1,2$ theories. With this notation in hand, it is easy to generalize the BCFW recursion; this is done in subsection \ref{subsec:recurs}.

\subsection{On-shell supersymmetry}
In figure \ref{n1n2multiplet}, we show the on-shell particle content of the $\cn=1,2$ vector multiplets labeling each particle by its helicity. Note, that in each case, CPT forces us to include two irreducible representations of the supersymmetry algebra in the physical theory. In contrast, the $\cn=4$ multiplet is its own CPT conjugate.
\FIGURE{
\epsfig{file=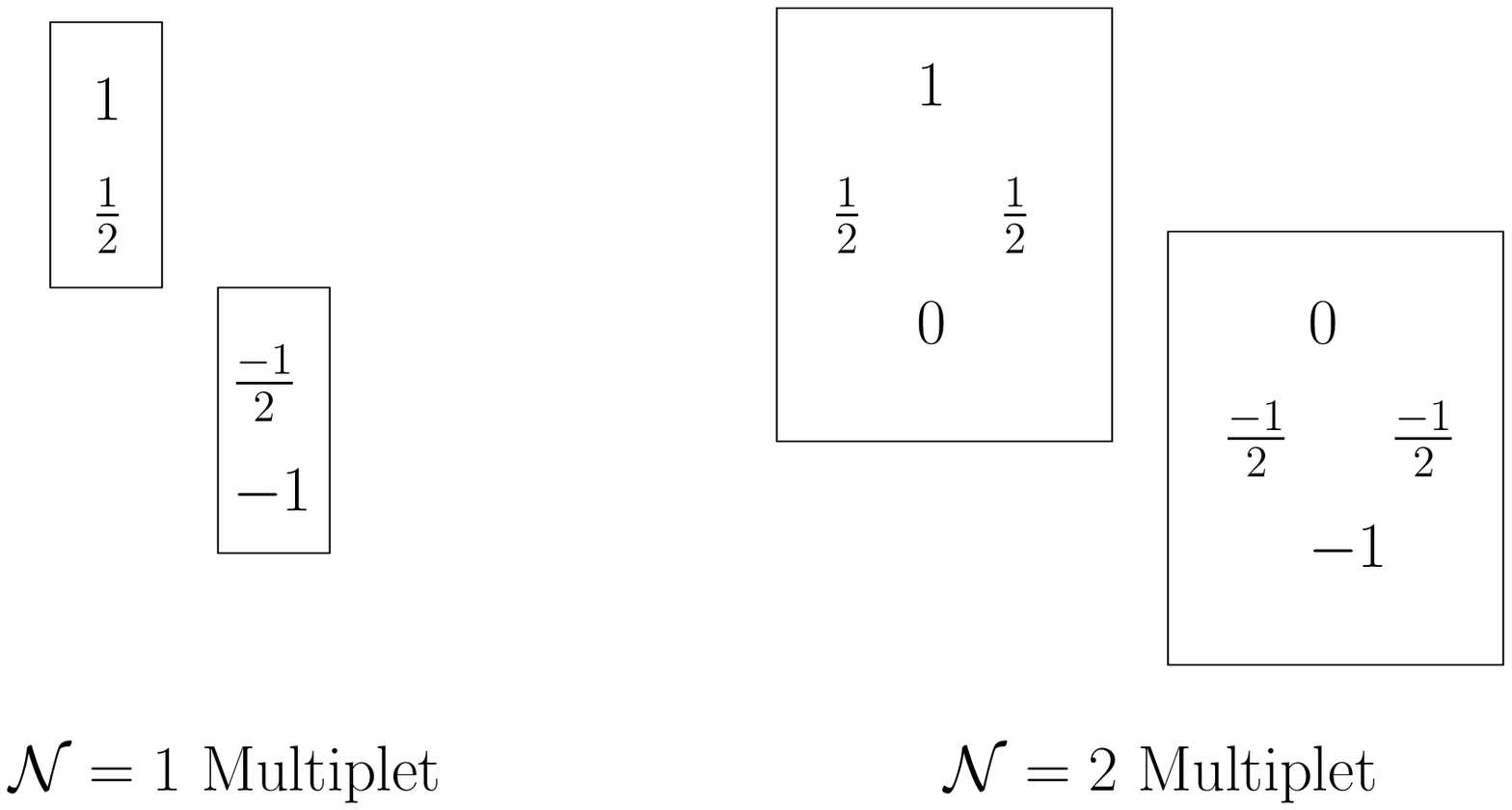,width=0.4\textwidth}\caption{The ${\cn=1}$ and ${\cn=2}$ vector multiplets have two irreps.}
\label{n1n2multiplet}
}

To keep track of the states in the multiplet, we will represent them using a number, $b$ and a Grassmann variable $\eta$. The number $b$, tells us which irreducible representation the state is in, while the Grassmann parameter will be useful for keeping track of the different states in a given irrep. For the vector multiplet, we take $b=1$ for the representation with positive helicity states and $b=-1$ for the representation with negative helicity states.

The classical theory has a $U(\cn)$ R symmetry that we will use for book-keeping (and no other) purposes. In the $\cn=2$ theory, writing $U(2) \sim U(1) \times SU(2)$,  $\eta$ transforms in a ${\bf 2}$ of the $SU(2)$ and with $U(1)$ charge $b$.
For the $\cn=1$ theory, we simply assign $\eta$ a $U(1)$ charge $b$.
 We emphasize that the transformation of a given Grassmann parameter under the R-symmetry is dependent on the corresponding value of $b$. 

Given two such Grassmann parameters, $\eta_{-1}$ linked to a particle with $b=-1$ and $\eta_1$ linked to a particle with $b=1$, we can form R-symmetry invariants 
\begin{equation}
\eta_{-1} \cdot \eta_1 = \sum_{I = 1}^{\cn} \eta_{-1}^I (\eta_1)_I.
\end{equation}
We follow the convention that $\eta_{-1}$ is naturally written with raised R-symmetry indices and $\eta_1$ is naturally written with lowered indices.

We define the state $|\{b,\eta,\l,\lb\}\rangle$ as follows:
\begin{equation}
\label{statedef}
|\{b,\eta,\l,\lb\}\rangle \equiv \left\{\begin{array}{l}e^{\langle \eta^I \cdot Q_I, w \rangle } |g^{-},\l, \lb \rangle, \quad \text {if}~ b=-1, \\ e^{\dotlb[ \eta_I \cdot \Qb^I, \wb]} |g^{+},\l,\lb \rangle, \quad  \text{if}~ b=1, \end{array} \right.
\end{equation}
where we have used $|g^{+}, \l, \lb \rangle$ ($|g^{-}, \l, \lb \rangle$) to denote a gauge boson with positive (negative) helicity and momentum $p_{\alpha \dot{\alpha}}=\l_{\alpha}\lb_{\dot{\alpha}}$. Here, the supersymmetry generators are normalized to satisfy
\begin{equation}
\{Q_{\alpha}, \Qb_{\dot{\alpha}}\} = P_{\alpha \dot{\alpha}}.
\end{equation}
The spinors $w, \wb$ satisfy  $\langle \l, w \rangle = 1 = [\lb, \wb]$.

Under a supersymmetry transformation,
\begin{equation}
\label{qsusy1}
e^{\dotl[Q_I,\zeta^I]}  |\{-1,\e,\l,\lb\}\rangle = |\{-1,\e - \dotl[\zeta, \l], \l, \lb\} \rangle.
\end{equation}
This is because we can write 
$\zeta = -\dotl[\zeta, \l] w + \dotl[\zeta, w] \l$ and the term proportional to
$\l$ does not modify the state at all.
Also,
\begin{equation}
\label{qsusy2}
e^{\dotl[Q_I,\zeta^I]} |\{1, \e, \l, \lb\} \rangle = e^{\dotl[\l, \zeta] \cdot \e} |\{1,\e,\l, \lb \} \rangle.
\end{equation}

Later, we will often need to consider products of two or more amplitudes with
intermediate particles that are summed over 
the supermultiplet. The sum can be written in a manifestly supersymmetric
form
\begin{equation}
\label{sumstates}
\begin{split}
\sum_{h \in {\rm multiplet}} &{\mathcal A}_{\rm left}(h,p) {\mathcal A}_{\rm right}(-h,-p) \\ 
\rightarrow \sum_{b<0} \int &\left[{\mathcal A}_{\rm left}(\{b,\mu_b,\l, \lb\}) {\mathcal A}_{\rm right}(\{-b,\mu_{-b},-\l,\lb\}) \right. \\
 + &\left.{\mathcal A}_{\rm right}(\{b,\mu_b,-\l, \lb\}) {\mathcal A}_{\rm left}(\{-b,\mu_{-b},\l,\lb\})\right] e^{\mu_{-b} \cdot \mu_{b}} \, (-1)^{\cn}  d^{\cn} \, \mu_{-b} \, d^{\cn} \, \mu_{b}.
\end{split}
\end{equation}
This form is invariant under supersymmetry since, under a supersymmetry 
transformation $e^{\dotl[Q, \zeta]}$, $\mu_b$ shifts according to \eqref{qsusy1}, while the amplitude
containing $-b$ picks up a phase according to \eqref{qsusy2}. If we shift $\mu_b$ back to its original
value, the transformation of the measure cancels this phase. The factor of $(-1)^{\cn}$ is required to ensure that the coefficient of $1$ in ${\mathcal A}_{\rm left}$ and ${\mathcal A}_{\rm right}$ contributes with a positive sign. We have included 
a sum over $b$ to account for the general case where we have 
multiplets other than the vector multiplet.

\subsection{Recursion Relations}
\label{subsec:recurs}
Consider a tree amplitude with $n$ particles
\begin{equation}
\label{treeamplitude}
{\mathcal A}^{\rm t}\left(\{b_1,\e_1 ,\l_1 ,\lb_1\}, \{b_2,\e_2,\l_2,\lb_2\} \ldots \{b_n,\e_n,\l_n,\lb_n\}\right).
\end{equation}

We will show that this amplitude may be calculated via recursion relations
provided at least two of the particles belong to the same irreducible 
representation in the vector multiplet. If all the external particles
are in the vector multiplet this must be true. Say, the two particles in 
question are $m_1$ and $m_2$. Now, depending on whether $b_{m_1} = b_{m_2} = 1$ or $b_{m_1} = b_{m_2} = -1$, we consider the following extension which is a function of a single parameter $z$.
\begin{equation}
\label{conditionalbcfw}
\begin{array}{|c|c|c|}\hline
\text{Case}&\multicolumn{2}{|c|}{\text{Extension}}\\\hline
\multirow{2}{*}{$b_{m_1}=b_{m_2}=-1$}
&\l_{m_1}(z) = \l_{m_1} + z \l_{m_2},\: \lb_{m_1}(z)=\lb_{m_1} &  
\lb_{m_2}(z)=\lb_{m_2}-z\lb_{m_1} \\
& \eta_{m_1}(z) = \eta_{m_1}(z) + z \eta_{m_2}& 
\l_{m_2}(z) = \l_{m_2},\: \eta_{m_2}(z) = \eta_{m_2}\\\hline
\multirow{2}{*}{$b_{m_1}=b_{m_2}=1$}&\l_{m_1}(z) = \l_{m_1},\:\lb_{m_1}(z) = \lb_{m_1} + z \lb_{m_2}, & \l_{m_2}(z) = \l_{m_2} - z \l_{m_1},
\lb_{m_2}(z)=\lb_{m_2} \\ 
&\eta_{m_1}(z) = \eta_{m_1}(z) + z \eta_{m_2}&\lb_{m_2}(z)=\lb_{m_2},\: \eta_{m_2}(z) = \eta_{m_2}\\\hline
\end{array}
\end{equation}

We now show that under this extension the amplitude \eqref{treeamplitude}
dies off as $\o[{1 \over z}]$ for large $z$.  
Consider the first possibility above where $b_{m_1}=b_{m_2}=-1$ (the proof easily generalizes to the other case). We perform a supersymmetry transformation $e^{-\dotl[Q, \chi]}$ on \eqref{treeamplitude} (extended according to \eqref{conditionalbcfw}) with parameter
\begin{equation}
\chi = {\l_{m_2} \eta_{m_1} - \l_{m_1} \eta_{m_2} \over \langle \l_{m_1}, \l_{m_2} \rangle}.
\end{equation}
Note that $\chi$ is a Lorentz spinor and also transforms under $U({\mathcal N})$ as explained above. 
From the definition \eqref{statedef}, the supersymmetry transformation leads us to
\begin{equation}
\label{aftersusy}
\begin{split}
&{\mathcal A}^{\mathrm t}\left(\{b_1,\eta_1,\l_1,\lb_1\},\ldots \{-1,\e_{m_1}(z),\l_{m_1}(z),\lb_{m_1}\}, \ldots \{-1,\e_{m_2},\l_{m_2},\lb_{m_2}(z)\},\ldots\{b_n,\eta_n,\l_n,\lb_n\}\right) \\ 
&= e^{\phi} {\mathcal A}^{\mathrm t}\left(\{b_1,\eta_1',\l_1,\lb_1\}, \ldots \{-1,0,\l_{m_1}(z),\lb_{m_1}\},\ldots
\{-1,0,\l_{m_2},\lb_{m_2}(z)\}, \ldots \{b_n,\eta_n',\l_n,\lb_n\}\right),
\end{split}
\end{equation}
where we have defined
\begin{equation}
\begin{split}
  \phi &= \sum_{b_i>0}  \langle \chi  \cdot \eta_i, \l_i \rangle, \\
\eta_i' &= \eta_i + \langle \chi, \l_i \rangle, {\rm~if~} b_i < 0
\end{split}
\end{equation}

Now, the tree-amplitude on the second line of \eqref{aftersusy} is just the scattering amplitude of two BCFW-extended negative helicity gauge bosons. If $b_{m_1} = b_{m_2} = 1$, we can reduce the scattering amplitude \eqref{treeamplitude} to one containing two BCFW-extended positive helicity gauge bosons. In either case, this amplitude vanishes as $\o[{1 \over z}]$ by the analysis of \cite{Cheung:2008dn}. This proves our result.

The amplitude in \eqref{treeamplitude}, when extended according to \eqref{conditionalbcfw}, has simple poles in $z$ whenever an intermediate particle goes on shell. Since it also vanishes at large $z$, it can be reconstructed from the residues at these poles. However, each of 
these residues is just the product of a tree amplitude on the left and the right. 

Let us describe this explicitly in a simple case. Consider color-ordered amplitudes in a pure $U(N)$ theory with $m_1 = 1, m_2 = n$ and $b_1 = b_n$. Then, we have the recursion relations
\begin{equation}
\label{n1n2recursion}
\begin{split}
&A^{\mathrm t}\left(\{b_{1},\eta_{1}(z), \l_1(z),\lb_1(z)\} \ldots \{b_{n},\eta_{n}, \l_n(z),\lb_n(z) \}\right) =\sum_{\substack{j=2 \\ b = \pm 1}}^{n-2}  \int {(-1)^{\cn} e^{\mu_1 \cdot \mu_{-1}} d^{\cn} \mu_1 d^{\cn} \mu_{-1} \over \left(p_1(z) + \sum_{i=2}^j p_i\right)^2} \\ 
&\times \left[A^{\mathrm t}\left(\{b_{1},\eta_{1}(z_p^j),\l_1(z_p^j), \lb_1(z_p^j)\} \ldots \{b_{j},\eta_{j},\l_j, \lb_j\}, \{b, \mu_b,\l_{\rm int}, \lb_{\rm int}\}\right)\right. \\
&\times \left. A^{\mathrm t}\left(\{-b,\mu_{-b},-\l_{\rm int}, \lb_{\rm int}\},\{b_{j+1},\l_{j+1}, \lb_{j+1}\} \ldots \{b_{n},\eta_{n},\l_{n}(z_p^j), \lb_{n}(z_p^j)\}\right) \right],
\end{split}
\end{equation}
where $\l_1,\lb_1,\l_n,\lb_n$ are extended according to \eqref{conditionalbcfw} and $z_p^j,p_{\rm int}$ are defined by
\begin{equation}
\left(p_1(z_p^j) + \sum_{i=2}^j p_i\right)^2 = 0,  \qquad  p_{\rm int} = p_1(z_p^j) + \sum_{i=2}^j p_i.
\end{equation}

We can use these recursion relations to work our way down to the three point amplitude. The 3-pt amplitude is a defining dynamical object that we discuss below.

\subsection{3-pt amplitude}
With three particles, and real non-collinear on-shell momenta, it is impossible to satisfy momentum conservation; however, this can be done with complex momenta. The 3-pt on-shell amplitude with complex external momenta occurs in intermediate calculations when we use \eqref{n1n2recursion} to construct amplitudes. 

We give the on-shell amplitude for scattering within the vector multiplet. 
If the three momenta are $p_i^{\dot{\alpha} {\alpha}} = \l_i^{\alpha} \lb_i^{\dot{\alpha}}$, with $i = 1 \ldots 3$, we must have 
\[
\langle \l_i,\l_j \rangle = 0 \quad {\rm or}\quad  [\lb_i, \lb_j] = 0.
\]
Now, if all the $b_i$ are equal, the amplitude is zero. There are two physically distinct cases, $b_1 = b_2 = -1, b_3 = 1$ and $b_1=b_2=1,b_3=-1$. In combination with the two possibilities above, there are four distinct possibilities to be considered. We consider, these in turn.
\begin{enumerate}
\item
If $\dotlb[\lb_i, \lb_j] = 0$ then the amplitude is non-zero only if 
at least two of the $b$ are -1. Consider the case where $b_1=b_2=-1, b_3 = 1$ (other cases are related to this by cyclic symmetry). Using a supersymmetry transformation $e^{\langle \chi \cdot Q \rangle}$ with the parameter 
\begin{equation}
\chi = {\e_1 \l_2 - \e_2 \l_1 \over \langle \l_1, \l_2 \rangle},
\end{equation}
we can convert the amplitude to a 3-gluon amplitude. This leads to the result
\begin{equation}
A^{\rm t}\left(\{-1,\e_1 ,\l_1 ,\lb_1\}, \{-1,\e_2,\l_2,\lb_2\},\{1,\e_3,\l_3,\lb_3\}\right)
= {\langle \l_1,\l_2 \rangle^3 \over \langle \l_2, \l_3 \rangle \langle \l_3, \l_1 \rangle} e^{ \langle \chi  \cdot \eta_3, \l_{3}\rangle}.
\end{equation}
\item
If $\dotl[\l_i, \l_j] = 0$, the amplitude is non-zero only if two of the $b_i$ are 1. Say $b_1=b_2=1, b_3 = -1$. Then, using a supersymmetry transformation by $e^{[\chi \cdot Q]}$ with
\begin{equation}
\chi = {\lb_2 \e_1 - \lb_1 \e_2 \over [ \lb_1, \lb_2 ]},
\end{equation}
we find
\begin{equation}
A^{\rm t}\left(\{1,\e_1 ,\l_1 ,\lb_1\}, \{1,\e_2,\l_2,\lb_2\},\{-1,\e_3,\l_3,\lb_3\}\right)
= {[ \lb_1,\lb_2 ]^3 \over [ \lb_2, \lb_3 ] [ \lb_3, \lb_1 ]} e^{ [ \chi  \cdot \eta_3, \lb_{3}]}.
\end{equation}
\end{enumerate}

Using these three point amplitudes and the relations \eqref{n1n2recursion}, it is possible to calculate any on-shell tree amplitude in the pure $\cn=1$ and $\cn=2$ gauge theories. 

\subsection{Chiral Multiplets}
\label{chiralmultiplets}
The formalism for vector multiplets described in the paper can be 
generalized to include other multiplets as well.

For example, the chiral multiplet in the $\cn=1$ theory consists of an irreducible representation of the $\cn=1$ supersymmetry algebra with helicities $-{1 \over 2}, 0$ and another irreducible representation 
with helicities ${1 \over 2},0$. We choose to build these representations
on top of the two scalars that we denote by $s^-$ and $s^+$. This ensures
that the amplitudes are always c-numbers. These scalars
are conjugates of each other and the superscript tells us the helicity
of the fermion that the scalar is paired with.

We can now extend the definition of \eqref{statedef} to include chiral multiplets,
\begin{equation}
\label{chiralstatedef}
|\{b,\eta,\l,\lb\}\rangle \equiv \left\{\begin{array}{l}e^{\langle \eta \cdot Q, w \rangle } |s^{+},\l, \lb \rangle, \quad \text {if}~ b=0^{+}, \\ e^{[ \eta \cdot \Qb, \wb]} |s^{-},\l,\lb \rangle, \quad  \text{if}~ b=0^{-}, \end{array} \right.
\end{equation}
The formulas of the section above hold provided we adopt the 
convention $0^{+} < 0$ (note the counterintuitive sign) and $0^+ \equiv -0^{-}$.

The coupling of the chiral multiplet with the vector multiplet is determined
by the 3-pt amplitudes
\begin{equation}
\begin{split}
&{\mathcal A}(\{0^-, \e_1, \l_1, \lb_1,i\},\{0^+, \e_2, \l_2, \lb_2,j\}, \{-1, \e_3, \l_3, \lb_3,a\}) \\
&=  \sqrt{2}  e^{\dotl[\chi \e_1, \l_1]}  {\dotl[ \l_2, \l_3 ] \dotl[\l_3, \l_1 ] \over \dotl[ \l_1, \l_2 ]} T^{a}_{j i},\\
\end{split}
\end{equation}
with
\begin{equation}
\chi = {\e_2 \l_3 - \e_3 \l_2 \over \dotl[\l_2, \l_3]},
\end{equation}
and we have explicitly displayed the color of the gauge boson $a$, and the 
 indices $i,j$ associated with the chiral multiplet.

The other non-zero 3-pt amplitude is
\begin{equation}
\begin{split}
&{\mathcal A}(\{0^-, \e_1, \l_1, \lb_1,i\},\{0^+, \e_2, \l_2, \lb_2,j\}, \{1, \e_3, \l_3, \lb_3,a\}) \\
&=    \sqrt{2}  e^{\dotlb[\chi \e_2, \lb_2]} { \dotlb[ \lb_2, \lb_3 ] \dotlb[\lb_3, \lb_1 ] \over \dotlb[ \lb_1, \lb_2 ]} T^{a}_{j i},
\end{split}
\end{equation}
with 
\begin{equation}
\chi = {\e_1 \lb_3 - \e_3 \lb_1 \over \dotlb[\lb_1, \lb_3]}.
\end{equation}

\section{One-Loop: Pure ${\cn=1}$ and ${\cn=2}$ Theories}
\label{sec:oneloop}
In any quantum field theory, one-loop amplitudes can be efficiently reconstructed from tree-level amplitudes. This utilizes the analytic structure of the S-matrix at one-loop.\footnote{In this discussion, we should remember that ordinary massless gauge theories have both UV and IR divergences. We define the S-matrix by working in $4 + 2 \epsilon$ dimensions} This subject has a long history and we refer the reader to \cite{Bern:1991aq,Bern:1993mq,Bern:1994zx,Bern:1994cg,Bern:1995db}. After the revival of interest in on-shell techniques, this work has been extended \cite{Britto:2004nc,Bern:2005cq,Berger:2006ci,Forde:2007mi,ArkaniHamed:2008gz}. This process also lends itself to easy automation \cite{Berger:2008sj}.

Briefly, any one-loop amplitude can be written as a sum of scalar boxes, triangles and bubbles and a rational remainder. 
\begin{equation}
\label{decompositiononeloop}
\begin{split}
{\mathcal A}^{1\ell} &=  \sum_{{\alpha_4}} \int { A_{\alpha_{4}}\over \prod_{i = 0}^{3}\left[(p + q^{\alpha_{4}}_i)^2 + i \epsilon\right]} \, {d^{4+2\epsilon} p \over (2 \pi)^{4 + 2\epsilon}} \\
&+ \sum_{\alpha_{3}} \int { B_{\alpha_{3}} \over \prod_{i = 0}^{2}\left[(p + q^{\alpha_{3}}_i)^2+i \epsilon \right]} \, {d^{4+2\epsilon} p \over (2 \pi)^{4 + 2\epsilon}} \\
&+ \sum_{\alpha_{2}} \int { C_{\alpha_2} \over \prod_{i = 0}^{1}\left[(p + q_i^{\alpha_2})^2 + i \epsilon\right]} \, {d^{4+2\epsilon} p \over (2 \pi)^{4 + 2\epsilon}}\\
&+ {\mathcal R} + {\rm O}\left(\epsilon\right).
\end{split}
\end{equation}
Here $A,B,C,{\mathcal R}$ are rational functions of the external kinematic invariants.  
The index $\alpha_{n}$ runs over different partitions of the external momenta into $n$ sets.  
It serves to remind us that in the expansion of any amplitude, there are {\it several} distinct boxes, triangle and bubbles. We can always take $q_0^{\alpha_n}=0$; $q_{i+1}^{\alpha_n}-q_i^{\alpha_n}$ is the sum of the external momenta in a set of the partition (as in \eqref{triangleqdef} below). 
We emphasize that the expansion above is correct as a Laurent series in $\epsilon$ up to terms of ${\rm O}(\epsilon)$. We will review some features of this decomposition below but we refer the reader to \cite{Bern:2005cq,Berger:2006ci,Forde:2007mi,Berger:2008sj} for a description of how these coefficients can be determined, using just tree amplitudes, in any gauge theory. 

In the maximally supersymmetric $\cn=4$ SYM theory, the coefficients $B,C$ above are zero i.e. there are no triangles or bubbles in the expansion of a $\cn=4$ amplitude. However, the expansion of pure $\cn=1,2$ theories is very similar to that of ordinary gauge theories in this basis. All the terms in \eqref{decompositiononeloop} are present in a generic amplitude. Let us briefly 
discuss how these terms arise. Our treatment follows \cite{Forde:2007mi,ArkaniHamed:2008gz}. 

\subsection{Triangles}
\label{subsec:triangles}
To calculate triangle coefficients, we first partition the external momenta into three sets that we denote by
\begin{equation}
\label{trianglepartition}
\{k_{\pi_{1}} \ldots k_{\pi_{i_1}}\}, \{k_{\pi_{i_1 + 1}} \ldots k_{\pi_{i_2}}\}, \{k_{\pi_{i_2 + 1}} \ldots k_{\pi_n}\}.
\end{equation}
Different possible partitions are indexed by $\alpha_3$. 
We take $q_0 = 0$ and define
\begin{equation}
\label{triangleqdef}
q_1 = \sum_{i=1}^{i_1} k_{\pi_i},  \qquad  q_2 = q_1 +  \sum_{i=i_1+1}^{i_2} k_{\pi_i}.
\end{equation}
Putting 3 lines on shell in a loop does not freeze the internal momenta; instead it leaves us with one free parameter. We fix this parameter by introducing $w_1, w_2$ s.t $w_i \cdot q_j = 0; w_i \cdot w_j = - \delta_{i j}$ and  solving the equations 
\begin{equation}
\label{threeonshell}
p^2 = (p + q_1)^2 = (p + q_2)^2 = 0, \:  p \cdot w_1 = z.
\end{equation}
This forces $p \cdot w_2 = \pm \sqrt{r^2 - z^2}$, for some $r$. We will call these two solutions $p^{\pm}$.
We calculate the three-cut
\begin{equation}
\label{threecutexplicit}
\begin{split}
&{\mathcal C}^{\pm}_{\alpha_3}(z) =  (-1)^{\cn} \sum_{\gamma^i=-1, 1} \int 
\prod_{i=1}^3 d^{\cn} \, \mu_{1}^i \, d^{\cn} \, \mu_{-1}^i \: \exp\{\sum_{i=1}^3  \mu_1^{i} \cdot \mu_{-1}^{i}\}\\
& \left[ {\mathcal A}^{\mathrm t}\left(\{\gamma^{1}, \mu_{\gamma^1}^{1},p^{\pm}\},\{b^{\pi_{1}},\eta^{\pi_{1}},k_{\pi_{1}}\} \ldots \{b^{\pi_{i_1}},\eta^{\pi_{i_1}},k_{\pi_{i_1}}\},\{-\gamma^{2}, \mu_{-\gamma^2}^{2},-p^{\pm}-q_1\}\right) \right. \\
& \left. {\mathcal A}^{\mathrm t}\left(\{\gamma^{2}, \mu_{\gamma_2}^{2},p^{\pm}+q_1\},\{b^{\pi_{i_1 + 1}},\eta^{\pi_{i_1 + 1}},k_{\pi_{i_1 + 1}}\} \ldots \{b^{\pi_{i_2}},\eta^{\pi_{i_2}},k_{\pi_{i_2}}\},\{-\gamma^{3}, \mu_{-\gamma^3}^{3},-p^{\pm}-q_2\}\right)\right.\\
&\left. {\mathcal A}^{\mathrm t}\left(\{\gamma^{3}, \mu_{\gamma^3}^{3},p^{\pm}+q_2\},\{b^{\pi_{i_2  + 1 }},\eta^{\pi_{i_2  + 1 }},k_{\pi_{i+2  + 1 }}\} \ldots \{b^{\pi_{n}},\eta^{\pi_{n}},k_{\pi_{n}}\},\{-\gamma^{1}, \mu_{-\gamma^1}^{1},-p^{\pm}\}\right) \right],
\end{split}
\end{equation}
The triangle coefficient, corresponding to the partition \eqref{trianglepartition} is then found through
\begin{equation}
\label{findtrianglefromthreecut}
B_{\alpha_3}={1 \over 4 \pi } \oint_{z=\infty} \sum_{\pm} {\mathcal C}_{\alpha_3}^{\pm}(z) \, {dz \over \sqrt{r^2 - z^2}}.
\end{equation}
In the $\cn=4$ theory, ${\mathcal C}^{\pm}(z)$ vanishes as ${\rm O}\left({1 \over z}\right)$ for large $z$ \cite{ArkaniHamed:2008gz} and hence the triangle coefficients vanish. However,
we can see that for $\cn=1,2$ theories, this is not the case.

For example, consider the term in \eqref{threecutexplicit}  where $\gamma^1 = \gamma^2=\gamma^3$. For each of the tree amplitudes in \eqref{threecutexplicit}, two momenta are going large. However, this is {\em not} the extension described in \eqref{bcfwextension}. In \eqref{bcfwextension}, the two momenta that were extended belonged to particles within the same supersymmetry multiplet i.e. particles with the same value of $b$. Thus, generically, \eqref{threecutexplicit} will not lead to a contribution that dies off for large $z$. This, in turn, leads to a non-zero triangle coefficient.

\subsection{Bubbles}
\label{subsec:bubble}
To calculate the scalar bubble coefficients, we divide the external momenta into two sets,
\begin{equation}
\label{bubblepartition}
\{k_{\pi_{1}} \ldots k_{\pi_{i_1}}\},\{k_{\pi_{i_1 + 1}} \ldots k_{\pi_n}\},
\end{equation}
with different possible partitions indexed by $\alpha_2$. 
With $q_0 = 0$, we define
\begin{equation}
\label{qforapartition}
q_1 = \sum_{i=1}^{i_1} k_{\pi_i}.
\end{equation}

It is convenient to analytically continue the external momenta till $q_1$ is time-like (for a technique of avoiding this, see \cite{Raju:2009yx}). Then, the solutions to the equations
\begin{equation}
\label{bubbleonshell1}
p^2 = (p+q_1)^2 = 0,
\end{equation}
lie on a sphere.  We parametrize this dependence by introducing auxiliary vectors $w_1,w_2$ with $w_i \cdot q_1 = 0; \: w_i \cdot w_j = -\delta_{i j}$ and setting
\begin{equation}
\label{bubbleonshell2}
p \cdot w_1 = \sqrt{q_1^2 \over 4} \cos \theta, \quad p \cdot w_2 = \sqrt{q_1^2 \over 4} \sin \theta \cos \phi.
\end{equation}
We denote the solutions to \eqref{bubbleonshell1}, \eqref{bubbleonshell2} by $p(\theta, \phi)$. At any given value of $\theta,\phi$, we can BCFW extend $p$ and $p+q_1$ through a vector $q$, such that $q \cdot p(\theta,\phi) = q \cdot (p(\theta,\phi) + q_1) = 0$. With a complex parameter $z$, we define 
\begin{equation}
\tilde{p} = p(\theta,\phi)+q z.
\end{equation}
The two-cut is then calculated through\footnote{\label{dependencespinors}Each individual tree amplitude in \eqref{twocut} depends on the decomposition of the cut-momenta into spinors. We are free to choose a convenient decomposition for one of the amplitudes; the decomposition for the other is then fixed as shown in \eqref{sumstates}.},
\begin{equation}
\label{twocut}
\begin{split}
&{\mathcal C}^{\pm}(\theta, \phi, z) =  \sum_{\gamma^i=-1,1} \int \prod_{i=1}^2
 d^{\cn} \, \mu_{1}^i d^{\cn} \, \mu_{-1}^i \: \exp\{\sum_{i=1}^{2} \mu_{1}^{i}\cdot \mu_{-1}^{i} \} \\
& \left[ {\mathcal A}^{\mathrm t}\left(\{\gamma^{1}, \mu_{\gamma^1}^1,\tilde{p} \},\{b^{\pi_{1}},\eta^{\pi_{1}},k_{\pi_{1}}\} \ldots \{b^{\pi_{i_1}},\eta^{\pi_{i_1}},k_{\pi_{i_1}}\},\{-\gamma^{2}, \mu_{-\gamma^2}^{2},-\tilde{p}-q_1\}\right) \right. \\
& \left. {\mathcal A}^{\mathrm t}\left(\{\gamma^{2}, \mu_{\gamma^2}^{2},\tilde{p}+q_1\},\{b^{\pi_{i_1 + 1}},\eta^{\pi_{i_1 + 1}},k_{\pi_{i_1 + 1}}\} \ldots \{b^{\pi_{n}},\eta^{\pi_{n}},k_{\pi_{n}}\},\{-\gamma^{1}, \mu_{-\gamma^1}^{1},-\tilde{p}\}\right)\right].
\end{split}
\end{equation}
The bubble coefficient, corresponding to the partition \eqref{bubblepartition} is given by
\begin{equation}
\label{bubbleint}
C_{\alpha_2} = \int \oint_{z = \infty} {\mathcal C}_{\alpha_2}(\theta,\phi,z) \, {d z \over 2 \pi i z}\, d \left(\cos \theta\right) \, {d\phi \over 4 \pi}.
\end{equation}
Once again, when $\gamma^1=\gamma^2$, two momenta are going large in each tree amplitude. However, our analysis of section \ref{sec:recrelation} tells us that this product of tree amplitudes does not die off at large $z$. Thus, generically, we expect pure $\cn=1$ and $\cn=2$ theories to contain bubbles in their one-loop expansion.

In fact this is clear from another perspective. In the expansion \eqref{decompositiononeloop}, bubbles are the only ultra-violet divergent diagrams. Since pure $\cn=1$ and $\cn=2$ theories do have UV-divergences at one-loop, they must contain bubble diagrams in their one-loop expansion. We discuss this in more detail below for the more interesting case of superconformal field theories.

\section{One Loop: Gauge Theories with Arbitrary Matter}
\label{sec:matter}
In the section above, we found that the structure of the tree-level and one-loop S-matrix in pure ${\cn=1}$ and ${\cn=2}$ gauge theories resembled that of pure Yang-Mills theory. In this section, we discuss the effect of the addition of matter (both fermionic and bosonic), in arbitrary representations, on the S-matrix.For simplicity, we will restrict ourselves to gluon scattering in such theories. 

The notion of a `color-ordered amplitude' is not very useful if we wish to 
consider gauge theories with matter in arbitrary representations. So, in this section, we will always deal with the full-amplitude, including all color-trace factors.

Tree-level gluon amplitudes in supersymmetric theories are unchanged by the addition of matter. At one-loop, with external gluons, we can write the full amplitude as
\begin{equation}
\label{oneloopmatterdecompose}
{\cal A}^{1 \ell} = {\cal A}^{1 \ell}_{\mathrm g} + {\cal A}^{1 \ell}_{\mathrm f}  + {\cal A}^{1 \ell}_{\mathrm s},
\end{equation}
where  ${\cal A}^{1 \ell}_{\mathrm g/f/s}$ denotes the contribution to the amplitude with only {\bf g}auge bosons, {\bf f}ermions or {\bf s}calars running in the loop. The pure gauge amplitude is given by ${\cal A}^{1 \ell}_{\mathrm g}$; we turn to a study of the other terms in \eqref{oneloopmatterdecompose}. 

At first sight the contribution of matter to the one-loop S-matrix may seem
horribly complicated. After all a one-loop diagram can involve an indefinite
number of gluon-matter interactions. This might suggest that there is no
simple way to capture the contribution of matter to triangles and bubbles.

On the other hand, bubbles are related
to the one-loop $\beta$ function. Matter in a given representation provides a well known universal contribution to the $\beta$ function that is dependent on the 
quadratic Index of the matter representation and nothing else! How 
does this remarkable result come out of an S-matrix analysis?

It turns out the true state of affairs is in between these extremes. 
Matter in a given representation provides a contribution to individual
bubble coefficients that depends, not only on the quadratic Index
but also on the 4$^{\rm th}$ order Indices. For any given bubble, the coefficients
of these Indices depend on the external states but not on the 
representation. The story for triangles is similar. The contribution of 
matter to a  triangle can depend on the higher Indices up to order {\em six}. These Indices are multiplied by rational coefficients
and when we change the representation, the Indices change but their coefficients
don't! 

For supersymmetric theories these statements must be modified. It turns
out that when we consider a $\cn=1$ chiral multiplet, the contribution of
fermions to the 4$^{\rm th}$ order Indices in bubbles exactly cancels the contribution
of scalars. Similarly, for triangles the two contributions to the 6$^{\rm th}$
order Indices cancel. So, a chiral multiplet contributes only via its quadratic
Index to bubbles and via the quadratic, 4$^{\rm th}$ and 5$^{\rm th}$ order  Indices to triangles.

It is interesting to understand the relation of these statements 
to the result about the $\beta$ function. The $\beta$ function is proportional to the sum of several bubble coefficients. Individually,
these bubble coefficients are not proportional to the tree amplitude 
and, in a non-supersymmetric theory, depend on the 4$^{\rm th}$ order Indices of the representation as well. 
However, when we add them all up,  we find a result that
just depends on the product of the tree-amplitude and the quadratic  
Index --- everything else cancels!

We discuss these results, first for bubbles and then for triangles. We start
by discussing scalars and then generalize our results to fermions and supersymmetric matter. 

\subsection{Many gluons and 2 matter particles}
\label{subsec:manygluon2matter}
We see from our analysis of the previous section that to understand the 
contribution of matter to bubbles and triangles in gluon scattering amplitudes,
we need to consider the tree amplitude with many gluons and 2 BCFW extended
matter particles. This amplitude is schematically shown in figure \ref{manygluon2matter}. $K = \sum k_i$ is the total momentum of the gluons. $P$ does
not diverge for large $z$ but may otherwise depend on $z$ as it, in fact, does when we consider the cut that contributes to a triangle. Since, $P + q z$ is on shell, $P \cdot q = \o[{1 \over z}]$ at large $z$.
\FIGURE{
\label{manygluon2matter}
\epsfig{file=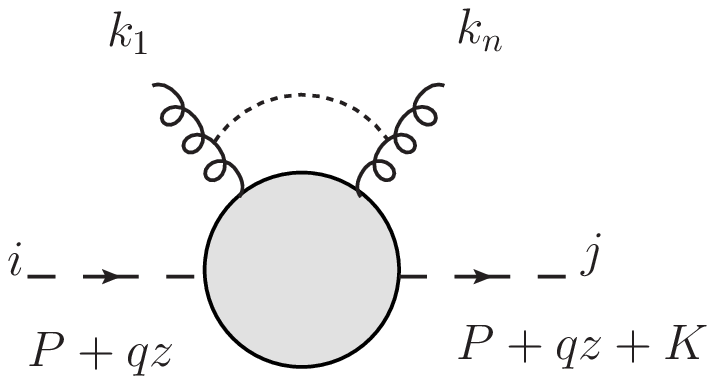, width=0.3\textwidth}\caption{Two BCFW extended matter particles and many gluons}}

Our first aim is to examine the possible color-structures that can emerge 
in the amplitude shown in figure \ref{manygluon2matter}. We will perform
our analysis for scalars first and then generalize it to include fermions.

 Our treatment here is similar to \cite{ArkaniHamed:2008yf}. To study the large $z$ behavior of the tree amplitudes in figure \ref{manygluon2matter}, we adopt the space-cone gauge of \cite{Chalmers:1998jb} that was used to analyze QCD amplitudes in \cite{Vaman:2005dt}. This was called the q-lightcone gauge in \cite{ArkaniHamed:2008yf}. We choose a gauge, so that the gauge field satisfies
\begin{equation}
\label{qlightgauge}
q \cdot A(l) = 0, \quad \text{for~} q \cdot l \neq 0.
\end{equation}
If we also choose the gauge-boson propagator to satisfy $q_{\mu} \Pi^{\mu \nu}(l) = 0$, for $q \cdot l \neq 0$ then, at large $z$, the tree-amplitude in Fig \ref{manygluon2matter} is dominated
by diagrams that have a small number of gluon-scalar vertices. This is because 
each scalar propagator comes with a factor of ${\rm O}\left({1 \over z}\right)$,
while factors of $z$ in the numerator can only come from interactions with 
 a gluon line that carries the momentum $K$.

\FIGURE{
\epsfig{file=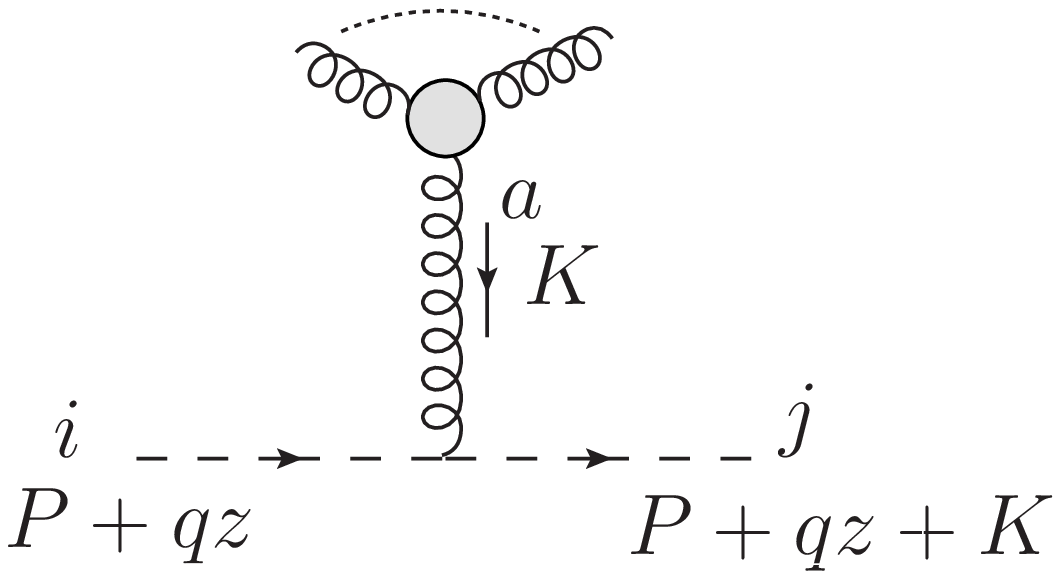,width=0.3\textwidth}\caption{Dominant diagram at large z}\label{dominantlargez}
}
Hence, in this gauge, the leading contribution to the tree amplitude in figure \ref{manygluon2matter} is given by the diagram shown in Fig \ref{dominantlargez}
This leads to a contribution
\[
 i \left[(2P+ K +2qz)\cdot A^a(K)\right]T^a_{ji}.
\]

In this and other diagrams below, the gluon lines that interact with the scalar  could come from external gluons. 
Alternately, they could come from a gluon propagator that connects to the
 external gluons through the blobs shown in the diagrams. 
Since these details are not important for counting powers of $z$, we have 
parametrized the interactions of the scalars with the gluon lines through
background vectors $A^a(l_i)$ carrying momentum $l_i$ and the color index $a$.

\FIGURE{
\epsfig{file=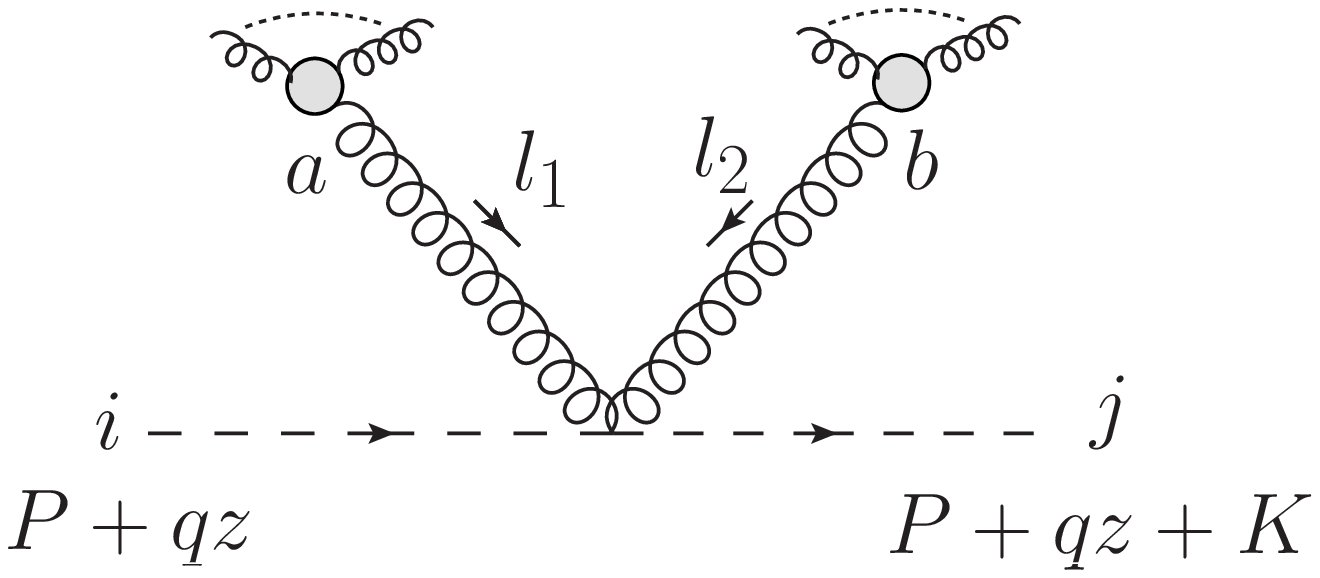,width=0.3\textwidth}\caption{Contact interaction at $\o[z^0]$}
}
Proceeding to $\o[z^0]$, we find that the contact interaction can also contribute. 
This leads to a contribution
\[
i \left[A^a(l_1)\cdot A^b(l_2)\right] \{T^a,T^b\}_{ji}.
\]

In fact, by drawing a few more such diagrams one quickly finds a pattern
emerging. Each diagram gives us a product of generators. Using the commutation
relations of the algebra, we can express this product in terms of completely
symmetrized products and the structure constants. We find that the
leading contribution to the symmetrized product of $p$ generators comes with the power $z^{2 - p}$. More precisely, an amplitude with $n$ gluons and two scalars --- $s^-$ carrying color index $i$ and $s^+$ carrying color index $j$ --- as shown in figure \ref{manygluon2matter} is of the form
\begin{equation}
\label{scalarpowers}
{\mathcal A}(s^-, s^+, \ldots) = \sum_{p=1}^n {c_{a_1 \ldots a_p} \over z^{p-2}} \left[T^{(a_1} \ldots T^{a_p)}\right]_{j i}.
\end{equation}
The $c$'s are rational functions of the external kinematic 
invariants, with the property that they do not diverge at large $z$, and are {\em independent of the representation} in which the scalars transform.
Here, we provide a formal proof of this statement. The reader who 
prefers diagrams in q-lightcone gauge is referred to appendix \ref{explicitchecks}
for further explicit calculations.

We will prove the statement \eqref{scalarpowers} via induction. Note, that with one gluon the amplitude is of the form above.  Now, assume that the form \eqref{scalarpowers} is true for amplitudes with up to $n$ gluons. We will show that this implies that it is true for an amplitude with $n+1$ gluons.

\FIGURE{
\label{fourterms}
\epsfig{file=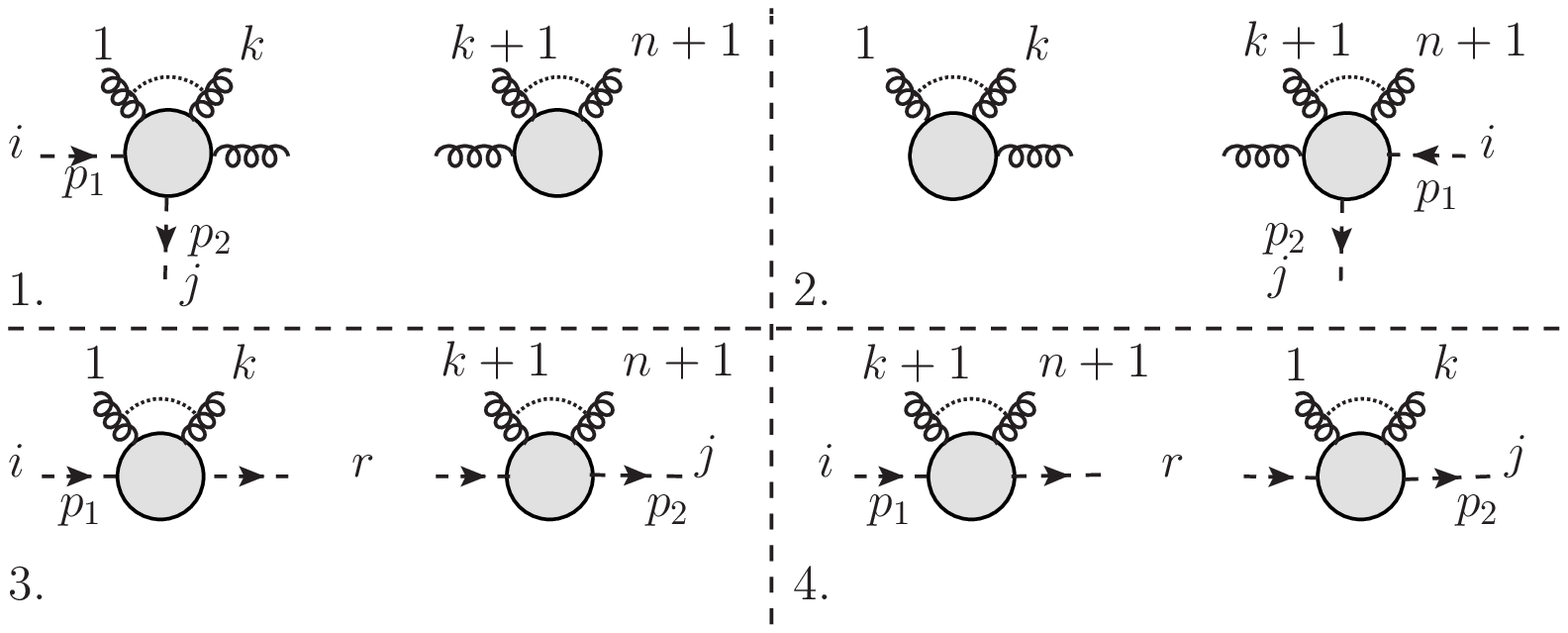}
\caption{Four terms in the recursion relation}
}
The amplitude with $n+1$ gluons can be calculated via BCFW recursion. Let us extend the 1$^{\rm st}$ and (n+1)$^{\rm th}$ gluon. So, we consider the BCFW extension $k_1 \rightarrow k_1 + t w, k_{n+1} \rightarrow k_{n+1} - t w$, as a function of the parameter $w$ (separate from $z$). We now find 4 types of terms. These are shown in figure \ref{fourterms} (where $p_1 \equiv P + q z, p_2 \equiv P + q z + K$).
\begin{itemize}
\item
Terms where the last gluon is paired with $n-k$ other gluons. The intermediate particle is a gluon and the other gluons and {\it both} the scalars are paired with the first  gluon.
\item
Terms where the first gluon is paired with $k-1$ other gluons. The intermediate particle is a gluon and the other gluons and {\it both} the scalars are paired with the last gluon.
\item
Terms where the $s^-$ is paired with the first gluon and $k-1$ other gluons and the intermediate cut particle is a scalar.
\item
Terms where the $s^-$ is paired with the last gluon and $n-k$ other gluons and the intermediate cut particle is a scalar.
\end{itemize}
The first two kinds of terms in the list above, clearly maintain the form \ref{scalarpowers}. Consider the third term. If we club the first gluon with $k-1$ other gluons and $s^-$, we find a pole when $w$ takes the value $w_0 = -{q \cdot (\sum_{i=1}^k k_i) \over q \cdot t} + \o[{1 \over z}]$.  So,
the third term gives us a contribution to the amplitude of the form
\begin{equation}
{1 + \o[{1 \over z}]\over 2 z q \cdot \sum_{i=1}^k k_i} \left[\sum_{p = 1}^k {c^1_{a_1 \ldots a_p} \over z^{p - 2}} T^{(a_1} \ldots T^{a_p)}_{r i} \sum_{m = 1}^{n+1-k} {c^2_{a_1 \ldots a_m} \over z^{m - 2}} T^{(a_1} \ldots T^{a_m)}_{j r}\right].
\end{equation}
To the left, we have the propagator factor as in \eqref{bcfwrecursion} and in the square bracket we have the product of the two amplitudes that, by hypothesis, 
have the form \eqref{scalarpowers}.
Now, the $c$'s in the expression above are rational functions of the momentum associated with $s^-$ --- $P + q z$ --- and the spinors associated
with the gluons. More precisely, the $c$'s can depend
on dot products of $P+q z$ with the external polarization vectors
and partial sums of the momenta that appear
in the amplitude.
Each of these terms is $\o[z]$ and its dependence on $P$ is subleading. The
only term we need to worry about is $(P + q z) \cdot (P + q z + K')$ where $K' = \sum_{i = 1}^k k_i + t w_0$.  However, we have
\begin{equation}
(P + q z) \cdot (P + q z + K') = (P + q z) \cdot K' = -{(K')^2 \over 2},
\end{equation}
which is also independent of $P$ to leading order. In particular, this means that to {\em leading order} in $z$, the coefficients $c^1, c^2$ are {\em independent of $P$}.
This will be important for us below. 

The fourth term involves a pole at 
$w_1 = -{q \cdot (\sum_{i=1}^k k_i) \over q \dot t} + \o[{1 \over z}]$. So $w_1 - w_0 = \o[{1 \over z}]$. This amplitude now gives us a term of the form
\begin{equation}
{-1 + \o[{1 \over z}]\over 2 z q \cdot \sum_{i=1}^k k_i} \left[\sum_{p = 1}^k {c^3_{a_1 \ldots a_p} \over z^{p - 2}} T^{(a_1} \ldots T^{a_p)}_{j r} \sum_{m = 1}^{n+1-k} {c^4_{a_1 \ldots a_m} \over z^{m - 2}} T^{(a_1} \ldots T^{a_m)}_{r i}\right].
\end{equation}
However, the coefficients $c^i$ must satisfy
\begin{equation}
c^1 - c^3 = \o[{1 \over z}] \quad c^2 - c^4 = \o[{1 \over z}].
\end{equation}
This is because they are both associated with the amplitude of a BCFW extended scalar shooting through a set of gluons. Up to terms of $\o[{1 \over z}]$, the gluons have the same momentum in two two cases since $w_1 = w_0 + \o[{1 \over z}]$. The subleading terms in the scalar momenta differ but as argued above, these only affect the $\o[{1 \over z}]$ terms in the $c$'s which are rational functions of the external 
scalar momenta.

This means that the amplitude with $n+1$ gluons is of the form,
\begin{equation}
\label{bcfwresult}
\begin{split}
&\sum_{p = 1}^k \sum_{m=1}^{n-k+1}{1 \over 2 q \cdot \sum_{i=1}^k k_i}\left[ {c^1_{a_1 \ldots a_p} c^2_{a_1 \ldots a_m} + \o[{1 \over z}] \over z^{p + m - 3}} [T^{(a_1} \ldots T^{a_m)}, T^{(a_1} \ldots T^{a_p)}]_{j i}\right] \\
&+ \o[{1 \over z^{p + m - 2}}]  \{T^{(a_1} \ldots T^{a_p)}, T^{(a_1} \ldots T^{a_m)}\}_{j i}.
\end{split}
\end{equation}
However, the commutator in \eqref{bcfwresult} can be expressed in terms of a symmetrized product of at most $p+m-1$ generators. This proves our result.

We can prove a very similar result for fermions. The only subtlety here
is that the fermionic amplitude is a rational function, not of the external
fermionic momenta, but of the spinors that we associate to the two matter lines in figure \ref{manygluon2matter}. We are interested in the case where this amplitude
comes from cutting a loop diagram. However, as we mentioned in footnote
\ref{dependencespinors}, in cutting a loop 
line we can choose to decompose the cut momenta $p$ into a convenient 
pair of spinors $p_{\alpha \dot{\alpha}} = \l_{\alpha} \lb_{\dot{\alpha}}$ 
provided that when we encounter $-p$ in another tree-amplitude
associated with the cut, we maintain consistency with \eqref{sumstates} by
decomposing it as $-p = (-\l_{\alpha}) \lb_{\dot{\alpha}}$. 

We use this freedom to make the choice that we decompose the
two  large BCFW momenta in figure \ref{manygluon2matter} as
\begin{equation}
\label{largezspinorchoice}
\begin{split}
(P + q z)_{\a \ad} &=  (\l z + \l_s^1)_{\a} (\lb + {\lb_s^1 \over z})_{\ad}   \\
-(P + q z + K)_{\a \ad} &=  (-\l z - \l_s^2)_{\a} (\lb + {\lb_s^2 \over z})_{\ad}.
\end{split}
\end{equation}
Here, $\l_s^i, \lb_s^i$ are some subleading terms that can depend on $z$
but must remain finite as $z \rightarrow \infty$. 
With the sign convention described above, this decomposition
 can be consistently
imposed on all the tree-amplitudes that appear in the calculation of a bubble
or a triangle coefficient.

Note that once we specify $\l_s^1, \lb_s^1$  then $\l_s^2$ and $\lb_s^2$
are completely fixed by the external gluon momenta. By an extension of the argument given for scalars above, the dependence of $c$ on  
$\l_s^1, \lb_s^1$ is subleading in $z$. 
With this observation, it is easy to repeat the proof above and show
that subject to the decomposition \eqref{largezspinorchoice}, fermionic
amplitudes also obey \eqref{scalarpowers}.

\subsection{Contribution of Matter to Bubbles}
\label{subsec:contribmatterbub}
We now have all the tools in hand to analyze the contribution of matter
to individual bubble coefficients.
We recall from section \ref{subsec:bubble} that a bubble coefficient is calculated by making a cut and taking the product of the two tree-amplitudes,
\begin{equation}
\label{bubbleintreeproduct}
C = \int d \Omega \oint_{z = \infty} {d z \over 2 \pi i z} \sum {\mathcal A}^{\rm t}_{\rm left} (p_1 + q z, -p_2 - q z, \ldots) {\mathcal A}^{\rm t}_{\rm right}(-p_1 - q z, p_2 + q z, \ldots).
\end{equation}
Here $p_1,p_2$ denote the two cut momenta and  $q \cdot p_1 = q \cdot p_2 = 0$. The sum is over the set of intermediate cut-particles (as, for example, in \eqref{twocut}) and the integral is over the phase-space associated with the cut.\footnote{Fermions contribute with a minus sign in \eqref{bubbleintreeproduct} because of the minus sign
associated with a closed fermion loop. In a supersymmetric theory, if we use 
the manifestly supersymmetric formalism described in the previous sections, this 
is automatic; otherwise we need to impose it by hand.}

From our analysis above, both ${\mathcal A}^{\rm t}_{\rm left}$ and ${\mathcal A}^{\rm t}_{\rm right}$ have the form of \eqref{scalarpowers}. Hence the contribution of scalars or fermions
to the bubble
coefficient is given by
\begin{equation}
\label{bubblexpandinc}
\begin{split}
C = (-1)^F \int d \Omega \oint_{z = \infty} {d z \over 2 \pi i z} 
&\left[ z^2 c^{l}_{a_1} c^{r}_{a_2} \tr(T^{a_1} T^{a_2}) \right. \\
&+ \left.\left(c^{l}_{a_1 a_2} c^{r}_{a_3 a_4} 
+ c^{r}_{a_1 a_2} c^{l}_{a_3 a_4} \right) \tr(T^{(a_1} T^{a_2)} T^{(a_3} T^{a_4)}) \right. \\
&+ \left. \left(c^{l}_{a_1 a_2 a_3} c^{r}_{a_4} 
+ c^{r}_{a_1 a_2 a_3} c^{r}_{a_4} \right) \tr(T^{(a_1} T^{a_2} T^{a_3)} T^{a_4})\right],
\end{split}
\end{equation}
where $c^l$ and $c^r$ are associated with  ${\mathcal A}^{\rm t}_{\rm left}$
and  ${\mathcal A}^{\rm t}_{\rm right}$ respectively, in the expansion
\eqref{scalarpowers}. 

We can interchange the partially symmetrized traces in 
\eqref{bubblexpandinc} for completely symmetrized traces 
at the cost of some structure constants. Hence, after 
doing the integral over $z$ and the phase space in \eqref{bubblexpandinc},
we find that the contribution of scalars/fermions in the representation $R_s/R_f$ can be written as
\begin{equation}
\label{bubbleintermsoftraces}
C_{s/f} = \sum_{n=2,4} \omega_{a_1 \ldots a_n} \tr_{R_s/R_f}(T^{(a_1} \ldots T^{a_n)}),
\end{equation}
where the $\omega$ are some coefficients that are independent of the representations $R_s/R_f$.

If we have a complex scalar running in the loop, \eqref{bubbleintreeproduct} tells us that the loop receives a separate contribution from its complex 
conjugate. This contribution is the same, in every respect,
except that we must modify the generators by $T^a \rightarrow -(T^a)^T$. To 
keep this manifest at every step we will adopt the slightly unusual 
convention that whenever we write $\tr_{R_s}$, we count 
a complex scalar and its conjugate as separate representations in $R_s$. 
This means, for example, that a scalar transforming in the fundamental
representation of $SU(N)$ will be represented using $R_s = {\bf N} + \overline{\bf N}$. This convention also allows us to treat complex and real (or pseudoreal) representations uniformly later. Moreover, it is immediately clear that the symmetrized product of three generators
never appears in \eqref{bubbleintermsoftraces} above since $R_s$ must be self-conjugate. With chiral fermions, the odd symmetrized products do not
vanish automatically. However, the symmetrized trace of 
three generators in $R_f$ must vanish for the theory to be anomaly free; so we
have excluded this contribution from \eqref{bubbleintermsoftraces}.

\subsubsection{A small detour}
\label{subsec:detourintoindices}
This expression can be put in a slightly nicer form that reveals its dependence
on various invariants of the algebra. We will take a small detour into 
group-theory to do this. So far our discussion has been valid for semi-simple 
groups. In this section, we focus on simple groups. Recall, that 
for any representation $R$,
\begin{equation}
\tr_R(T^a T^b) = I_2(R) \kappa^{a b},
\end{equation}
where $\kappa^{a b}$ is the Killing form. The coefficient $I_2(R)$ is called the (quadratic) Index and is commonly encountered in the 
calculation of the one-loop $\beta$ function. Similarly, for $SU(N)$ it is conventional to define
\begin{equation}
{1 \over 2} \tr_F\left(T^a \{T^b T^{c}\}\right) \equiv d^{a b c},
\end{equation}
in the fundamental representation. Then, for any representation $R$
\begin{equation}
{1 \over 2} \tr_R\left(T^{a} \{T^b,T^c\}\right) = I_3 (R) d^{a b c}.
\end{equation}
$I_3(R)$ --- the third-order Index --- is commonly called the anomaly. 

In exactly the same way, for the fourth order trace we have \cite{okubo:8}\footnote{For the exceptional algebras, $A_1$ (SU(2)) and $A_2$ (SU(3)), $d^{a b c d}$ can be taken to be zero. For $D_4$ (SO(8)), there is an additional independent fourth order tensor. We refer the reader to \cite{okubo:8} for a discussion} 
\begin{equation}
\tr_R\left(T^{(a} T^b T^{c} T^{d)}\right) = I_4(R) d^{a b c d} + I_{2,2}(R) \kappa^{(a b} \kappa^{c d)},
\end{equation}
where $d^{a b c d}$ are some suitably defined completely symmetric reference
tensors. The coefficients $I_4(R)$ and $I_{2,2}(R)$ are called the {\em fourth order Indices} of the representation. We refer the reader to appendix \ref{app:solveprtrace} for further group theoretic details.

Substituting this expansion into \eqref{bubbleintermsoftraces}, we find
that we can write
\begin{equation}
\label{bubbleintermsofindices}
C = I_2(R) \omega_{a_1 a_2} \kappa^{a_1 a_2}
+ I_4(R) \omega_{a_1 a_2 a_3 a_4} d^{a_1 a_2 a_3 a_4} +  I_{2,2} (R) \omega_{a_1 a_2 a_3 a_4} \kappa^{(a_1 a_2}\kappa^{a_3 a_4)}.
\end{equation}
In this equation all the dependence on the representation $R$ has been captured
by the three indices $I_2, I_4, I_{2,2}$. Hence, the contribution 
of matter in a given representation to individual bubble coefficients 
can depend on up to the fourth order Indices of the representation!

This is related to the result that matter in a representation $R$ contributes
to the one-loop $\beta$ function through the second order Index. The only
ultra-violet divergent diagrams in \eqref{decompositiononeloop} are the bubbles. However, any one-loop amplitude receives contributions from several distinct bubble coefficients. Each bubble coefficient corresponds to a partition of the external momenta as in \eqref{bubblepartition}. With $q_{\alpha_2}$ defined in \eqref{qforapartition} for a given partition indexed by $\alpha_2$, the ultra-violet divergent piece of a one-loop amplitude is
\begin{equation}
\label{bubblemultuv}
{\mathcal A}^{1 l}_{UV} = \sum {C_{\alpha_2} \over 32 \pi^2} \ln {\Lambda^2 \over q_{\alpha_2}^2}.
\end{equation}
While each $C_{\alpha_2}$ is individually important for the one-loop amplitude,
the total UV-divergence depends only on the sum $\sum_{\alpha_2} C_{\alpha_2}$. In fact
 this sum must be proportional to the tree amplitude 
for scattering of the external gluons \cite{ArkaniHamed:2008gz}.

Hence, while individual bubble coefficients depend on the higher order 
Indices and are not necessarily proportional to the tree-amplitude the 
sum of all bubble coefficients is proportional to the product of the tree
amplitude and the second order Index! In fact, the extra UV divergences
that appear in 
individual bubble
coefficients are a bit like the `spurious poles' \cite{Hodges:2009hk} 
that appear in tree amplitudes; they cancel when we add all contributing
terms. We verify this in section \ref{sec:examples} but it would be nice to have a 
direct proof of this fact.

What about scalars in the adjoint representation? Is it true that, at least, for this kind of matter the bubble coefficient is proportional to the tree-amplitude? From our analysis in this subsection, we cannot answer this question. However, our explicit calculations in section \ref{subsec:22scattering} will show us that, even with adjoint matter, the contribution of matter to a bubble coefficient is not proportional to the tree-amplitude. Miraculously, when we add up all bubble coefficients, the ultra-violet divergent term at one-loop is proportional to the tree-amplitude! 
\subsection{Contribution of Matter to Triangles}

Now, we consider the contribution of matter to triangles. Recall, from section \ref{subsec:triangles} that the triangle coefficient is calculated by making a 3-cut and calculating the product of three tree-amplitudes
\begin{equation}
\label{triangleinttreeproduct}
B =  \sum_{\pm} \oint_{z = \infty} {d z \over 4 \pi \sqrt{r^2 - z^2}} A^{\rm t}(p_1^{\pm}(z), -p_2^{\pm}(z), \ldots) A^{\rm t}(-p_1^{\pm}(z), p_3^{\pm}(z), \ldots) A^{\rm t}(p_2^{\pm}(z), -p_3^{\pm}(z), \ldots).
\end{equation}
Here, as described in more detail in section \ref{subsec:triangles}, $p_i^{\pm}(z)$ are null-vectors and for large $z$, $p_i^{\pm}(z) \rightarrow q^{\pm} z$ where $q^{\pm}$ are also null.

We can see from a simple extension of the analysis that we performed above that
for the contributions of scalars in a representation $R_s$, or of fermions
in a representation $R_f$, to a triangle coefficient we can write
\begin{equation}
\label{triangleintermsoftraces}
B_{s/f} = \sum_{n=2,4,5,6} \omega_{a_1 \ldots a_n} \tr_{R_s/R_f}(T^{(a_1} \ldots T^{a_n)}).
\end{equation}
Hence, the contribution to matter to triangles can involve up to the sixth order
Indices of the matter representation.

\subsection{Simplifications in Supersymmetric Theories}
\label{subsec:simplesusy}
We now describe a remarkable simplification that occurs in supersymmetric
theories.
In theories, with at least $\cn=1$ supersymmetry,
 in each individual bubble coefficient,
 the contribution of fermions to  
the fourth order Indices exactly cancels the contribution of scalars. 
Similarly, in each individual triangle coefficient, the contribution
of fermions to the sixth order Indices exactly cancels the contribution
of scalars.

This means that for supersymmetric theories, each individual
bubble and triangle coefficient can be written as
\begin{equation}
\label{susytraceexpansion}
C = \omega^{C}_{a_1 a_2} I_2(R) \kappa^{a_1 a_2}, \quad
B = \sum_{n=2,4,5} \omega^{B}_{a_1 \ldots a_n} \tr_{R}(T^{(a_1} \ldots T^{a_n)}).
\end{equation}
If the chiral multiplets are in self-conjugate representations (this is automatic if they are part of ${\cn=2}$ hypermultiplets), then the term with $n=5$ is also absent in \eqref{susytraceexpansion}.

This happens because, subject to the scaling choice \eqref{largezspinorchoice},
the coefficients for fermions and scalars in the expansion \eqref{scalarpowers} satisfy
\begin{equation}
\label{susycsame}
c^s - c^f = \o[{1 \over z}].
\end{equation}
We sketch a proof of this using supersymmetry. This proof is completed
in appendix \ref{explicitchecks} which also contains some explicit calculations 
that verify \eqref{susycsame} for the coefficients of symmetrized products 
involving up to 3 generators.

First, notice that 
we can embed the amplitude of figure \ref{manygluon2matter}
in a supersymmetric theory. Since all our external particles are gluons, 
the gauginos are just spectators. With the supersymmetric 
version of the chiral multiplet introduced in section \ref{chiralmultiplets},
we can write this amplitude as
\begin{equation}
\label{susymattermanygluons}
{\mathcal A}^{\rm t}(\{0^-, \eta_1, \l_1(z), \lb_1(z)\}, \{0^+, \eta_2,  \l_2(z), \lb_2(z)\} \ldots \{b_i, 0, \l_i, \lb_i\}, \ldots), 
\end{equation}
where  the $b_i=\pm 1$ represent gluons with either negative or positive helicity. The spinors $\l_1, \lb_1$ and $\l_2, \lb_2$ are chosen according to
 \eqref{largezspinorchoice}.

Now, consider a supersymmetry transformation $e^{\dotl[Q,\chi]}$, with 
\begin{equation}
\chi = {\eta_2 w \over z}, \quad \langle w, {\l_2 \over z} \rangle = 1.
\end{equation}
$w$ is not uniquely defined above since we can add any multiple of $\l_2$ to $w$ but this ambiguity will not affect us below provided we ensure that for large $z$, $w \rightarrow \o[1]$. This implies $\chi \rightarrow \o[{1 \over z}].$

Demanding that the amplitude in \eqref{susymattermanygluons} 
be invariant under this supersymmetry transformation 
leads to 
\begin{equation}
\label{susysteps1}
\begin{split}
&{\mathcal A}^{\rm t}(\{0^-, \eta_1, \l_1(z), \lb_1(z)\}, \{0^+, \eta_2,  \l_2(z), \lb_2(z)\} \ldots \{b_i, 0, \l_i, \lb_i\}, \ldots)\\
&= e^{\dotl[Q, \chi]} {\mathcal A}^{\rm t}(\{0^-, \eta_1, \l_1(z), \lb_1(z)\}, \{0^+, \eta_2,  \l_2(z), \lb_2(z)\} \ldots \{b_i, 0, \l_i, \lb_i\}, \ldots)\\
&= e^{\phi} {\mathcal A}^{\rm t}(\{0^-, \eta_1, \l_1(z), \lb_1(z)\}, \{0^+, 0,  \l_2(z), \lb_2(z)\} \ldots \{b_i, \eta_i', \l_i, \lb_i\}, \ldots),
\end{split}
\end{equation}
where 
\begin{equation}
\phi = {\dotl[\l_1, w] \over z} \eta_2 \eta_1 = \eta_2 \eta_1 \left(1 + \o[{1 \over z}]\right), \quad  \eta_i' = -\eta_2 {\dotl[ w, \l_i] \over z},~{\rm if}~ b_i = -1.
\end{equation}
We can now 
series expand the last line of \eqref{susysteps1} in the $\eta_i'$. The leading
term in this expansion has only gluons and no gauginos. Since, particle $2$
is now a scalar and since there are no external gauginos, particle $1$ must
also be a scalar. 
The term proportional to $\eta_i'$  involves a gaugino
in position $i$ and leads to an amplitude involving 
a chiral multiplet scalar, a negative helicity chiral multiplet fermion 
and a negative helicity vector multiplet gaugino. However, this amplitude is suppressed by ${1 \over z}$!
\begin{equation}
\begin{split}
&{\mathcal A}^{\rm t}(\{0^-, \eta_1, \l_1(z), \lb_1(z)\}, \{0^+, \eta_2,  \l_2(z), \lb_2(z)\} \ldots \{b_i, 0, \l_i, \lb_i\}, \ldots) \\
&= \left[1 + \eta_2 \eta_1\left(1 + \o[{1 \over z}]\right)\right] {\mathcal A}^{\rm t}(\{0^-, 0, \l_1(z), \lb_1(z)\}, \{0^+, \eta_2,  \l_2(z), \lb_2(z)\} \ldots \{b_i,\eta_i', \l_i, \lb_i\}, \ldots) \\
&=  \left[1 + \eta_2 \eta_1\left(1 + \o[{1 \over z}]\right)\right] {\mathcal A}^{\rm t}(\{0^-, 0, \l_1(z), \lb_1(z)\}, \{0^+, 0,  \l_2(z), \lb_2(z)\} \ldots \{b_i,0, \l_i, \lb_i\}, \ldots) \\ &+ {R \over z}.
\end{split}
\end{equation}
If $b_3 = -1$, then $R$ includes a term
 \[
 {\eta_2 \eta_1 \over z} {\mathcal A}^{\rm t}(\{f^-, \l_1(z), \lb_1(z)\}, \{s^+,  \l_2(z), \lb_2(z)\} ,\{\psi^-, \l_3, \lb_3\},\{g^{\pm}, \l_4, \lb_4\}\ldots),
 \]
and similar terms for each negative $b_i$. To complete the proof, we need 
to show that $R$ does not contain a contribution to the $k^{\rm th}$ symmetrized
product that scales faster than ${1 \over z^{k-2}}$. We show this in appendix \ref{explicitchecks}.

The argument above proves our result, since the coefficient of $\eta_2 \eta_1$ in \eqref{susymattermanygluons} gives us the amplitude with fermions while the coefficient
of $1$ gives us the amplitude with scalars. 

We should emphasize here that it is not the case that the scalar and fermion
amplitudes are equal. In fact, the $\o[{1 \over z}]$ terms in \eqref{susycsame}
can and do contribute to both bubble and triangle coefficients. However,
in the consideration of the fourth order Indices for bubbles, as \eqref{bubblexpandinc} tells us, it is only the leading terms in $c$ that are important. 
Similarly, while calculating the contribution to the sixth order Indices
in a triangle coefficient it is only the leading terms in $c$ that 
can contribute. Since, the fermion loop comes with a negative sign, this implies
that the matter contribution of a chiral multiplet to bubble coefficients
does not depend on the fourth-order Indices. Similarly, the matter contribution
of a chiral multiplet to triangle coefficients does not depend on the sixth
order Indices.

\section{The Next to Simplest Quantum Field Theories}
\label{sec:nexttosimplest}
In the previous section we argued that the contribution of matter in a given representation to individual bubble and triangle coefficients is dependent
on a small set of group-theoretic invariants. Now, we also know that 
there are no triangles or bubbles in the $\cn=4$ theory. However,
we can think of the $\cn=4$ theory itself as a gauge theory with matter
in the adjoint representation. The contribution of matter is such
that it exactly cancels out the contribution of the gauge bosons to bubbles
and triangles. This leads to an interesting possibility: what if
we find a representation that {\em mimics} the higher order
Indices of the adjoint representation? We could then replace the adjoint
fields of the ${\cn=4}$ theory with fields in this representation and 
still get a simple S-matrix!

Let us state this a little more explicitly. Let us call 
the contribution of scalars in a representation ${\rm R}_s$ to
the bubble coefficient 
$C^{s}({\rm R}_s, k_1, \ldots k_n, r_1, \ldots r_n)$ and the contribution
to the triangle coefficient $B^{s}({\rm R}_s, k_1, \ldots k_n, r_1, \ldots r_n)$  The notation emphasizes that this 
coefficient depends on the external momenta $k_i$ (we have suppressed the helicity dependence), the colors of the external gluons $r_i$ and the representation ${\rm R}_s$. We denote the contributions of gluons and fermions by $C^{g}, B^{g}$, $C^{f}$ and $B^{f}$.

We know from the previous section that 
\begin{equation}
\begin{split}
C_{s/f}(R_{s/f}, k_i, r_i) &= \sum_{n=2,4} \omega^C_{a_1 \ldots a_n}\left(k_i, r_i\right) \tr_{R_s/R_f}(T^{(a_1} \ldots T^{a_n)}), \\
B_{s/f}(R_{s/f}, k_i, r_i) &= \sum_{n=2,4,5,6} \omega^B_{a_1 \ldots a_n}\left(k_i, r_i \right) \tr_{R_s/R_f}(T^{(a_1} \ldots T^{a_n)}),
\end{split}
\end{equation}
where $\omega^B$ and $\omega^C$ are independent of $R_s$ and $R_f$ but depend
on $k_i,r_i$. However we also know, from the fact that there are no bubbles
or triangles in the $\cn=4$ theory that 
\begin{equation}
\label{nobubn4theory}
\begin{split}
&C^{g}(k_i, r_i) + 4 C^{f}({\rm A},k_i, r_i) + 6 C^{s}({\rm A},k_i, r_i) = 0,\\
&B^{g}(k_i, r_i) + 4 B^{f}({\rm A},k_i, r_i) + 6 B^{s}({\rm A},k_i, r_i) = 0,
\end{split}
\end{equation}
where we have denoted the adjoint representation by ${\rm A}$. 
So, we wish to find representations $R_s, R_f$ with the property\footnote{Demanding the equality of the unordered products up to order $p$ is the same as demanding the equality of the symmetrized products up to order $p$.}
\begin{equation}
\label{nonsusycond}
\begin{split}
\tr_{R_s}\left(\Pi_{i=1}^n T^{a_i} \right) &= 6 \tr_{A}\left(\Pi_{i=1}^n T^{a_i}\right), \forall n \leq p\\
\tr_{R_f}\left(\Pi_{i=1}^n T^{a_i}\right) &= 4 \tr_{A}\left(\Pi_{i=1}^n T^{a_i} \right), \forall n \leq p.
\end{split}
\end{equation}
If we can find solutions for $p=6$ that would give us a theory that is free of
triangles and bubbles. If we can find a solution for $p=4$ that would still
be sufficient to give us a theory without bubbles.

These conditions simplify for supersymmetric theories. For a chiral multiplet $\chi$ in a representation $R_{\chi}$ we know that
\begin{equation}
\begin{split}
C_{\chi}(R_{\chi}, k_i, r_i) &= \omega^C_{a_1 a_2 }\left(k_i, r_i\right) \tr_{R_{\chi}}(T^{a_1} T^{a_2}), \\
B_{\chi}(R_{\chi}, k_i, r_i) &= \sum_{n=2,4,5} \omega^B_{a_1 \ldots a_n}\left(k_i, r_i \right) \tr_{R_{\chi}}(T^{(a_1} \ldots T^{a_n)}),
\end{split}
\end{equation} 
Furthermore, thinking of the $\cn=4$ theory as a $\cn=1$ theory with 3 chiral multiplets and denoting
the vector multiplet by $V$, we know that
\begin{equation}
\label{nobubn4theoryn1language}
\begin{split}
&C^{V}(k_i, r_i) + 3 C^{\chi}({\rm A},k_i, r_i) = 0,\\
&B^{V}(k_i, r_i) + 3 B^{\chi}({\rm A},k_i, r_i) = 0.
\end{split}
\end{equation}
So, in supersymmetric theories we need to put the chiral multiplet
in a representation $R_{\chi}$ that satisfies
\begin{equation}
\label{susycond}
\tr_{R_\chi}\left(\Pi_{i=1}^n T^{a_i} \right) 
= 3 \tr_{A}\left(\Pi_{i=1}^n T^{a_i}\right), \forall n \leq p.
\end{equation}
Now a solution for $p=2$ is enough to give us a theory without bubbles while
a solution for $p=5$ is enough to give us a theory with neither bubbles
nor triangles. For the convenience of the reader we have summarized 
all the different conditions above in Table \ref{simpleconds}.

\TABLE{
\label{simpleconds}
\begin{tabular}{|c|c|c|}\hline
\multicolumn{3}{|l|}{Condition ({\bf C}): ${\rm Tr}_{\rm R}(\Pi_{i=1}^n T^{a_i}) =  m\,{\rm Tr}_{\rm adj}(\Pi_{i=1}^n T^{a_i}),~ n \leq p$} \\ \hline
Non-susy theories have&only boxes & no bubbles \\ \hline
if $R_f$ satisfies {\bf C} with&p=6, m=4&p=4,m=4\\\hline
and $R_s$ satisfies {\bf C} with&p=6, m=6&p=4,m=6.\\\hline
Susy theories have&only boxes & no bubbles \\ \hline
if $R_{\chi}$ satisfies {\bf C} with&p=5, m=3&p=2,m=3.\\\hline
\end{tabular}
\caption{Conditions for the S-matrix to simplify}
}

These equations for the absence of bubbles and triangles have a nice
interpretation in terms of the Indices that we discussed earlier. The
representations $R_s,R_f$ and $R_{\chi}$ that we have been talking about 
above can be decomposed into irreducible representations
\begin{equation}
R = \bigoplus n_i^s R_i.
\end{equation}
Since Indices are additive (this is just because $\tr_{R_1 \oplus R_2} = \tr_{R_1} + \tr_{R_2}$), \eqref{nonsusycond} or \eqref{susycond} are equivalent to a set of linear Diophantine equations in the variables $n_i$ that involve the higher-order Indices of the irreducible representations $R_i$! 

We now present some solutions to these equations. Some further group-theoretic
details are provided in appendix  \ref{app:solveprtrace}. 

\subsection{Theories with only Boxes}
\label{onlyboxes}
We have been able to find solutions to \eqref{susycond} for $p \leq 5$. 

Consider the $\cn=2$, $SU(N)$ (for $N \geq 3$) theory
 with a hypermultiplet that transforms in the symmetric tensor representation of $SU(N)$ and another 
hypermultiplet that transforms as the anti-symmetric tensor. In ${\cn=1}$
language, this leads to
\begin{equation}
\label{n2susybubvanish}
\begin{split}
{\rm R}_{\chi} &= \cdot {\bf adj} \oplus \cdot {\bf sym}\oplus {\bf asym} \oplus {\rm {\overline{\bf sym}}} \oplus  {\rm \overline{\bf asym}}. 
\end{split}
\end{equation}
We provide more details
in appendix \ref{app:solveprtrace} but the reader can verify that, with this choice of representations, \eqref{susycond} is satisfied. So, this theory
has neither bubbles nor triangles in its one-loop gluon S-matrix!

Note that gluon amplitudes, at one-loop, are not sensitive to the superpotential. As far as such amplitudes are concerned, a ${\cn=1}$ theory with the
same spectrum as the theory above will also be free of triangles and bubbles.

To take a more exotic example, consider a gauge theory based on the gauge-group $G_2$. The adjoint of $G_2$ is the $[{\bf 14}]$. If we consider the ${\cn=1}$ theory, with a chiral multiplet in the representation
\begin{equation}
\label{g2withonlyboxes}
R_{\chi} = 3 \cdot [{\bf 7}] \oplus [{\bf 27}],
\end{equation}
then also \eqref{susycond} is satisfied! So this gauge theory also has only
boxes in its one-loop S-matrix!

Finally, we should mention that commonly studied superconformal theories like
 the $\cn=2$, $SU(N)$ theory with $2 N$ hypermultiplets 
or superconformal quiver gauge theories that are dual to orbifolds of $AdS_5 \times S^5$ \cite{Kachru:1998ys} generically contain fundamental (or bi-fundamental) matter in $SU(N)$. Individual triangle coefficients in such theories do not vanish since with scalars and fermions in a combination of just the fundamental and anti-fundamental representations of $SU(N)$, \eqref{susycond} cannot
be satisfied for $p \geq 2$.  However, at large $N$, these theories
do provide {\em approximate} solutions to \eqref{susycond} as discussed in \ref{subsec:largeN}.

\subsection{Theories without Bubbles}
\label{onlytriangles}
It is much easier to find theories without bubbles. First, note that the
condition for supersymmetric theories to be free of bubbles, which is 
\eqref{susycond} with $p = 2$, is just the condition for the one-loop $\beta$
function to vanish. So, any supersymmetric theory with vanishing one-loop
$\beta$ function is free of bubbles at one-loop. This includes, of course,
the $\cn=2$, $SU(N)$ theory with $2 N$ hypermultiplets.

More interestingly, we can also find non-supersymmetric
examples of theories that do not have bubbles. For non-supersymmetric
theories the condition that individual bubble coefficients
 vanish is considerably stronger
than the condition for the one-loop $\beta$ function to vanish. 

For example, consider the $SU(2)$ theory with the following content
\begin{equation}
\label{su2bubvanish}
R_s = 14 \cdot [{\bf 2}] + [{\bf 4}],\quad R_f = 4 \cdot [{\bf 3}].
\end{equation}
This is a $SU(2)$ gauge theory with 7 complex scalar doublets and one scalar in the  ${3 \over 2}$ representation. We impose a reality condition on this last field. More explicitly, if we write this field as a symmetric 3-index tensor $\phi^{i_1 j_1 k_1}$ in a 2 dimensional space, then the reality condition is
$\phi^*_{i_2 j_2 k_2} = \epsilon_{i_1 i_2} \epsilon_{j_1 j_2} \epsilon_{k_1 k_2} \phi^{i_1 j_1 k_2}$. In addition, this theory has 4 adjoint Weyl fermions.
Since \eqref{nonsusycond} is met for $p \leq 4$, this theory does not have bubbles at one-loop!

Then we can derive other examples from the two supersymmetric theories
that are free of both triangles and bubbles. For example, we could replace
all the fermions in \eqref{n2susybubvanish} with adjoint fermions. 
After including the adjoint fermions from the vector multiplet, this would lead to a theory with 
\begin{equation}
\begin{split}
{\rm R}_{s} &= 2 \cdot {\bf adj} \oplus 2 \cdot {\bf sym}\oplus 2 \cdot {\bf asym} \oplus 2 \cdot {\rm {\overline{\bf sym}}} \oplus 2 \cdot {\rm \overline{\bf asym}} .\\
{\rm R}_{f} &= 4 \cdot {\bf adj}
\end{split}
\end{equation}
Proceeding this way we can derive many other examples of non-supersymmetric
theories without bubbles from the two theories of section \ref{onlyboxes}. 

We have not yet found representations ${\rm R}_s$ and ${\rm R}_f$ that would make triangles vanish for non-supersymmetric theories. However, it is possible
to systematically search for such solutions and we provide a method to 
do so in appendix \ref{app:solveprtrace}.

\subsection{Large N}
\label{subsec:largeN}
The results about 1-loop gluon amplitudes that we have presented above hold 
exactly, without any need to take the planar limit. However, 
when the gauge group becomes large, it is possible to find new approximate
solutions to \eqref{susycond}. 

For 
example, for $SU(N)$, at large N, the generators of the adjoint
can be written as
\begin{equation}
T_{\rm adj}^a = T_{\rm f}^a \otimes 1 + 1 \otimes (T_{\rm a f}^a),\quad T_{\rm a f}^a = -(T_{\rm f}^a)^T,
\end{equation}
and so,
\begin{equation}
\label{largeNadjrelation}
\tr \left(T_{\rm adj}^{(a_1} \ldots T_{\rm adj}^{a_{2 n})} \right) = 2 N \tr \left(T_{\rm f}^{(a_1} \ldots T_{\rm f}^{a_{2 n})} \right) + \ldots,
\end{equation}
where the $\ldots$ indicate terms that are subleading in $N$.

In particular, this tells us that the superconformal quiver gauge theories
 that are dual to orbifolds of AdS$_5 \times S^5$ are free of triangles
and bubbles in the {\em planar} limit. This had to be, since these
theories are {\em daughters} of the ${\cn=4}$ theory and their
planar correlation functions can be obtained from the ${\cn=4}$ theory \cite{Bershadsky:1998cb,Erlich:1998gb,Schmaltz:1998bg}. Note that \eqref{largeNadjrelation} also implies that, at large $N$, gluon amplitudes in the ${\cn=2}$, $SU(N)$ 
theory with $2 N$ hypermultiplets are also free of triangles or bubbles.

\section{Examples}
\label{sec:examples}
We now provide several examples of concrete amplitudes. This somewhat 
lengthy list performs two roles. It helps to check the claims that we have
made but also serves to elucidate the beautiful structure described above.

\FIGURE{
\epsfig{file=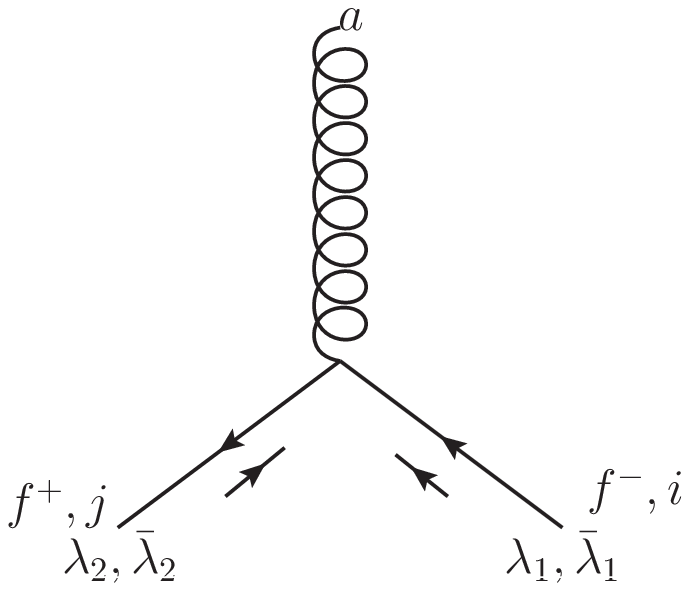,width=0.25\textwidth}\caption{Fermion-gluon interaction}\label{threepointfermion}}
This section is divided in three parts. In subsection \ref{treeamplitudes},
we provide explicit calculations of tree amplitudes with 2 matter particles
(scalars or fermions) and gluons. In subsection \ref{subsec:22scattering},
we use these tree amplitudes to calculate the triangle and bubble coefficients
that appear in  $2 \rightarrow 2$ gluon scattering. In subsection \ref{elucidate},
we summarize the results of  these calculations. The reader
can skip directly to \ref{elucidate} and then turn back to \ref{treeamplitudes} and \ref{subsec:22scattering} for details.

\subsection{Tree Amplitudes with 2 matter particles}
\label{treeamplitudes}
Here, we list the formulas for amplitudes involving two scalars or two fermions and up to 3 external gluons. We have stripped off a factor of 
$-i \left(-g_{YM}\right)^{n-2}$ from the $n$ point amplitude below.  

\subsubsection{3-pt Amplitudes}
\FIGURE[l]{
\epsfig{file=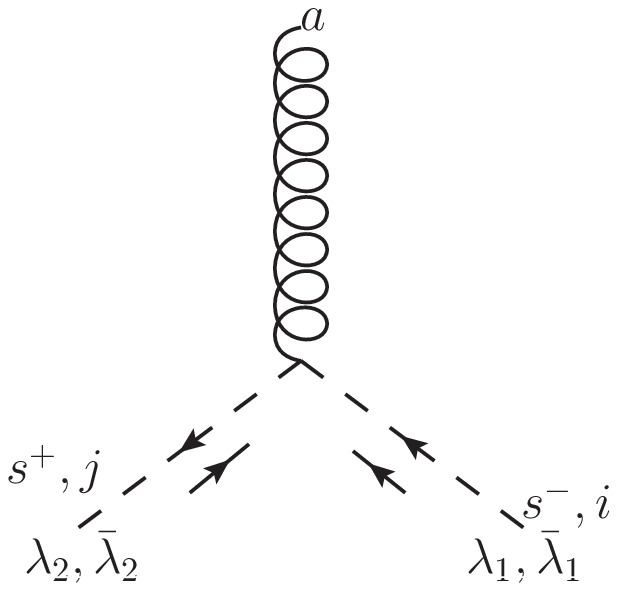,width=0.25\textwidth}\caption{Scalar-gluon
interaction} \label{threepointscalar}}

We use the following 3-pt amplitudes. For fermions, figure \ref{threepointfermion} evaluates to
\begin{equation}
\label{fermi3pt}
\begin{split}
{\mathcal A}^{\rm t}(f^-, f^+, g^-) &= \sqrt{2} {\dotl[ \l_3, \l_1 ]^2 \over \dotl[ \l_1, \l_2 ]} T^{a}_{j i}, \\
{\mathcal A}^{\rm t}(f^-, f^+, g^+) &= \sqrt{2} {[ \lb_3, \lb_2 ]^2 \over [ \lb_2, \lb_1 ]} T^{a}_{j i}.
\end{split}
\end{equation}

For scalars, we have the figure \ref{threepointscalar}
\begin{equation}
\label{scalar3pt}
\begin{split}
{\mathcal A}^{\rm t}(s^-, s^+, g^-) &= \sqrt{2} {\dotl[ \l_2, \l_3 ] \dotl[\l_3, \l_1 ] \over \dotl[ \l_1, \l_2 ]} T^{a}_{j i}, \\
{\mathcal A}^{\rm t}(s^-, s^+, g^+) &= \sqrt{2} {[ \lb_2, \lb_3 ][\lb_3, \lb_1] \over [ \lb_1, \lb_2 ]} T^{a}_{j i}.
\end{split}
\end{equation}

For completeness, we list the gluon 3-pt amplitudes
\begin{equation}
\begin{split}
{\mathcal A}^{\rm t}(g^-, g^-, g^+) &= \sqrt{2} {\dotl[ \l_1, \l_2 ]^3 \over \dotl[\l_2, \l_3] \dotl[\l_3, \l_1 ]} f^{1 2 3}, \\
{\mathcal A}^{\rm t}(g^-, g^-, g^+) &= \sqrt{2} {[ \lb_1, \lb_2 ]^3 \over [\lb_2, \lb_3] [\lb_3, \lb_1 ]} f^{1 2 3}, 
\end{split}
\end{equation}
in conventions where the structure constants are defined by
\begin{equation}
\left[T^a, T^b\right] = f^{a b c}\, T^{c}.
\end{equation}
\subsubsection{4-pt Amplitudes}
The four point amplitudes for 2 fermions and gluons of the same helicity vanishes; the amplitude with gluons of opposite helicity is
\begin{equation}
\label{fermi4pt}
\begin{split}
{\mathcal A}^{\rm t}(f^-, f^+, g^+, g^-) = {2 {\dotl[\l_1, \l_4]}^2 \over \dotl[\l_3, \l_4] \dotl[\l_1, \l_2]} \left( {\dotl[\l_2, \l_4] \over \dotl[\l_2, \l_3] } \left(T^a T^b\right)_{j i} - {\dotl[\l_1, \l_4] \over \dotl[ \l_1, \l_3] } \left( T^b T^a \right)_{j i} \right),
 \end{split}
 \end{equation}
where $g^+$ carries color $a$ and $g^-$ carries color $b$.
The four point amplitude with 2 scalars and gluons of opposite helicities is
\begin{equation}
\label{scalar4pt}
\begin{split}
{\mathcal A}^{\rm t}(s^-, s^+, g^+, g^-) = {2 \dotl[\l_4, \l_1] \dotl[\l_2, \l_4] \over \dotl[\l_3, \l_4] \dotl[\l_1, \l_2]} \left({\dotl[\l_2, \l_4] \over \dotl[\l_2, \l_3]} \left(T^a T^b \right)_{j i} - {\dotl[\l_1,\l_4] \over \dotl[\l_1,\l_3]}  \left(T^b T^a \right)_{j i}\right).
\end{split}
\end{equation}
\subsubsection{5-pt Amplitudes}
Finally, we turn to the 5-pt amplitude. With two gluons of positive helicity and one of negative helicity, this is
\begin{equation}
\label{fiveptfermion}
\begin{split}
&{\mathcal A}^{\rm t}(f^-,f^+, g^+, g^+, g^-)\\ &= \frac{-2 \sqrt{2}  \dotl[\l_1,\l_5]{}^3}{\dotl[\l_1,\l_2] \dotl[\l_3,\l_5]   \dotl[\l_4,\l_5] \dotl[\l_1, \l_3]} \left(\frac{\dotl[\l_2,\l_5] (T^{b}T^{c}T^{a})_{ji}}{\dotl[\l_2,\l_4]}-\frac{\dotl[\l_1,\l_5] (T^{c}T^{b}T^{a})_{ji}}{\dotl[\l_1,\l_4]}\right)\\
&+\frac{2 \sqrt{2}   \dotl[\l_2,\l_5]  \dotl[\l_1,\l_5]{}^2}{\dotl[\l_1,\l_2]   \dotl[\l_4,\l_5] \dotl[\l_3,\l_5] \dotl[\l_2, \l_3]}  \left(\frac{\dotl[\l_2,\l_5] (T^{a}T^{b}T^{c})_{ji}}{\dotl[\l_2,\l_4]}-\frac{\dotl[\l_1,\l_5] (T^{a}T^{c}T^{b})_{ji}}{\dotl[\l_1,\l_4]}\right)\\
&+\frac{2 \sqrt{2} f^{a b d} \dotl[\l_1,\l_5]{}^2    }{\dotl[\l_1,\l_2] \dotl[\l_3,\l_4] \dotl[\l_3,\l_5]}  \left(\frac{\dotl[\l_2,\l_5] (T^{d}T^{c})_{ji}}{\dotl[\l_2,\l_4]}
-\frac{\dotl[\l_1,\l_5] (T^{c}T^{d})_{ji}}{\dotl[\l_1,\l_4]}\right),
\end{split}
\end{equation}
where the gluons carry colors $a,b,c$ respectively.
The scalar amplitude, with the same choice of gluon helicities, is,
\begin{equation}
\begin{split}
&{\mathcal A}^{\rm t}(s^-,s^+, g^+, g^+, g^-)\\
&\frac{-2 \sqrt{2} \dotl[\l_2,\l_5] \dotl[\l_1,\l_5]{}^2}{\dotl[\l_1,\l_2] \dotl[\l_1, \l_3] \dotl[\l_3,\l_4] \dotl[\l_4,\l_5]}  \left(\frac{\dotl[\l_1,\l_5] (T^{c}T^{b}T^{a})_{ji}}{\dotl[\l_1,\l_4]}-\frac{\dotl[\l_2,\l_5] (T^{b}T^{c}T^{a})_{ji}}{\dotl[\l_2,\l_4]}\right) \\
&+\frac{2 \sqrt{2} \dotl[\l_2,\l_5]{}^2  \dotl[\l_1,\l_5]}{\dotl[\l_1,\l_2] \dotl[\l_2, \l_3] \dotl[\l_3,\l_5] \dotl[\l_4,\l_5]} \left(\frac{\dotl[\l_1,\l_5] (T^{a}T^{c}T^{b})_{ji}}{\dotl[\l_1,\l_4]}-\frac{\dotl[\l_2,\l_5] (T^{a}T^{b}T^{c})_{ji}}{\dotl[\l_2,\l_4]}\right) \\
&-\frac{2 \sqrt{2} f^{a b d} \dotl[\l_1,\l_5]   \dotl[\l_2,\l_5]  }{\dotl[\l_1,\l_2] \dotl[\l_3,\l_4] \dotl[\l_3,\l_5]}  \left(\frac{\dotl[\l_2,\l_5] (T^{d}T^{c})_{ji}}{\dotl[\l_2,\l_4]}-\frac{\dotl[\l_1,\l_5] (T^{c}T^{d})_{ji}}{\dotl[\l_1,\l_4]}\right).
\end{split}
\end{equation}
Note, that we have
\begin{equation}
\label{scalfermilinkhol}
{\mathcal A}^{\rm t}(f^-, f^+, g^+, g^+, g^-)= -{\dotl[\l_1, \l_5] \over \dotl[\l_2, \l_5]} {\mathcal A}^{\rm t}(s^-, s^+, g^+, g^+, g^-).
\end{equation}
The amplitudes with the other choice of helicities for the gluons may be obtained from the expressions given above by parity and charge conjugation. 

For example, the scalar amplitude with the opposite helicities for gluons can be obtained by turning all the $\l$'s above into $\lb$'s.
\begin{equation}
\begin{split}
&{\mathcal A}^{\rm t}(s^-,s^+, g^-, g^-, g^+)\\
&=\frac{-2 \sqrt{2} \dotlb[\lb_2,\lb_5] \dotlb[\lb_1,\lb_5]{}^2}{\dotlb[\lb_1,\lb_2] \dotlb[\lb_1, \lb_3] \dotlb[\lb_3,\lb_4] \dotlb[\lb_4,\lb_5]}  \left(\frac{\dotlb[\lb_1,\lb_5] (T^{c}T^{b}T^{a})_{ji}}{\dotlb[\lb_1,\lb_4]}-\frac{\dotlb[\lb_2,\lb_5] (T^{b}T^{c}T^{a})_{ji}}{\dotlb[\lb_2,\lb_4]}\right) \\
&+\frac{2 \sqrt{2} \dotlb[\lb_2,\lb_5]{}^2  \dotlb[\lb_1,\lb_5]}{\dotlb[\lb_1,\lb_2] \dotlb[\lb_2, \lb_3] \dotlb[\lb_3,\lb_5] \dotlb[\lb_4,\lb_5]} \left(\frac{\dotlb[\lb_1,\lb_5] (T^{a}T^{c}T^{b})_{ji}}{\dotlb[\lb_1,\lb_4]}-\frac{\dotlb[\lb_2,\lb_5] (T^{a}T^{b}T^{c})_{ji}}{\dotlb[\lb_2,\lb_4]}\right) \\
&-\frac{2 \sqrt{2} f^{a b d} \dotlb[\lb_1,\lb_5]   \dotlb[\lb_2,\lb_5]  }{\dotlb[\lb_1,\lb_2] \dotlb[\lb_3,\lb_4] \dotlb[\lb_3,\lb_5]}  \left(\frac{\dotlb[\lb_2,\lb_5] (T^{d}T^{c})_{ji}}{\dotlb[\lb_2,\lb_4]}-\frac{\dotlb[\lb_1,\lb_5] (T^{c}T^{d})_{ji}}{\dotlb[\lb_1,\lb_4]}\right).
\end{split}
\end{equation}
The fermionic amplitude with the opposite helicities requires us to use CP. This interchanges $f^+$ and $f^-$. C takes the representation matrices $T^a \rightarrow -(T^a)^T$. To maintain the convention that the negative helicity fermion is associated with $\l_1, \lb_1, i$, we need to interchange the labels $\lb_1, i \leftrightarrow \lb_2,j$; finally, we need to add a minus sign that is associated with performing CP on the fermion pair.
This leads to the amplitude, 
\begin{equation}
\label{scalfermilinkantihol}
{\mathcal A}^{\rm t}(f^-, f^+, g^-,g^-,g^+) = {\dotlb[\lb_2, \lb_5] \over \dotlb[\lb_1, \lb_5]} {\mathcal A}^{\rm t}(s^-,s^+, g^+, g^+, g^-).
\end{equation}

\subsubsection{Checking Structure}
\label{subsubsec:checkingstructure}
Let us now check that the structures that we have claimed do appear in these amplitudes. We wish to verify that these amplitudes can be written
in the form \eqref{scalarpowers} when the matter momenta are BCFW extended. We will check this for the five point amplitude. The amplitude \eqref{fiveptfermion} can be written as 
\begin{equation}
\begin{split}
&{\mathcal A}^{\rm t}(f^-,f^+, g^+, g^+, g^-) = \kappa \left( {\dotl[\l_2, \l_5] \over \dotl[\l_2, \l_3]} - {\dotl[\l_1, \l_5] \over \dotl[\l_1, \l_3]}\right) \left( {\dotl[\l_2, \l_5] \over \dotl[\l_2, \l_4]} - {\dotl[\l_1, \l_5] \over \dotl[\l_1, \l_4]}\right) T^{(a}T^{b}T^{c)} \\
&+ \kappa \left( {\dotl[\l_2, \l_5] \over \dotl[\l_2, \l_3]} + {\dotl[\l_1, \l_5] \over \dotl[\l_1, \l_3]} + 2 {\dotl[\l_4, \l_5] \over \dotl[\l_3,\l_4]}\right) \left( {\dotl[\l_2, \l_5] \over \dotl[\l_2, \l_4]} - {\dotl[\l_1, \l_5] \over \dotl[\l_1, \l_4]}\right) {\{[T^a,T^b],T^c\} \over 4} \\
&+ \kappa \left( {\dotl[\l_2, \l_5] \over \dotl[\l_2, \l_3]} + {\dotl[\l_1, \l_5] \over \dotl[\l_1, \l_3]} \right) \left( {\dotl[\l_2, \l_5] \over \dotl[\l_2, \l_4]} - {\dotl[\l_1, \l_5] \over \dotl[\l_1, \l_4]}\right) {\{[T^a,T^c],T^b\} \over 4} \\
&+  \kappa \left( {\dotl[\l_2, \l_5] \over \dotl[\l_2, \l_3]} - {\dotl[\l_1, \l_5] \over \dotl[\l_1, \l_3]}\right) \left( {\dotl[\l_2, \l_5] \over \dotl[\l_2, \l_4]} + {\dotl[\l_1, \l_5] \over \dotl[\l_1, \l_4]}\right) {\{[T^b,T^c],T^a\} \over 4} \\
&+  \kappa \left[ {\dotl[\l_2, \l_5] \over \dotl[\l_2, \l_3]} \left(2 {\dotl[\l_2, \l_5] \over \dotl[\l_2, \l_4]} +  {\dotl[\l_1, \l_5] \over \dotl[\l_1, \l_4]}\right) +  {\dotl[\l_1, \l_5] \over \dotl[\l_1, \l_3]} \left(2 {\dotl[\l_1, \l_5] \over \dotl[\l_1, \l_4]} +  {\dotl[\l_2, \l_5] \over \dotl[\l_2, \l_4]}\right)\right. \\ 
&\left. +  3 {\dotl[\l_4, \l_5] \over \dotl[\l_3,\l_4]}  \left( {\dotl[\l_2, \l_5] \over \dotl[\l_2, \l_4]} +  {\dotl[\l_1, \l_5] \over \dotl[\l_1, \l_4]}\right) \right]  {[[T^a,T^b],T^c] \over 6} \\
&+  \kappa \left( {\dotl[\l_2, \l_5] \over \dotl[\l_2, \l_3]} \left( {\dotl[\l_2, \l_5] \over \dotl[\l_2, \l_4]} +  2 {\dotl[\l_1, \l_5] \over \dotl[\l_1, \l_4]}\right) +  {\dotl[\l_1, \l_5] \over \dotl[\l_1, \l_3]} \left( {\dotl[\l_1, \l_5] \over \dotl[\l_1, \l_4]} +  2 {\dotl[\l_2, \l_5] \over \dotl[\l_2, \l_4]}\right) \right)  {[[T^c,T^a],T^b] \over 6}.
\end{split}
\end{equation}
In this equation
\begin{equation}
\kappa = {2 \sqrt{2} \dotl[\l_1, \l_5]^2  \over 
\dotl[\l_1, \l_2] \dotl[\l_3, \l_5]\dotl[\l_4, \l_5]}.
\end{equation}
In this form, it is clear that if $\l_1,\l_2$ grow large as in \eqref{largezspinorchoice} then the amplitude takes the form \eqref{scalarpowers}.
The reader might be concerned about the $\dotl[\l_1, \l_2]$ term that appears in $\kappa$ which seems to depend on the subleading pieces of $\l_1, \l_2$.  However, we can write, 
\begin{equation}
\dotl[\l_1, \l_2] = \dotl[\l_1, \l_3] { \dotl[\l_5, \l_1] \dotlb[\lb_5, \lb_3] + \dotl[\l_4, \l_1] \dotlb[\lb_4, \lb_3] \over 2(k_1 + k_4 + k_5) \cdot k_3},
\end{equation}
which shows that this term does not, in fact, depend on the subleading pieces in $\l_1, \l_2$.
In fact, subject to the condition \eqref{largezspinorchoice}, we can write
\begin{equation}
\begin{split}
\l_2 &= - \l_1 + \l_3 { \dotl[\l_5, \l_1] \dotlb[\lb_5, \lb_3] + \dotl[\l_4, \l_1] \dotlb[\lb_4, \lb_3] \over 2(k_1 + k_4 + k_5) \cdot k_3}, \\
\lb_2 &= {(k_1 + k_4 + k_5) \cdot k_3 \over k_1 \cdot k_3 } \lb_1 -  \lb_3 { \dotl[\l_5, \l_3] \dotlb[\lb_5, \lb_1] + \dotl[\l_4, \l_3] \dotlb[\lb_4, \lb_1] \over 2 k_1  \cdot k_3}.
\end{split}
\end{equation}
Since the scalar amplitude can be written
\begin{equation}
{\mathcal A}^{\rm t}(s^-,s^+, g^+, g^+, g^-) = -{\dotl[\l_2, \l_5] \over \dotl[\l_1, \l_5]}  {\mathcal A}^{\rm t}(f^-,f^+, g^+, g^+, g^-).
\end{equation}
The structure that we have advertised holds for this amplitude as well. Moreover, with the choice \eqref{largezspinorchoice}, we see that the coefficients $A$ in \eqref{scalarpowers} is the same for fermions and scalars to leading order in $z$.

\subsection{A One Loop Example: $2 \rightarrow 2$ scattering}
\label{subsec:22scattering}
We now explicitly calculate a triangle and a bubble coefficient in 
a $2 \rightarrow 2$ scattering process in a gauge theory with matter. 

Without loss of generality, we choose the following initial momenta
\begin{equation}
\begin{split}
k_1 &= y^2 (1,1,0,0),  \quad  k_2 = y^2 (1,-1, 0,0), \\
k_3 &= y^2 (-1, \cos \theta, \sin \theta, 0), \quad k_4 = y^2 (-1,-\cos \theta, -\sin \theta, 0). 
\end{split}
\end{equation}
We will denote $x \equiv  e^{i \theta}$. We also choose the external helicities to be $h_1 = h_3 = 1, h_2 = h_4 = -1$.
Writing $k_i \sigma^{\mu}_{\alpha \dot{\alpha}} = \left(\l_i\right)_{\alpha} \left(\lb_i\right)_{\dot{\alpha}}$, with the Minkowski space condition $\l_i^* = \pm \lb_i$, we find the following spinor decomposition of the momenta
\begin{equation}
\begin{split}
\l_1 &= y \{1, 1\}; \lb_1 = y\{1, 1\}; \quad
\l_2 = y \{1, -1\}; \lb_2 = y \{1, -1\}; \\
\l_3 &= y \{i, -i x\}; \lb_3 = y\{i, -i/x\}; \quad
\l_4 = y \{i, i x\}; \lb_4 = y \{i, i/x\}.
\end{split}
\end{equation}

The tree-amplitude for this choice of momenta is
\begin{equation}
\label{treeamp4part}
\begin{split}
{{\mathcal A}^{\rm t} \over 4} &= -\frac{(x+1)^4}{4 (x-1)^2 x} \left[\tr_F\left(T^a T^b T^c T^d\right) + \tr_F\left(T^b T^a T^d T^c \right) \right] \\ &+ 
\frac{(x+1)^2}{4 x} \left[\tr_F\left(T^a T^b T^d T^c\right) + \tr_F\left(T^b T^a T^c T^d\right) \right] \\ &+ 
\frac{ (x+1)^2}{(x-1)^2}\left[\tr_F\left(T^a T^d T^b T^c\right) + \tr_F\left(T^d T^a T^c T^b \right)\right].
\end{split}
\end{equation}

\subsubsection{A Triangle Coefficient}
\label{trcoeff}

Let us calculate the coefficient of the triangle diagram shown in figure \ref{trn2theory}. There are two possibilities when the intermediate momenta are put on shell and extended in the form explained above.
\[
\begin{array}{l|l|l|l}
&p_{31} & p_{24} & p_{43} \\ \hline &&& \\
p^{+}&(z \l_4 + \l_3) \lb_3 &  \l_4 (z \lb_3 - \lb_4) & z \l_4 \lb_3  \\ &&& \\
p^{-}&\l_3 (-z \lb_4 + \lb_3) & (-z \l_3 - \l_4) \lb_4 & -z \l_3 \lb_4  \\
\hline
\end{array}
\]

Using formula \eqref{triangleinttreeproduct}, the contribution of gluons, fermions and scalars to the triangle coefficient is
\begin{equation}
\label{contribtrcoeff}
\begin{split}
{B^{g}} &=  \frac{32 y^4 \left(x^4+2 x^3+10 x^2+2 x+1\right)}{ \left(x+1\right)^4}\left[\tr_A \left(T^a T^b T^c T^d\right) + \tr_A \left(T^b T^a T^d T^c\right)\right],\\
{-B^{f}} &= -\frac{16 y^4 \left(x-1\right)^4}{\left(x+1\right)^4} \tr_{R_f}\left(T^a T^b T^c T^d\right) -\frac{256 y^4 x^2}{\left(x+1\right)^4}\tr_{R_f}\left(T^b T^a T^d T^c\right),\\
{B^{s}} &= \frac{-32 y^4 x \left(x-1\right)^2}{\left(x+1\right)^4} \left[\tr_{R_s}\left(T^a T^b T^c T^d\right) + \tr_{R_s}\left(T^b T^a T^d T^c\right)\right],
\end{split}
\end{equation}
where the subscript below the trace indicates that it is representation dependent. 

\subsubsection{Triangles in Supersymmetric Theories}
\FIGURE{
\label{trn2theory}
\epsfig{width=0.3\textwidth,file=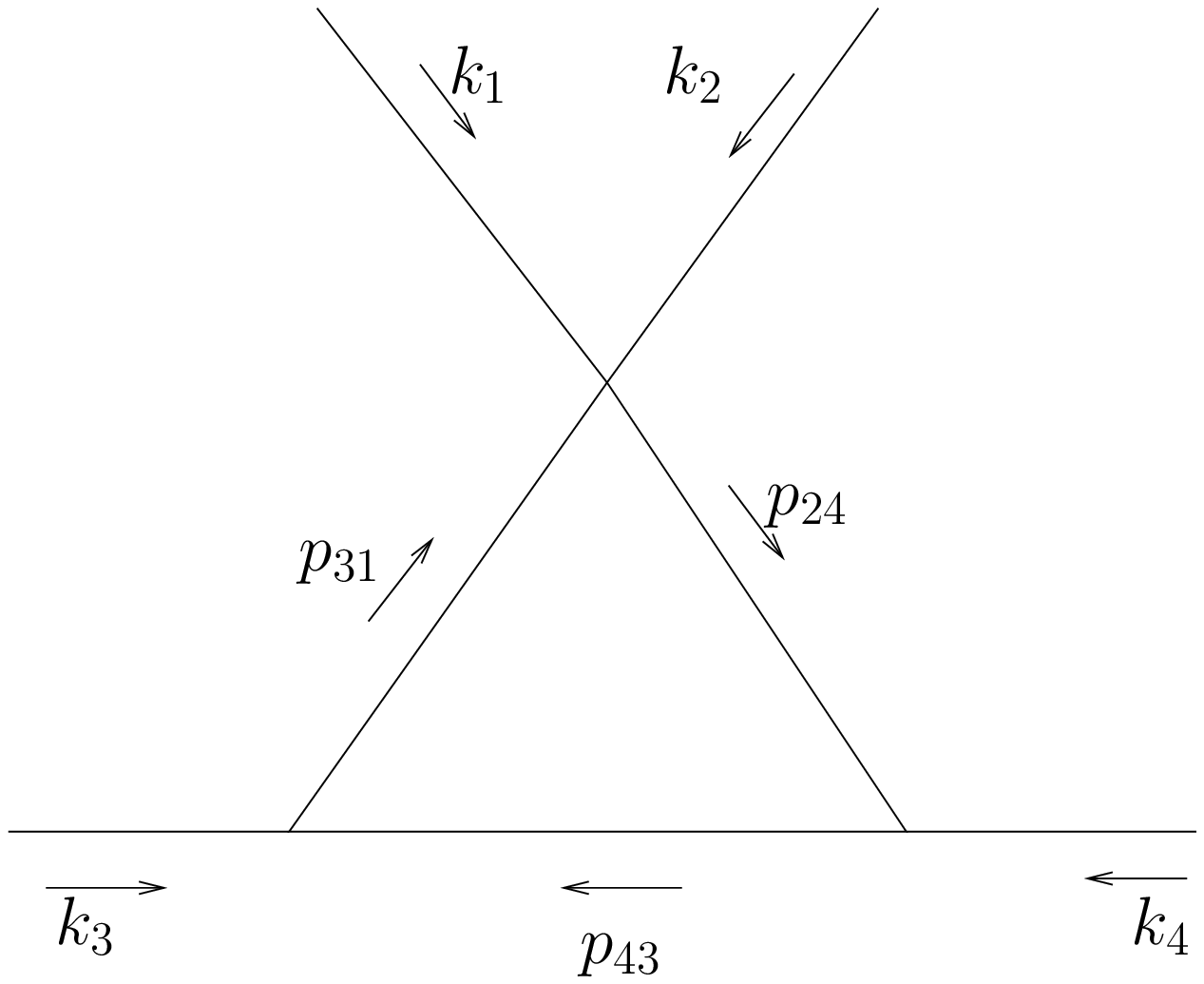}
\caption{A triangle diagram for $2 \rightarrow 2$ scattering}
}
Here, we show via a direct calculation that the contribution of a
$\cn=1$ chiral multiplet (or a $\cn=2$ hypermultiplet) to the triangle
depends on the 4$^{\rm th}$ order Indices of its representation. Just to make
sure that nothing we are doing is an artifact of our coordinate choice,
we will work with completely general external spinors $\l_i, \lb_i$. 

Consider the configuration of  \ref{trn2theory}. We have 4 gluons with spinors $\l_i, \lb_i$. The helicities are $1,-1,1,-1$ respectively.  The cut momenta must be one of two choices $(p_{3 4})_{\alpha \dot{\alpha}} = A_1 (\l_3)_{\alpha} (\lb_4)_{\dot{\alpha}}$ or $p_{3 4} = A_2 (\l_4)_{\alpha} (\lb_3)_{\dot{\alpha}}$ (where A is a constant that we will fix below). 

According to the prescription detailed above we now need to find two unit vectors $w_1, w_2$ with $w_i \cdot k_3 = w_i \cdot k_4 = 0$ and set $p_{3 4} \cdot w_1 = z$.
We can choose
\begin{equation}
\begin{split}
(w_1)_{\alpha \dot{\alpha}} &= {{1 \over \dotl[\l_3, \l_4]} (\l_3)_{\alpha} (\l_4)_{\dot{\alpha}} + {1 \over \dotlb[\lb_4, \lb_3]} (\l_4)_{\alpha} (\l_3)_{\dot{\alpha}}}, \\
(w_2)_{\alpha \dot{\alpha}} &= {{-i \over \dotl[\l_3, \l_4]} (\l_3)_{\alpha} (\l_4)_{\dot{\alpha}} + {i \over \dotlb[\lb_4, \lb_3]} (\l_4)_{\alpha} (\l_3)_{\dot{\alpha}}}.
\end{split}
\end{equation}
This sets $A_1 = {z \over \dotl[\l_3, \l_4]}$ and $A_2 = {z \over \dotl[\lb_4, \lb_3]}$. However, note that in the first case, $p_{3 4} \cdot w_2 = i z$, whereas in the second case $p_{3 4} \cdot w_2 = - i z$. Hence, the variable $r$ defined below \eqref{threeonshell} is zero. This means that we can freely rescale $z$,
since  \eqref{findtrianglefromthreecut} tells us to look at the $z^0$ term in the product of three tree amplitude. We will use this to set $A_1 = A_2 = z$ above.

Consider the first possibility above with $(p_{3 4})_{\alpha \dot{\alpha}} = (\l_3)_{\alpha} (\lb_4)_{\dot{\alpha}}$. We will first analyze the fermionic contributions and then the bosonic contributions.

The 4-pt amplitude involves a negative helicity fermion with spinors $z \l_3 + \l_4, \lb_4$, a positive helicity fermion with spinors $-z \l_3, \lb_4 - {\lb_3 \over z}$ a positive helicity gluon with $\l_1, \lb_1$ and a negative helicity gluon with $\l_2, \lb_2$. The amplitude with this combination can be obtained using
formula \eqref{fermi4pt} and we find that it is
\begin{equation}
{\cal A}_1 = {-2 \over z} {\dotl[z \l_3 + \l_4, \l_2]^2 \over \dotl[\l_1, \l_2] \dotl[\l_4, \l_3]} \left[{\dotl[\l_3, \l_2] \over \dotl[\l_3, \l_1]} (T^a T^b)_{j i} - {\dotl[z \l_3 + \l_4, \l_2] \over \dotl[z \l_3 + \l_4, \l_1]} (T^b T^a)_{j i}\right].
\end{equation}
The two three-point amplitudes are 
\begin{equation}
\begin{split}
{\cal A}_2 &= \sqrt{2} z \dotl[\l_4, \l_3] T^d_{i k}, \\
{\cal A}_3 &= \sqrt{2} z \dotlb[\lb_3, \lb_4] z T^c_{k j}.
\end{split}
\end{equation}
Since, we are looking for the coefficient of $\tr{\left(T^{(a} T^b T^c T^{d)}\right)}$, the color-factors above are not important and we find
\begin{equation}
\label{fermonehel}
\begin{split}
{\cal A}_1 {\cal A}_2 {\cal A}_3 &= {- 4 z \dotlb[\lb_3, \lb_4] \dotl[z \l_3 + \l_4, \l_2]^2 \over \dotl[\l_1, \l_2]} \left[{\dotl[\l_3, \l_2] \over \dotl[\l_3, \l_1]} - {\dotl[z \l_3 + \l_4, \l_2] \over \dotl[z \l_3 + \l_4, \l_1]} \right] \tr(T^{(a}T^b T^c T^{d)})\\ &+ {\rm lower~order~traces.}
\end{split}
\end{equation}
Another contribution comes from reversing all the fermion lines. With this,
the amplitudes are
\begin{equation}
{\cal A}_1 = {-2 \over z} {\dotl[z \l_3, \l_2]^2 \over \dotl[\l_1, \l_2] \dotl[\l_4, \l_3]} \left[{\dotl[\l_3, \l_2] \over \dotl[\l_3, \l_1]} (T^a T^b)_{i j} - {\dotl[z \l_3 + \l_4, \l_2] \over \dotl[z \l_3 + \l_4, \l_1]} (T^b T^a)_{i j}\right].
\end{equation}
The two three-point amplitudes are 
\begin{equation}
\begin{split}
{\cal A}_2 &= -\sqrt{2} z \dotl[\l_4, \l_3] T^d_{k i}, \\
{\cal A}_3 &= -\sqrt{2} z \dotlb[\lb_3, \lb_4] z T^c_{j k}.
\end{split}
\end{equation}
Adding the contributions from the amplitudes above and including the minus
sign for the fermion loop, we find that the contribution to the three-cut
from fermions is:
\begin{equation}
\begin{split}
-{\mathcal C}_f(z) &= - 4 z {\dotlb[\lb_3, \lb_4]  \left[\dotl[z \l_3 + \l_4, \l_2]^2 + z^2 \dotl[\l_3, \l_2]^2 \right] \over \dotl[\l_1, \l_2]} \left[ {\dotl[\l_3, \l_2] \over \dotl[\l_3, \l_1]} - {\dotl[z \l_3 + \l_4, \l_2] \over \dotl[z \l_3 + \l_4, \l_1]} \right] \tr(T^{(a}T^b T^c T^{d)}) \\ &+ {\rm lower~order~traces.}
\end{split}
\end{equation}
The contribution for scalars is very similar. The reader can easily work
out using \eqref{scalar4pt} that the contribution of scalars to the 4$^{\rm th}$
order trace is 
\begin{equation}
\begin{split}
{\mathcal C}_s &= - 4 z {\dotlb[\lb_3, \lb_4] \over \dotl[\l_1, \l_2]} \left[\dotl[z \l_3 + \l_4, \l_2]\dotl[z \l_3, \l_2] \right] \left[ {\dotl[\l_3, \l_2] \over \dotl[\l_3, \l_1]} - {\dotl[z \l_3 + \l_4, \l_2] \over \dotl[z \l_3 + \l_4, \l_1]} \right] \tr(T^{(a}T^b T^c T^{d)})\\ &+ {\rm lower~order~traces.}
\end{split}
\end{equation}
The sum of the two contributions is
\begin{equation}
\begin{split}
{\mathcal C}_s + {\mathcal C}_f &=   4 z {\dotlb[\lb_3, \lb_4] \over \dotl[\l_1, \l_2]} \left[\dotl[z \l_3 + \l_4, \l_2] -   \dotl[z \l_3, \l_2] \right]^2 \left[ {\dotl[\l_3, \l_2] \over \dotl[\l_3, \l_1]} - {\dotl[z \l_3 + \l_4, \l_2] \over \dotl[z \l_3 + \l_4, \l_1]} \right] \tr(T^{(a}T^b T^c T^{d)}) \\ &+ {\rm lower~order~traces.}
\end{split}
\end{equation}
We can expand out the last bracket to $\o[{1 \over z}]$ to find that the contribution to the 4$^{\rm th}$ order trace in the triangle coefficient is
\begin{equation}
2(B_s^4 + B_f^4) = -4 {\dotlb[\lb_3, \lb_4] \dotl[\l_4, \l_2]^2 \over \dotl[\l_1, \l_2]} \left[ {\dotl[\l_4, \l_2] \over \dotl[\l_3, \l_1]} - {\dotl[\l_3, \l_2] \dotl[\l_4, \l_1] \over \dotl[\l_3, \l_1]^2} \right].
\end{equation}
This combination does not vanish for generic external momenta. 

\FIGURE{
\epsfig{file=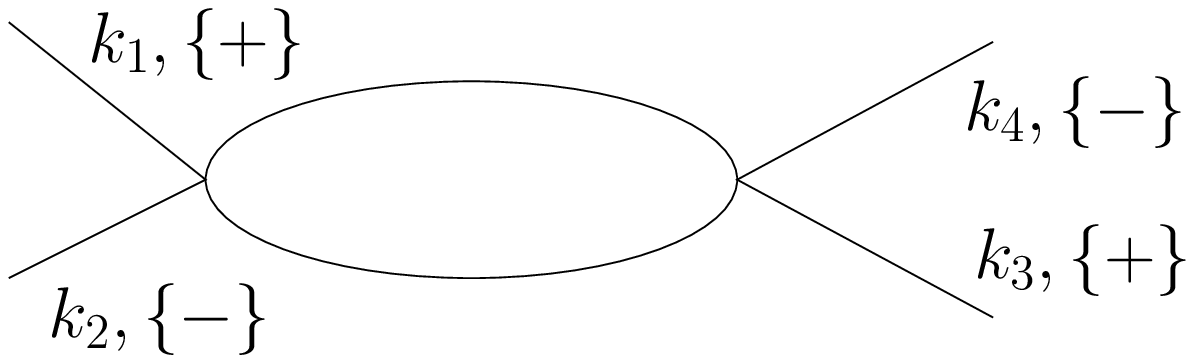,width=0.4\textwidth}
\label{bubblen2theory}
\caption{A bubble diagram for $2 \rightarrow 2$ scattering}
}
\subsubsection{A Bubble Coefficient}
\label{bubcoeff}
With the same initial momenta as above, we will calculate the coefficient of the bubble diagram 
shown in \ref{bubblen2theory}.

Using \eqref{bubbleintreeproduct} and integrating over the phase space\footnote{The prescription of \cite{ArkaniHamed:2008gz} can lead to singular phase
space integrals. This is avoided by using the prescription of \cite{Raju:2009yx}}
leads
to the following contribution from gluons, fermions and scalars.
\begin{equation}
\label{bubble1}
\begin{split}
{C^{g}_1 \over 4} &= 
\frac{11 x^4+32 x^3+138 x^2+32 x+11}{12 x (x+1)^2}\left[\tr_A\left(T^a T^b T^c T^d\right) +\tr_A\left(T^b T^a T^d T^c \right) \right] \\
&- \frac{11 (x+1)^2}{12 x}\left[ \tr_A\left(T^a T^b T^d T^c\right) +  \tr_A\left(T^b T^a T^c T^d\right)\right], \\
{-C^{f}_1 \over 4} &=\frac{x^4-14 x^3+18 x^2-14 x+1}{6 x (x+1)^2} \tr_{R_f}\left(T^a T^b T^c T^d\right)-\frac{(x+1)^2}{6 x} \tr_{R_f}\left(T^a T^b T^d T^c\right)\\
&-\frac{(x+1)^2}{6 x} \tr_{R_f}\left(T^b T^a T^c T^d\right)+ \frac{x^4+10 x^3+66 x^2+10 x+1}{6 x (x+1)^2} \tr_{R_f}\left(T^b T^a T^d T^c\right),\\
{C^{s}_1 \over 2} &=  -\frac{x^4+16 x^3-66 x^2+16 x+1}{12 x (x+1)^2}\left[\tr_{R_s}\left(T^a T^b T^c T^d\right) + \tr_{R_s}\left(T^b T^a T^d T^c\right) \right]\\
&+\frac{(x+1)^2}{12 x}\left[\tr_{R_s}\left(T^a T^b T^d T^c\right) + \tr_{R_s}\left(T^b T^a T^c T^d\right)\right]. 
\end{split}
\end{equation}
Note that the fermionic contribution in \eqref{bubble1} includes a finite coefficient of $\tr\left(\left[T^a, T^b\right]\left\{T^c, T^d\right\}\right)$ although anomaly cancellation requires the trace itself to vanish.

Let us also calculate the other non-zero bubble coefficient for this amplitude, shown in \ref{secondbubblen2theory} 
suffices to give us the right $\beta$ functions. Evaluating this bubble diagram, we find that the contribution of the gluons, scalars and fermions is as follows. 
\begin{equation}
\label{bubble2}
\begin{split}
{{\mathcal C}^{g}_2 \over 4} &=
\frac{2 \left(7 x^4+10 x^3+54 x^2+10 x+7\right)}{3 \left(x^2-1\right)^2} 
\left[\tr_A\left(T^a T^b T^c T^d\right) + \tr_A\left(T^b T^a T^d T^c\right)\right] \\
&-\frac{11 (x+1)^2}{3 (x-1)^2} \left[\tr_A\left(T^a T^d T^b T^c\right) + 
\tr_A\left(T^a T^c T^b T^d\right) \right], \\
{-{\mathcal C}^{f}_2 \over 4}&=\frac {11 x^4 - 16 x^3 + 42 x^2 - 16 x + 11} {3 \left (x^2 - 1 \right)^2}\tr_{R_f}\left(T^a T^b T^c T^d\right) -\frac{2 (x+1)^2}{3 (x-1)^2} \tr_{R_f}\left(T^a T^d T^b T^c\right) \\
&-\frac{2 (x+1)^2}{3 (x-1)^2} \tr_{R_f}\left(T^a T^c T^b T^d\right) -\frac{x^4+16 x^3-66 x^2+16 x+1}{3 \left(x^2-1\right)^2}\tr_{R_f}\left(T^b T^a T^d T^c\right), \\
{{\mathcal C}^{s}_2 \over 2} &= \frac{2 \left(x^4-14 x^3+18 x^2-14 x+1\right)}{3 \left(x^2-1\right)^2}\left[\tr_{R_s}\left(T^a T^b T^c T^d\right) + \tr_{R_s}
\left(T^b T^a T^d T^c \right)\right] \\
&+ \frac{(x+1)^2}{3 (x-1)^2}\left[\tr_{R_s}\left(T^a T^d T^b T^c\right) + \tr_{R_s}\left(T^a T^c T^b T^d \right)\right].
\end{split}
\end{equation}
\subsubsection{Beta Function}
\FIGURE{
\label{secondbubblen2theory}
\epsfig{width= 0.4\textwidth,file=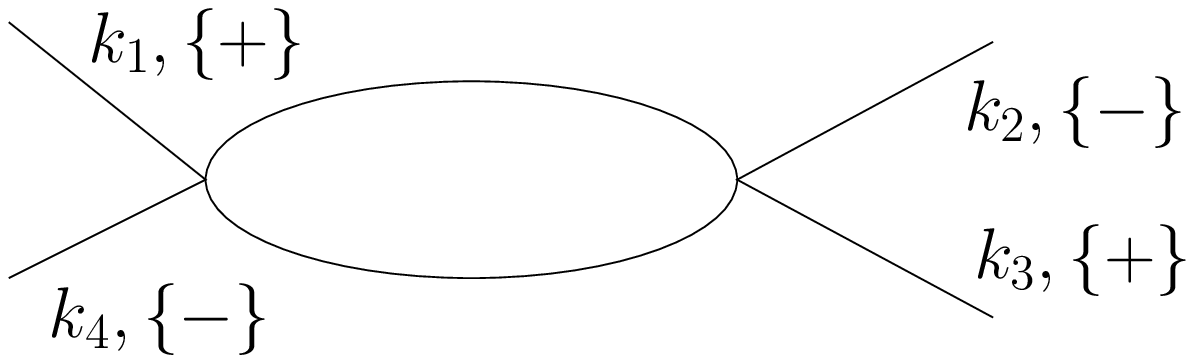}
\caption{The other bubble diagram}
}
Note that although the two bubble coefficients \eqref{bubble1} and \eqref{bubble2} are, individually, quite complicated they give us exactly the right $\beta$
function when they are added together. In the $SU(N)$ theory, using $I_2(F) = {1 \over 2}$, we see 
\begin{equation}
\begin{split}
&C^{g}_1 + C^{g}_2 = {-22 \over 3} N_c M_{\rm tree},\\
&C^{f}_1 + C^{f}_2 = {4 \over 3} I_2(R_f) M_{\rm tree},\\
&C^{s}_1 + C^{s}_2 = {1 \over 3} I_2(R_s) M_{\rm tree}.
\end{split}
\end{equation}
Restoring powers of $g_{\rm YM}$ and recalling that each bubble coefficient
multiplies a UV divergent term as in \eqref{bubblemultuv}, this leads to the $\beta$ function equation,
\begin{equation}
\label{betafunc}
{d g_{\rm YM} \over d \ln \Lambda} = {g_{\rm YM}^3 \over 16 \pi^2} \left({-11 N_c \over 3} + {2 \over 3} I_2(R_f) + {1 \over 6} I_2(R_s)\right).
\end{equation}
We remind the reader of our conventions for $R_s$ explained at the end of section \ref{subsec:contribmatterbub}.

\subsection{The Moral of the Story}
\label{elucidate}
We will now use the calculations above to illustrate several of the claims made
in the text.
\begin{enumerate}
\item
The checks of \ref{subsubsec:checkingstructure} show that tree amplitudes
with many gluons and two BCFW extended scalars or fermions 
can be expanded in a series like \eqref{scalarpowers} where the
symmetrized product of $k$ generators is multiplied by $z^{2-k}$.
Moreover, as \eqref{scalfermilinkhol} and \eqref{scalfermilinkantihol} tell 
us, under the choice \eqref{largezspinorchoice}, the relation
\eqref{susycsame} is satisfied.
\item
Turning now to the one-loop calculation, we find that the contribution of gluons, scalars and fermions to the triangle coefficient \eqref{contribtrcoeff} is complicated; these contributions do not seem to be simply related
to the tree amplitudes \eqref{treeamp4part}. Nevertheless, the relation \eqref{nobubn4theory} holds which is reassuring since it implies that the triangle coefficient vanishes in the ${\cn=4}$ theory, for any choice of external momenta and colors.
\item
Note that for generic ${\cn=1}$ and ${\cn=2}$ theories this coefficient does not vanish. For example, in the ${\cn=2}$, $SU(2)$ theory with $4$ hypermultiplets, taking all colors to be the same, this coefficient evaluates to $B = 48 y^4,$
with our choice of external momenta.
\item
However, for the theories described in section \ref{onlyboxes}, the triangle
coefficient vanishes for any choice of the external colors! While we had only
4 external gluons in this example, we have numerically verified this no-triangle property for up to 12 external gluons.
\item
The contribution of gluons, scalars and fermions to the bubble coefficient \eqref{bubble1} is also
quite complicated and is not proportional to the tree amplitude \eqref{treeamp4part}. 
Once again, for the $\cn=4$ theory, this coefficient vanishes because of \eqref{nobubn4theory}.
\item
However, this coefficient also vanishes in the Seiberg-Witten theory
and in fact in any theory that has at least $\cn=1$ supersymmetry and
vanishing one-loop $\beta$ function.
This coefficient also vanishes for the non-supersymmetric
$SU(2)$ theory described in section \ref{onlytriangles}.
\item
Finally, note that when we add the two bubble coefficients \eqref{bubble1} and
\eqref{bubble2} we get exactly the right contribution to the $\beta$
function as shown in \eqref{betafunc}.
Note, that while there are massive cancellations when we consider the net UV-divergence, all the terms in individual bubble coefficients are important for the scattering amplitude.
\end{enumerate}

\section{Conclusions and Discussion}
\label{sec:conclusions}
\begin{itemize}
\item
In this paper, we started by introducing a version of on-shell superspace for gauge theories with $\cn=1$ or $\cn=2$ supersymmetry. 
This representation led to new recursion relations for tree amplitudes in these theories that are described in section \ref{subsec:recurs}.
\item
In section \ref{sec:oneloop}, we used 
these recursion relations to show that the one-loop S-matrix of pure 
$\cn=1,2$ theories contains both triangles and bubbles.
\item
In section \ref{sec:matter}, we discussed gauge theories with 
arbitrary matter content. We found that, in non-supersymmetric theories,
bubble coefficients are sensitive only to the quadratic and fourth
order Indices of the matter representation. Triangle coefficients are sensitive
to the quadratic, fourth, fifth and sixth order Indices. 
\item
In \ref{subsec:simplesusy}, we argued that in supersymmetric theories, cancellations between fermions and bosons
lead to simplifications in the statements above. A ${\cn=1}$ 
chiral multiplet, in a given representation, contributes to bubbles 
only through its quadratic Index. It contributes to triangles through
its quadratic, fourth and fifth order Indices. 
\item
So demanding the absence of bubbles and triangles in a theory
leads to a set of linear Diophantine equations that involve
the higher order Indices of the matter representation. These equations are discussed 
in section \ref{sec:nexttosimplest}.
\item
In section \ref{onlyboxes}, we presented some new examples of theories that 
have only boxes and no triangles or bubbles. These include
the ${\cn=2}$ $SU(N)$ theory with a hypermultiplet in the symmetric tensor and 
another hypermultiplet transforming in the anti-symmetric tensor. 
We also provided some examples of theories without bubbles in \ref{onlytriangles}. At large $N$ it is possible to find other approximate solutions to these 
Diophantine equations and this is described in \ref{subsec:largeN}.
\end{itemize}

A natural question that arises from this analysis is whether the ${\cn=2}$ 
theory, with symmetric and anti-symmetric tensor hypermultiplets described
above, shares any of the other nice properties of the ${\cn=4}$ theory. An
immediately encouraging observation comes from calculating the 
central charges $a$ and $c$ in this theory. In the free-theory, we 
find (see \cite{Duff:1993wm,Gubser:1998vd} and references there) that
\begin{equation}
\begin{split}
a &= {2 n_v + n_h \over 12} = {3 N^2 - 2 \over 12} = {N^2 \over 4} 
- {1 \over 6}\\
c &= {5 n_v + n_h \over 24} = {6 N^2 - 5 \over 24}  = {N^2 \over 4} 
- {5 \over 24}
\end{split}
\end{equation}
So, in the large $N$ limit, $a$ and $c$ are the same up to subleading terms of $\o[{1 \over N^2}]$! This is a necessary condition for this theory
to have a gravity dual \cite{Aharony:1999ti}.

After this paper first appeared on the arXiv, we learned of the 
work \cite{Ennes:2000fu}, where this ${\cn=2}$ 
theory was studied; its gravity
dual was shown to be an orientifold of AdS$_5 \times S^5$ by T-dualizing
the brane construction found in \cite{Park:1998zh}. In particular, 
using the arguments of planar equivalence described in \cite{Armoni:2003gp},
this implies that planar correlation functions of this theory can be
obtained from the ${\cn=4}$ theory. Hence, many nice properties of the
planar limit of ${\cn=4}$ theory, such as dual superconformal invariance,
descend to this theory. 

We emphasize though, that this observation does {\it not subsume} our results,
since our analysis does not rely on large $N$. At
finite $N$, other daughters of the ${\cn=4}$ theory do posses triangles and
bubbles as emphasized in \ref{subsec:largeN} but the special ${\cn=2}$ theory described above does not.

It would also be interesting to extend the analysis in this paper
to higher loops. In an appropriately chosen scheme, the beta function for ${\cn=1}$ and ${\cn=2}$ theories is one-loop exact. What simplifications
does this imply for the S-matrix of ${\cn=1}$ and ${\cn=2}$ theories
at higher-loops?

Moving away from supersymmetry, we would also like to find some examples of
non-supersymmetric theories without triangles or learn of a proof that 
there are no such theories.
Finally, it would be very interesting to extend the study in this paper to supergravity theories with less than $\cn=8$ supersymmetry.

\acknowledgments
We would like to thank  Shamik Banerjee, R. Loganayagam, Ashoke Sen and 
especially Rajesh Gopakumar for several very helpful discussions.

\appendix
\section*{Appendices}
\section{Mimicking the Adjoint}
\label{app:solveprtrace}

In this appendix, we indicate how one may go about systematically looking for solutions to 
\eqref{nonsusycond} and \eqref{susycond}. Stated abstractly, the problem we are concerned with, is as follows. Is there any group, G 
with a representation R (not necessarily irreducible) that has the following property:
\begin{equation}
\label{mimicadjointapp}
{\rm Tr}_{\rm R}(\Pi_{i=1}^n T^{a_i}) =  m\,{\rm Tr}_{\rm adj}(\Pi_{i=1}^n T^{a_i}),~ n \leq p,
\end{equation}
where $T^{a_i}$ are generators of the group and ${\rm Tr}_{\rm R}$ means trace in the representation $R$.  For non-supersymmetric
theories, the S-matrix simplifies if we can find solutions to \eqref{mimicadjointapp} for the parameters shown in Table \eqref{nonsusyconds}.
\TABLE{
\label{nonsusyconds}
\begin{tabular}{|c|c|c|}\hline
&Only Boxes & No Bubbles \\ \hline
$R_f$ satisfies \eqref{mimicadjointapp} with&p=6, m=4&p=4, m=4\\\hline
$R_s$ satisfies \eqref{mimicadjointapp} with&p=6, m=6&p=4, m=6\\\hline
\end{tabular}
\caption{Conditions for non-supersymmetric theories}
}

For supersymmetric theories, these conditions simplify to those of Table \ref{susyconds}.
\TABLE{
\label{susyconds}
\begin{tabular}{|c|c|c|}\hline
&Only Boxes & No Bubbles \\ \hline
$R_{\chi}$ satisfies \eqref{mimicadjointapp} with&p=5, m=3&p=2, m=3\\\hline
\end{tabular}
\caption{Conditions for supersymmetric theories}
}
Tables \ref{susyconds} and \ref{nonsusyconds} contain the same 
information as Table \ref{simpleconds}.
The representation $R_s$ in Table \ref{nonsusyconds} must be self-conjugate. The $\cn=4$ theory provides a trivial solution to the constrains of Table \ref{nonsusyconds} with ${\rm R}_f = 4 \cdot {\bf adj}, {\rm R}_s = 6 \cdot {\bf adj}$ and to the constraints of Table \ref{susyconds} with $R_{\chi} = 3 \cdot {\bf adj}$.  We are interested in replacing some or all of the adjoints by other representations.

First, let us elaborate on the statement that \eqref{mimicadjointapp}
is the same as demanding the equality of {\em symmetrized} traces up to order $p$. It is clear that \eqref{mimicadjointapp} implies
the equality of the symmetrized traces on both sides. However, the converse is also true. This is because any trace of $n$ generators can be written 
as a sum of symmetrized traces of products of up to $n$ generators with coefficients that depend only on the structure-constants. For example
\begin{equation}
{\rm Tr}(T^a T^b T^c) = {1 \over 2} \left[{\rm Tr}(\{T^a,T^b\}T^c) + {\rm Tr}([T^a,T^b]T^c)\right] = {\rm Tr}(T^{(a}T^bT^{c)}) + {f^{a b d} \over 2} {\rm Tr}(T^d T^c).
\end{equation}

As we described in section \ref{subsec:detourintoindices}, in a simple
Lie algebra, a symmetrized trace of a product of generators can be expanded in terms of the invariant tensors of the algebra multiplied by numbers 
that are called Indices. Two commonly encountered examples of this property
are the statements that
\begin{equation}
\tr_R(T^a T^b) = I_2(R) \kappa^{a b}, \quad
{1 \over 2} \tr_R\left(T^a \{T^b,T^c\}\right) = I_3 (R) d^{a b c}.
\end{equation}
where $\kappa^{a b}$ is the Killing form and $d^{a b c}$ is a suitable 
defined completely symmetric tensor. The coefficient $I_2(R)$ is called the 
quadratic Index and the third order Index $I_3(R)$ is called the anomaly. 
In general, we have
\begin{equation}
\tr_R\left(T^{(a_1} T^{a_2} \ldots T^{a_n)}\right) = I_n (R) d^{a_1 a_2 \ldots a_n} + {\rm products~of~lower~order~tensors}.
\end{equation}
The coefficients that appear in this expansion are called Indices.
A systematic study of the higher order Indices and their relation to 
higher-order Casimir invariants was undertaken in the early eighties by Okubo and Patera \cite{okubo:219,okubo:2722,okubo:8,okubo:2382}. We also refer the reader to the detailed paper \cite{vanRitbergen:1998pn}.

These Indices are additive
\begin{equation}
I_n({\rm R}_1 \oplus  {\rm R}_2) =  I_n({\rm R}_1) +  I_n({\rm R}_2).
\end{equation}
This implies that if we write $R = \bigoplus n_i R_i$, where the $R_i$ are irreducible representations, \eqref{mimicadjointapp} is equivalent to 
a set of linear Diophantine equations in the variables $n_i$ that involve the
higher order Indices of the representations $R_i$. 

Using the detailed formulas for Indices and Casimir invariants, up to fourth order given in \cite{okubo:8} and the relations between the different Indices
at each order given in \cite{vanRitbergen:1998pn}, we can check
the solution \eqref{n2susybubvanish}. For the group $SU(n+1)$ it is useful
to work in terms of $n+1$ orthogonal weights, $o_i$, rather than the $n$ weights in the Dynkin basis. The relation between these two bases is given in 
\eqref{anorth}. In this orthogonal basis, the adjoint, symmetric tensor and anti-symmetric tensor are given by
\begin{equation}
\begin{split}
&{\rm \bf adj:}~ o_1=-o_{n+1}=1;\quad o_i = 0, ~{\rm otherwise}\\
&{\rm \bf sym:}~ o_1 = {2 n \over n+1};\quad o_{i \geq 2} = {-2 \over n+1} \\
&{\rm \bf asym:}~ o_1 = o_2 = {n-1 \over n+1}; \quad o_{i \geq 3} = {-2 \over n+1}.
\end{split}
\end{equation}
We will also need the half-sum of positive roots in this basis:
\begin{equation}
\rho_i = {n \over 2} + 1 - i.
\end{equation}
For each representation, we define the auxiliary quantities
\begin{equation}
\begin{split}
&\tilde{c}_2 = \sum_{i=1}^{n+1} (o_i + \rho_i)^2 -  \sum_{i=1}^{n+1} (\rho_i)^2,\\
&\tilde{c}_4 = \sum_{i=1}^{n+1} (o_i + \rho_i)^4 -  \sum_{i=1}^{n+1} (\rho_i)^4,\\
&\tilde{c}_{22} = \left[\sum_{i=1}^{n+1} (o_i + \rho_i)^2\right]^2 -  \left[\sum_{i=1}^{n+1} (\rho_i)^2\right]^2,
\end{split}
\end{equation}
while the dimensions of the three representations are
\begin{equation}
d_{\rm ad} = (n+1)^2 - 1,~d_{\rm sy} = {(n+1)(n+2) \over 2},~d_{\rm as} = { n (n+1) \over 2}.
\end{equation}
$\tilde{c}_2$ is the famous quadratic Casimir; the fourth order 
Casimir is not unique but in the convention of  \cite{okubo:8}, it is given
by a linear combination of $\tilde{c}_{22}$ and $\tilde{c}_4$.
 
The Indices and the Casimirs differ by a factor of the dimension of the representation. Using the explicit formulas
in \cite{okubo:8} and \cite{vanRitbergen:1998pn}  the reader can check that the Indices of \eqref{n2susybubvanish} mimic the Indices of the adjoint up to fourth order, because the 
following relations hold:
\begin{equation}
\begin{split}
&d_{\rm ad} \tilde{c}_2^{\rm ad} = d_{\rm sy} \tilde{c}_2^{\rm sy} + d_{\rm sy} \tilde{c}_2^{\rm sy}, \\
&d_{\rm ad} (\tilde{c}_2^{\rm ad})^2 = d_{\rm sy} (\tilde{c}_2^{\rm sy})^2 + d_{\rm as} (\tilde{c}_2^{\rm as})^2, \\
&d_{\rm ad} \tilde{c}_{22}^{\rm ad} = d_{\rm sy} \tilde{c}_{22}^{\rm sy} + d_{\rm as} \tilde{c}_{22}^{\rm as},\\
&d_{\rm ad} \tilde{c}_{4}^{\rm ad} = d_{\rm sy} \tilde{c}_{4}^{\rm sy} + d_{\rm as} \tilde{c}_{4}^{\rm as}.
\end{split}
\end{equation} 
The conjugate representations in \eqref{n2susybubvanish} are required 
to make the 
representation anomaly free and to make the fifth order Indices vanish as 
they do for the adjoint

We can also check \eqref{g2withonlyboxes}. According to \cite{Okubo:1978qe}, for any generator $X$ of $G_2$, in any representation $\lambda$, we have
\begin{equation}
\tr(X^4) = K(\lambda) \tr(X^2)^2 = {d_{ad} \over 2 (2 + d_{ad}) d_{\lambda}} \left(6 - 
{c_2^{ad} \over c_2^{\lambda}} \right) \tr(X^2)^2,
\end{equation}
where $d_{ad}, d_{\lambda}$ are the dimensions of the adjoint and $\lambda$ and $c_2$ is the quadratic Casimir. In light of this, all we need to do to verify \eqref{g2withonlyboxes} is check that
\begin{equation}
\begin{split}
3 I_2({\bf 14}) &= 3 I_2({\bf 7}) + I_2({\bf 27}),  \\
3 K({\bf 14}) I_2({\bf 14})^2 &=  3 K({\bf 7}) I_2({\bf 7})^2 + K({\bf 27}) I_2({\bf 27})^2.
\end{split}
\end{equation}
These equations are true, since if we normalize $I_2({\bf 7}) = 1$, then we have
\begin{equation}
\begin{split}
I_2({\bf 7}) &= 1, \quad I_2({\bf 14}) = 4, \quad I_2({\bf 27}) = 9, \\
c_2^{\bf 7} &= 2, \quad c_2^{\bf 14} = 4, \quad c_2^{\bf 27} = {14 \over 3}.
\end{split}
\end{equation}

Finally, it is useful to rephrase \eqref{mimicadjointapp} as follows. Define
\begin{equation}
t \equiv  \sum_{a=1}^{\rm dim(G)} x_a T^a.
\end{equation}
Then, \eqref{mimicadjointapp} is equivalent to demanding
\begin{equation}
\label{mimicsymmetrized}
{\rm Tr}_R t^n = m \, {\rm Tr}_{\rm adj} \, t^n, \quad n = 1 \ldots p.
\end{equation}

This problem can be simplified even further. Equation \eqref{mimicsymmetrized}
is equivalent to demanding that
\begin{equation}
\label{mimicexponent}
{\rm Tr}_R e^{\l t} = m \, {\rm Tr}_{\rm adj} \, e^{\l t} + \o[\l^0] + {\rm O}(\l^{p+1}).
\end{equation}
On both sides of \eqref{mimicexponent}, we now have a group character. The $\o[\l^0]$ term enters because the two representations might have different 
dimensions. It is well known that we can always rotate $e^t$ via conjugation into an element of 
the maximal torus
\begin{equation}
{\rm Tr}\left(e^{\l t}\right) = {\rm Tr}\left(e^{\l H}\right),
\end{equation}
where $H$ is a member of the Cartan subalgebra and the statement is independent
of the representation in which the trace is taken.
Hence, if we consider the group character
\begin{equation}
\chi_{\rm R} (\theta_i) \equiv {\rm Tr}_{\rm R}\, e^{\sum_{i=1}^n\theta_i T^i},
\end{equation}
(where we emphasize that the sum in the exponent runs only over the elements 
of the Cartan subalgebra), then \eqref{mimicadjointapp} is equivalent to:
\begin{equation}
\chi_{\rm R}(\l \theta_i) = m \chi_{\rm adj}(\l \theta_i) + \o[\l^0] + {\rm O}(\l^{p+1}).
\end{equation}
This equation is computationally easier to check than \eqref{mimicadjointapp}. 
The reader can use this to check the validity of the solutions \eqref{g2withonlyboxes} and \eqref{su2bubvanish}.

\subsection{Character Formulae}
For the readers convenience, we list here the characters for the various infinite series of simple Lie groups -- $A_n, B_n, C_n, D_n$.

\subsubsection{$A_n$}
This is the group $SU(n+1)$. The representations of $SU(n+1)$ can be conveniently encoded in Young diagrams. We will use $(d_1, \ldots d_n)$ to indicate the Young diagram with $d_n$ columns of length $n$. For the purpose of calculating the character, it is convenient to go to the so-called 
orthogonal basis \cite{fuchs1997sla}. We define the $n+1$ numbers $o_1, \ldots o_{n+1}$ by
\begin{equation}
\label{anorth}
o_j = \sum_{i=j}^n d_i - {1 \over n+1} \sum_{i=1}^n i d_i.
\end{equation}
If we choose the Cartan subalgebra to comprise generators $T^i$ (which, when acting on the highest weight state have eigenvalues $d_i$), then the character
\begin{equation}
\chi_R(\vec{\theta}) = \tr_R e^{\theta_i T^i}, 
\end{equation}
is conveniently encoded in the variables
\begin{equation}
\phi_1 = \theta_1; \quad \phi_{1 < i \leq n} = \theta_{i}-\theta_{i-1}; \quad \phi_{n+1} = -\theta_n.
\end{equation}
With this definition,
\begin{equation}
\chi_R = {\det\left[\exp\left({\phi_i (o_j + {n \over 2} + 1 - j)}\right)\right] \over \det\left[\exp\left({\phi_i ({n \over 2} + 1 - j)}\right)\right]}.
\end{equation}

\subsubsection{$B_n$ }
$B_n$ is the group $SO(2 n +1)$. The orthogonal basis is a natural basis to use here.  We write the highest weights of a representation in  the orthogonal basis as $o_1 \ldots o_n$; these are the weights under rotations in orthogonal half-planes in $R^{2 n +1}$.   These  weights are positive and must be either all integral or all half-integral. Furthermore, if $j > i$, then $o_i > o_j$. (We refer the reader to \cite{fuchs1997sla} for the conversion to the Dynkin basis.)
With this choice of basis for the Cartan subalgebra, the character is
\begin{equation}
\label{bnchar}
\chi_R =  {\det\left[\sinh\left({\theta_i (o_j + n + {1 \over 2} - j)}\right)\right] \over \det\left[\sinh\left({\theta_i (n + {1 \over 2} - j)}\right)\right]}.
\end{equation}

\subsubsection{$C_n$}
$C_n$ is the group $Sp(2 n)$.  In the orthogonal basis, the Cartans of $C_n$, in the fundamental representation, are $2n \times 2n$ matrices with elements ${\rm diag}(i \sigma_2, 0, 0 \ldots ), {\rm diag}(0, i \sigma_2, 0, 0, \ldots), \ldots$, where each $0$ is shorthand for a $2 \times 2$ matrix. The weights  in the orthogonal basis, $o_i$, must be all integral. Furthermore,just as in the case of $B_n$, if $j > i$, then $o_i > o_j$. 
The character is then given by
\begin{equation}
\label{cnchar}
\chi_R = {\det\left[\sinh\left({\theta_i (o_j + n + 1 - j)}\right)\right] \over \det\left[\sinh\left({\theta_i (n + 1 - j)}\right)\right]}.
\end{equation} 

\subsubsection{$D_n$}
This is the group $SO(2 n)$. Once again we write the weights in an orthogonal
basis as $o_1, \ldots o_n$. If $j > i$, $|o_i| > |o_j|$. The weight $o_n$ can be negative whereas all others must be positive. As for $B_n$, the $o_i$ must
be either all integral or all half-integral. The formula for the character is
\begin{equation}
\label{dnchar}
\chi_R = {\det\left[\sinh\left(\theta_i (h_j + n - j)\right)\right]+\det\left[\cosh\left(\theta_i (h_j + n - j)\right)\right] \over \det\left[\cosh\left(\theta_i (n - j)\right)\right]}.
\end{equation}

\section[A Loose End]{A Loose End and Some Explicit Calculations}
\label{explicitchecks}
\subsection{The amplitude with many gluons and a gaugino}
We will now complete the proof outlined in the section \ref{subsec:simplesusy}
using induction. We have already shown that subject to the condition \eqref{largezspinorchoice}, both scalar and fermion amplitudes of the type shown in figure \ref{manygluon2matter} can be expanded as shown in \eqref{scalarpowers}. We now wish to prove two additional statements. First that, in supersymmetric theories, 
\begin{equation}
c^s_{a_1 \ldots a_n} - c^f_{a_1 \ldots a_n}= \o[{1 \over z}],
\end{equation}
and second that the contribution of an amplitude with 1 negative 
helicity gaugino, a negative helicity chiral multiplet fermion 
and a chiral multiplet scalar to the n$^{\rm th}$ symmetrized product of 
generators can grow no faster than ${1 \over z^{n-2}}$.

Let us assume that the first statement is true for amplitudes involving
up to $n$ gluons. We will refer to the statement as $P(n)$. Let us 
also assume that the second statement is true for amplitudes with $1$ gaugino
and $n$ gluons. We will refer to this as $Q(n)$. Then the 
supersymmetry argument given in the text shows that $ Q(n) \Rightarrow P(n+1)$. We will now show that $P(n) \wedge Q(n) \Rightarrow Q(n+1)$. 

\FIGURE{
\epsfig{file=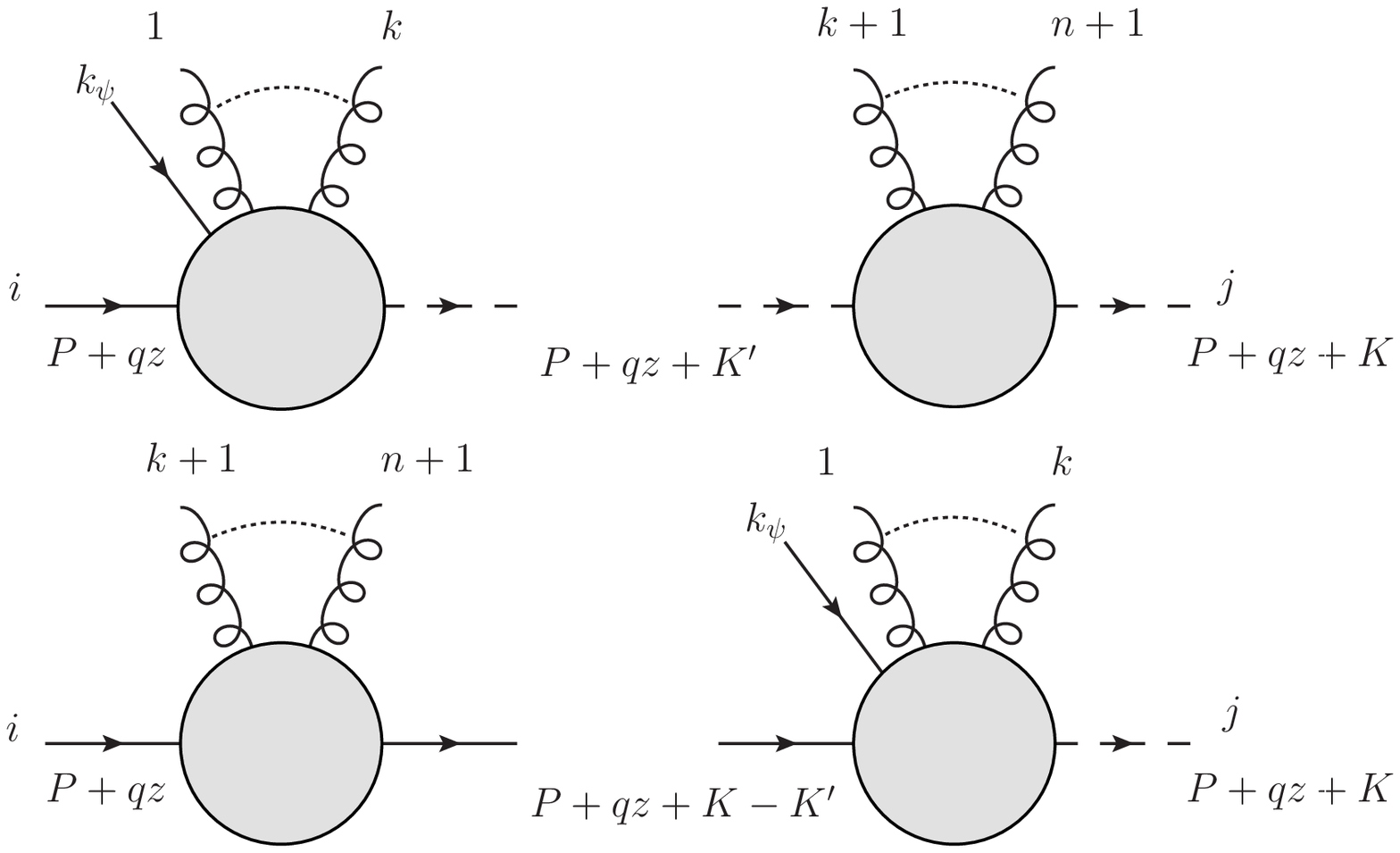,width=0.8\textwidth}\caption{BCFW for the amplitude with 1 gaugino}\label{onegaugino}}

The amplitude with 1 gaugino and  $n+1$ gluons can be computed 
via BCFW recursion. There are two interesting terms in this recursion. These are shown in the figure
\ref{onegaugino}.
By the hypotheses $P(n)$ and $Q(n)$, the first line leads to
\[
\left[\sum_{p=1}^{n-k+1} \sum_{q=1}^{k+1} { c^2_{a_1 \ldots a_p} T^{(a_1} \ldots T^{a_p)} \over z^{p-2}} {c^1_{a_1 \ldots a_q} T^{(a_1} \ldots T^{a_q)} \over z^{q-2}} \right]{1 + \o[{1 \over z}] \over 2 q z \cdot K'},
\]
wit $K' = k_{\psi} + \sum_{i = 1}^k k_i$.
The second line leads to
\[
\left[\sum_{p=1}^{n-k+1}\sum_{q=1}^{k+1} {c^4_{a_1 \ldots a_q} T^{(a_1} \ldots T^{a_q)} \over z^{q-2}} {c^3_{a_1 \ldots a_p} T^{(a_1}\ldots T^{a_p)} \over z^{p-2}} \right]{-1+ \o[{1 \over z}] \over 2 q z  \cdot K'}.
\]
However, by hypothesis $P(n)$,
\begin{equation}
c^3 - c^2 = \o[{1 \over z}].
\end{equation}
Also,
\begin{equation}
c^4 - c^1 = \o[{1 \over z}],
\end{equation}
since both $c^4$ and $c^1$ are analytic functions of the external spinors
and the external gluon momenta 
which only differ in the subleading terms in the two cases. Adding the 
two terms, we now get a commutator and subleading terms
\[
\begin{split}
&\left[\sum_{p,q} { c^1_{a_1 \ldots a_p} c^2_{a_1 \ldots a_q} + \o[{1 \over z}]\over z^{p + q -3 }} [T^{(a_1} \ldots T^{a_p)},  T^{(a_1} \ldots T^{a_q)}] \right]{1 \over 2 q  \cdot K'}  \\
&+ \o[{1 \over z^{p+q-2}}]  \{T^{(a_1} \ldots T^{a_p)},  T^{(a_1} \ldots T^{a_q)}\}.
\end{split}
\]
Since the commutator above can be written as a symmetrized product of at most
$p+q-1$ generators, we have shown that $P(n) \wedge Q(n) \Rightarrow Q(n+1)$.
This completes our proof.

\subsection{Explicit Checks}

\TABLE{
\caption{Dominant diagrams at large $z$}
\label{domtable}\centering
\begin{tabular}{|m{0.45\textwidth}|m{0.45\textwidth}|}
\hline
\begin{center} Scalars \end{center} & \begin{center} Fermions \end{center} \\ \hline
\begin{center}\includegraphics*[height=2.4cm]{largez1.eps}\end{center}&\begin{center}\includegraphics*[height=2.4cm]{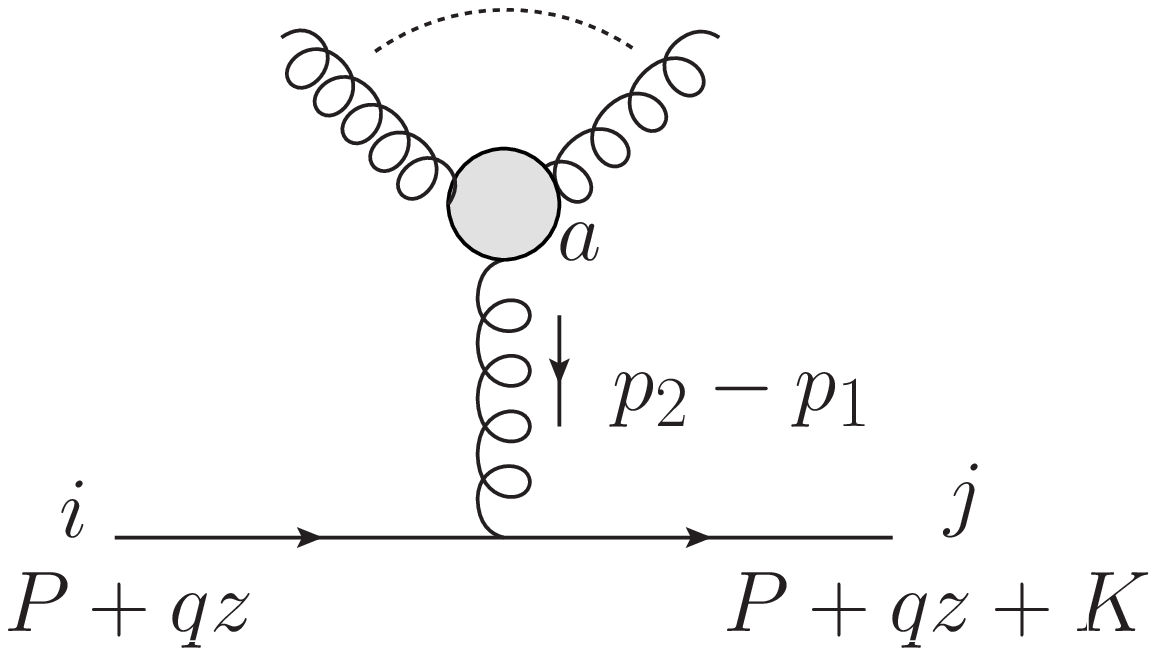}  \end{center}\\ \hline
\begin{center} \includegraphics*[height=2.4cm]{largez3.eps} \end{center}&\begin{center}\includegraphics*[height=2.4cm]{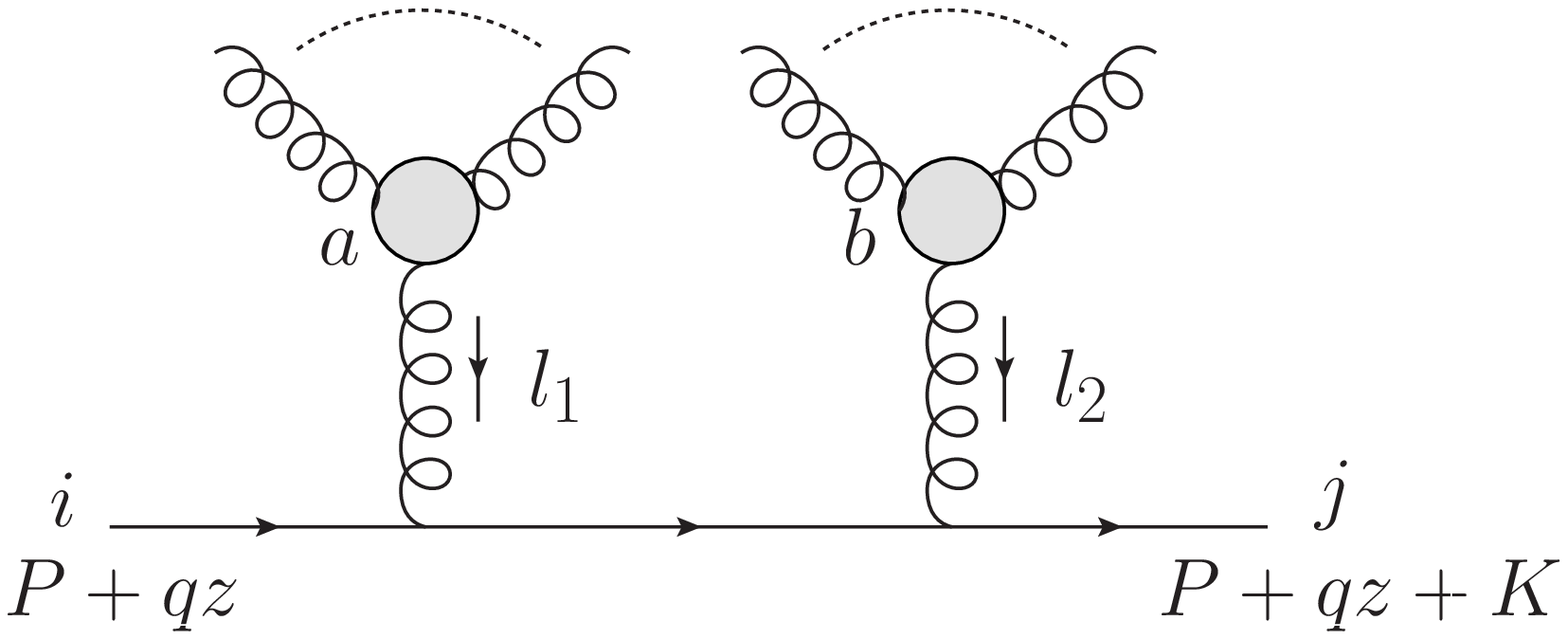} \end{center}\\\hline
\begin{center} \includegraphics*[height=2.4cm]{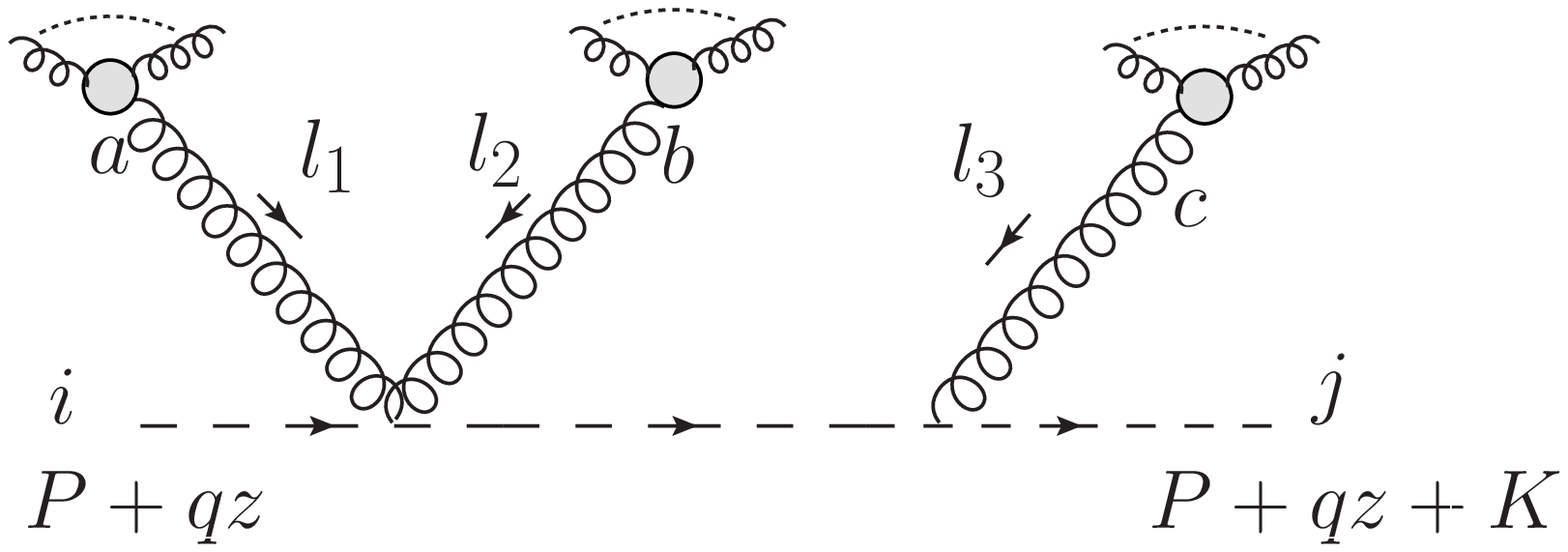} \end{center}&\begin{center}\includegraphics*[height=2.4cm,width=7.0cm]{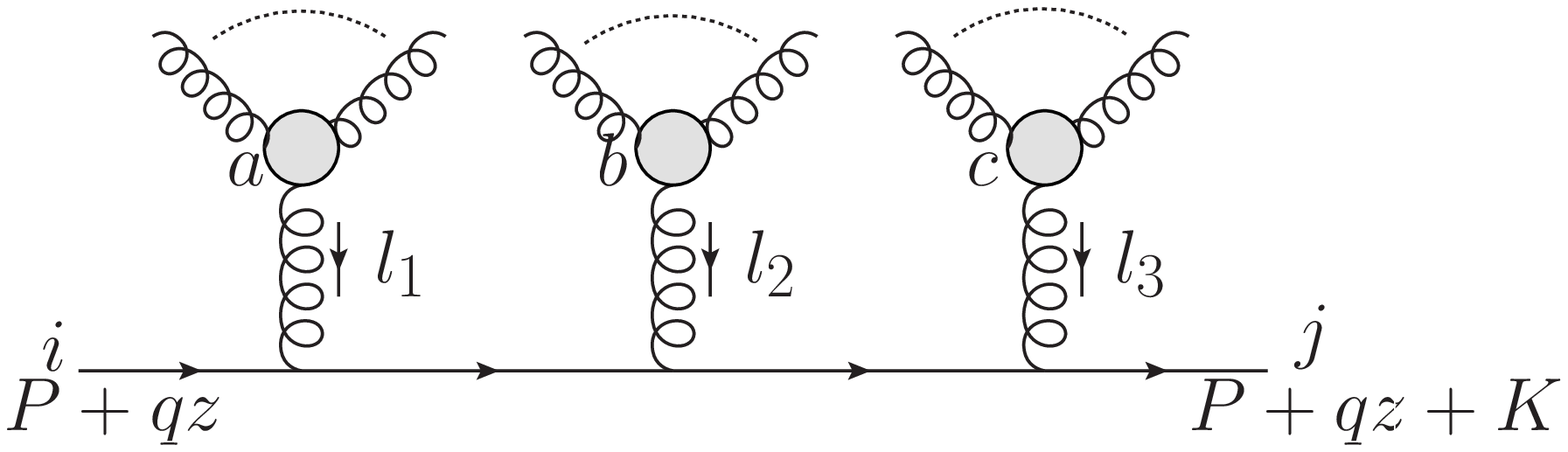} \end{center}\\\hline
\begin{center} \includegraphics*[height=2.2cm]{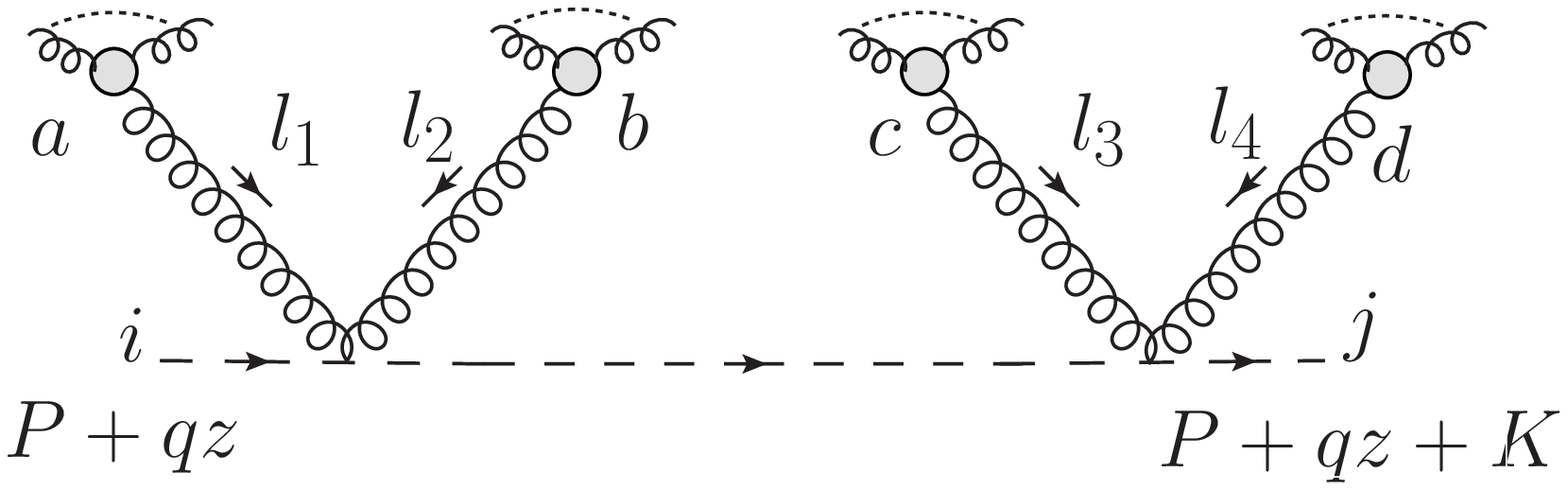} \end{center}&\begin{center}\includegraphics*[height=2.2cm,width=7.0cm]{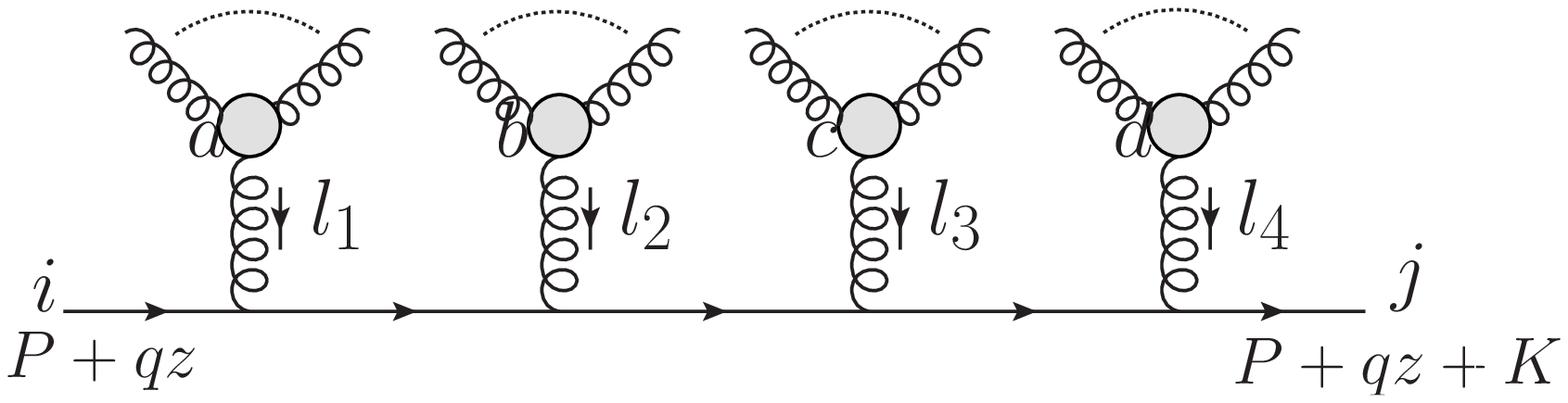} \end{center}\\\hline
\end{tabular}}

We now perform explicit checks of \eqref{susycsame} by drawing Feynman diagrams
in q-lightcone gauge. As we described in section \eqref{subsec:manygluon2matter}, the advantage of this gauge is that it lets us write down all diagrams
that contribute to the symmetrized product of generators at some order. In 
Table \ref{domtable}, we have drawn these diagrams for both scalars and fermions. The diagrams on the first line give us the leading coefficient of $T^a$; the diagrams on the second line give us the leading contribution to the 
symmetrized product of two generators. The diagrams on the third and fourth line
give us the leading contribution to the symmetrized product of three 
generators. We emphasize that the gluon lines shown in these diagrams
do not necessarily belong to external gluons. They could as well 
come come from a gluon propagator that connects to the
 external gluons through the blobs shown in the diagrams. As in \ref{subsec:manygluon2matter}, we 
have parametrized the interactions of the matter fields with the gluon lines through background vectors $A_i$. These carry momentum $l_i$ and to lighten
the notation we take the color-index also to be $i$. Also,  $K = \sum l_i$ and
we always keep the decomposition \eqref{largezspinorchoice} in mind. 
Our objective here is to verify \eqref{susycsame} using explicit
calculations in q-lightcone gauge. We put $g = 1$ below but keep track of 
all other factors.

\subsubsection*{Single Generator}
The coefficient $c^1$ in \eqref{scalarpowers} is dominated
by diagrams that have a single gluon-matter vertex. The first line of 
Table \ref{domtable} shows this interaction for both fermions
and scalars. The scalar diagram evaluates to 
\begin{equation}
{\cal M}^s = \left(2 i q z \cdot A_1 + \o[1]\right)T^1_{j i}.
\end{equation}
The fermionic amplitude is (reading the Feynman diagram in the first line of Table \ref{domtable} against the line)
\begin{equation}
{\cal M}^f = i T^1_{j i} \lb_{\dot{\alpha}} A_1^{\dot{\alpha} \beta} \l_{\beta} z  = 2 i q z \cdot A_1 T^1_{j i} + \o[1].
\end{equation}
 Note, that the $\lb$ on the left, comes from the cut-line on the right. In principle,
we could have chosen to decompose the momentum here into different spinors than the 
choice we made for the cut-line on the left. We would then not have got an 
answer that matched with the scalar answer. This is the relevance of condition \eqref{largezspinorchoice}.

\subsubsection*{Symmetrized Product of 2 Generators}
With two gauge-boson matter vertices we can get a term with a symmetrized
product of two generators. The scalar and fermion 
amplitudes are dominated by the diagrams on the second line of 
Table \ref{domtable} and we find
\begin{equation}
{\cal M}^s = i A_1 \cdot A_2  \{T^1, T^2\}_{j i}.
\end{equation}
Note, that the q-lightcone gauge implies that it must be possible to write $A_1$ and $A_2$ as
\begin{equation}
A_1^{\alpha \dot{\alpha}} = \l^{\alpha} \ab_1^{\dot{\alpha}} + \lb^{\dot{\alpha}} a_1^{\alpha}.
\end{equation}
where $ a_1$ and $\ab_1$ are some spinors. 
Summing over the two arrangements of the gluon lines, we find the large z 
fermionic contribution
\begin{equation}
{\mathcal M}_f = -i (\lb_{\dot{\alpha}} \ab_1^{\dot{\alpha}} \l^{\beta} (l_1)_{\beta \dot{\gamma}} \lb^{\dot{\gamma}} a_2^{\nu} \l_{\nu}) 
{ T^1 T^2 \over q \cdot l_1 z} + 1 \leftrightarrow 2 + \o[{1 \over z}].
\end{equation}
Note the advantage of the q-lightcone gauge here. the momentum running
between lines 1 and 2 is actually $P + q + l_1$. However,
the $q$-lightcone gauge tells us that if we decided to contract $A_1$ and $A_2$
with the $q$ in the center, then we would pick up the subleading pieces $\lb_s^2$ and $\l_s^1$ on the right and the left. This would give us an overall power of ${1 \over z}$. 
The contribution to the symmetric term is obtained by adding the coefficients
of the two color-factors above. This is
\begin{equation}
\begin{split}
{\mathcal M}_f &= -i {\{T^1,T^2\}\over 2} \left( \dotl[a^1, \l] \dotlb[\lb, \ab^2] + \dotl[a^2, \l] \dotlb[\lb, \ab^1] + \o[{1 \over z}] \right) + \ldots \\
&= \left(i A_1 \cdot A_2 + \o[{1 \over z}]\right)\{T^1,T^2\} + \ldots
\end{split}
\end{equation}
where the $\ldots$ denote terms proportional to $[T^1, T^2]$. So the contribution to the second order symmetrized product from scalars and fermions is again the same.

Note that while the contribution to the symmetric product is the same as that of the scalar, the contribution to the commutator term is not. However, the commutator is effectively a single-generator term 
this contribution is subleading to the leading single-generator term that we
evaluated above which is $\o[z]$.
\subsubsection*{Symmetrized Product of 3 Generators}
We now turn to the diagram with 3 external lines. We consider three gluon lines carrying momenta $l_i$. This is somewhat more complicated than above. For the scalar, summing
the diagram in line 3 of table \ref{domtable} and the other permutations of the gluon lines, we get
\begin{equation}
\begin{split}
{\mathcal M}^s &= i^3 {\left((2 P + l^3) \cdot A_3\right) \left(A_1 \cdot A_2\right) \over q z \cdot l^3} \{T^2, T^1\} T^3 + 
i^3 {\left((2 (P + l^1 + l^2) + l^3) \cdot  A_3\right) \left(A_1 \cdot A_2\right) \over q z \cdot(P + l^1 + l^2)} T^3 \{T^1,T^2\} \\
&+ {\rm 2~permutations} + \o[{1 \over z^2}] \\
&= \left[{2 i \over z} {(l_1 + l_2) \cdot A_3 \over q \cdot l_3} + {\rm 2~permutations} + \o[{1 \over z^2}]\right]T^{(1}T^2T^{3)}  + \ldots
\end{split}
\end{equation}
On the other hand defining $\chi_{\alpha \dot{\alpha}} = (\l z + \l_s^1)_{\alpha} (\lb + {\lb_s^2 \over z})_{\dot{\alpha}}$, the fermionic amplitude can be conveniently written as a trace
\begin{equation}
\begin{split}
{\mathcal M}_f &= i^5 {\tr\left[(\chi \cdot \sigma) (A_3 \cdot \bar{\sigma}) \left((P + q z + l^1 + l^2)\cdot \sigma\right)(A_2 \cdot \bar{\sigma})\left((P + q z + l^1)\cdot \sigma\right)(A_1 \cdot \bar{\sigma})\right] \over (2 q z \cdot l^1) (2 q z \cdot (l^1 + l^2))} \, T^3 T^2 T^1 \\
&+ {\rm 6~permutations} + \o[{1 \over z^2}].
\end{split}
\end{equation}
While the $\sigma$ matrix algebra is somewhat complicated here, it is quite 
easy to check with the help of a computer that the contribution
to the completely symmetrized third order product is the same to leading order
\begin{equation}
{\mathcal M}^f - {\mathcal M}^s = \o[{1 \over z^2}] T^{(1}T^2 T^{3)} + \ldots
\end{equation}
\subsubsection*{4 External Lines} 
At first sight, the diagram on the last lines of Table \ref{domtable} might
seem confusing. This seems to give a term with 4 generators but only comes
with a $\o[{1 \over z}]$. However, there are six other diagrams that
contribute to the same 4$^{\rm th}$ order symmetrized product that comes from
this diagram. In particular, the diagram shown
gives a contribution 
\begin{equation}
\label{fourlinediagrams}
i^3 \left[{(A_1 \cdot A_2) (A_3 \cdot A_4) \over 2 z}\right] \left[ {\{T^3, T^4\} \{T^1, T^2 \} \over q \cdot (l_1 + l_2)}\right] + \o[{1 \over z^2}].
\end{equation}
However, there is another diagram for which  the first bracket is the same but the term in the second bracket is different. This arises from permuting the 4-pt vertices in the figure. Adding these terms, we find that the contribution to the amplitude is given by
\begin{equation}
\label{sixpermutations}
\begin{split}
i^3 \left[{(A_1 \cdot A_2) (A_3 \cdot A_4) \over 2 z}\right] \times \left[ {\{T^3 T^4\} \{T^1 T^2 \} \over q \cdot (l_1+l_2)} 
+  {\{T^1 T^2\} \{T^3 T^4 \} \over q \cdot (l_3 + l_4)}\right] + \o[{1 \over z^2}].
\end{split}
\end{equation}
We now note that
\begin{equation}
q \cdot (l_1 + l_2 + l_3 + l_4) = 0,
\end{equation}
because both the BCFW-extended legs are null. This means that the term in \eqref{sixpermutations} simplifies to
\begin{equation}
\left[{(A_1 \cdot A_2) (A_3 \cdot A_4) \over z}\right] {\left[ \{T^3 T^4\}, \{T^1 T^2 \}\right] \over (q \cdot (l_1+l_2))} + \o[{1 \over z^2}].
\end{equation}
However the term above can be written as a symmetrized trace of up to 3 generators. As we expect, this comes with a factor of $\o[{1 \over z}]$. 
The proof that the fermionic term shown on the last line of Table \ref{domtable} gives the same leading contribution to the completely
symmetric part of the amplitude is very similar to the proof given 
for the case with 2-external lines and we will not repeat it here.

Note that \eqref{fourlinediagrams} does give a  $\o[{1 \over z^2}]$ contribution to the 4$^{\rm th}$ order symmetrized product even after we permute the gluon 
lines. We also need to consider 
another set of diagrams involving one 4-pt vertex and 2 3-pt vertices. The
sum of these two sets of diagrams matches the leading fermionic contribution to the 4$^{\rm th}$ order product.
In a very similar spirit, one can check that the diagrams with 5 and 6 external lines that give a contribution to the 4$^{\rm th}$ order symmetrized 
product are the same, to leading order, for fermions and scalars.

\section[A Tree Amplitude]{Feynman Diagrams vs BCFW: An Explicit Tree Amplitude}
In this appendix, we perform a simple calculation that shows, how our new recursion relations work in detail.  
We will consider a $2 \rightarrow 2$ scattering processes in a $\cn=1$ gauge theory without matter.

We take our particles to have momenta $k_i$ (marked as ingoing) with a spinor decomposition $k_i^{\alpha \dot{\alpha}} = \l_i^{\alpha}\lb_i^{\dot{\alpha}}$, with $i = 1 \ldots 4$. We take $b_1 = b_2 = 0, b_3 = b_4 = 1$. We will label the 
first two Grassmann parameters by $\xi^1,\xi^2$, and the third and fourth by $\eta1,\eta^2$. Our analysis here is valid for a general gauge group with structure constants $f^{abc}$.

In this example, we will show --- through a direct evaluation of Feynman diagrams --- that this scattering amplitude has the right behavior when the first two momenta are BCFW extended.

We perform the BCFW extension
\begin{equation}
\label{bcfwexample}
\l_1 \rightarrow \l_1 + z \l_2,\: \lb_2 \rightarrow \lb_2 - z \lb_1, \: \xi_1 \rightarrow \xi_1 + z \xi_2,
\end{equation}
and expand the scattering amplitude in a power series in the Grassmann parameters. Keeping only the non-vanishing terms, this leads to
\begin{equation}
\begin{split}
&{\mathcal A}^{\mathrm t}(\xi^1(z),\xi^2,\eta^1,\eta^2) = {\mathcal A}^{\mathrm t}(g^{-},g^{-},g^{+},g^{+})+\xi^{1}\eta^{2}{\mathcal A}^{\mathrm t}(f^{-},g^{-},g^{+},f^{+}) +\xi^{1}\eta^{1}{\mathcal A}^{\mathrm t}(f^{-},g^{-},f^{+},g^{+})\\ &+\xi^{1}\xi^{2}\eta^{1}\eta^{2}{\mathcal A}^{\mathrm t}(f^{-},f^{-},f^{+},f^{+}) 
+ \xi^{2}\eta^{2}\left[ {\mathcal A}^{\mathrm t}(g^{-},f^{-},g^{+},f^{+}) +z{\mathcal A}^{\mathrm t}(f^{-},g^{-},g^{+},f^{+})\right]\\
&+\xi^{2}\eta^{1}\left[ {\mathcal A}^{\mathrm t}(g^{-},f^{-},f^{+},g^{+})+z{\mathcal A}^{\mathrm t}(f^{-},g^{-},f^{+},g^{+})\right],
\end{split}
\end{equation}
where $f^{\pm}$ denotes a fermion with helicity ${\pm}{1 \over 2}$ and $g^{\pm}$ denotes a gluon with helicity $\pm 1$.
Every amplitude above has an implicit dependence on $z$ through the spinors which have been extended in \eqref{bcfwexample}.  We see that  for ${\mathcal A}^{\rm t}(\xi^1,\xi^2,\eta^1,\eta^2)$ to vanish in the large $z$ limit non-trivial cancellations are required. For example 
${\mathcal A}^{\rm t}(g^{-},f^{-},f^{+},g^{+})+z{\mathcal A}^{\rm t}(f^{-},g^{-},f^{+},g^{+})$ should vanish as z goes to infinity. We now show that this is indeed the case.
\FIGURE{
\epsfig{file=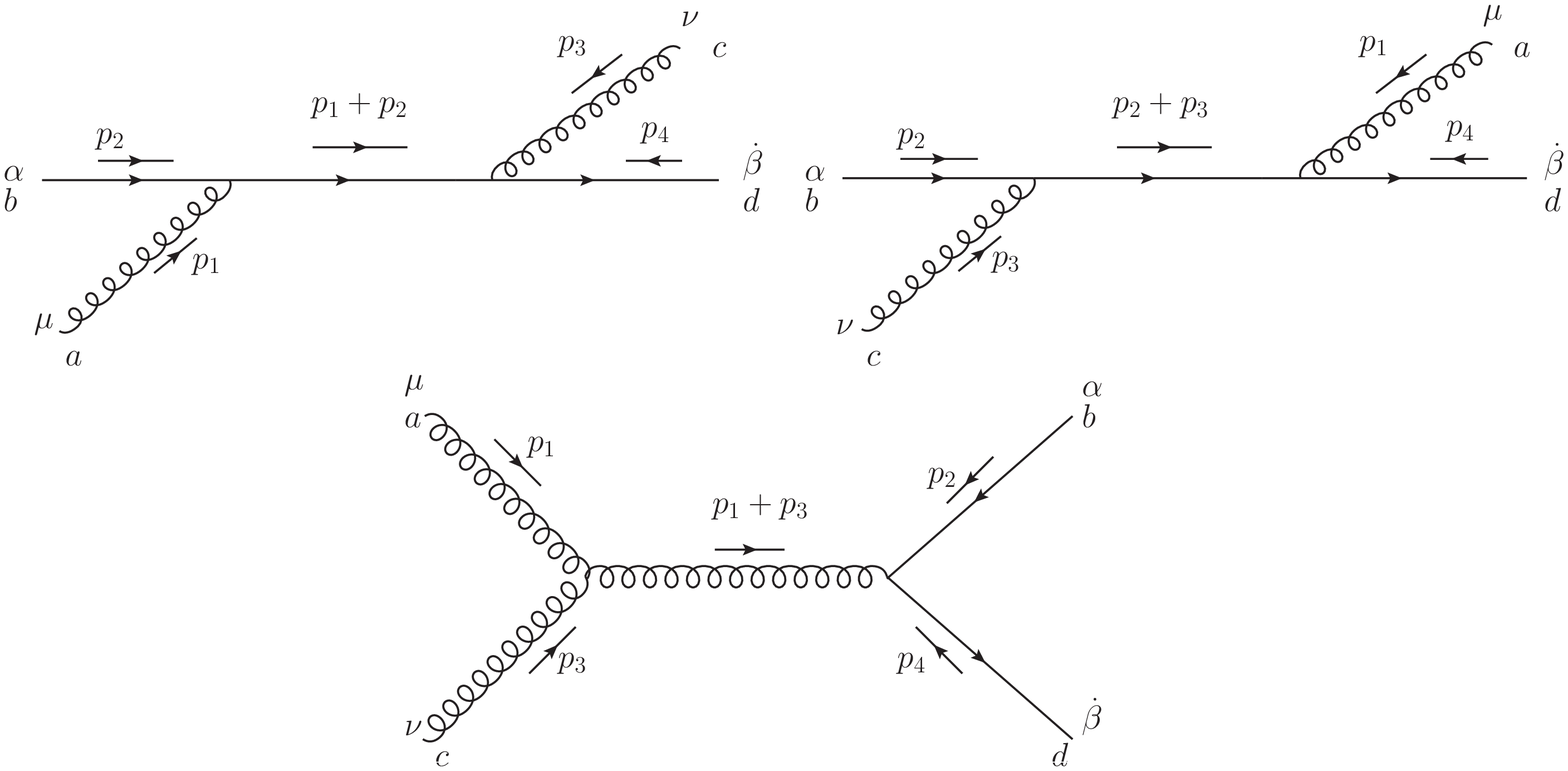,width=\textwidth}\caption{Feynman diagrams: 2 fermions to 2 gluons}
\label{ffgg}
}

The amplitude for ${\mathcal A}^{\rm t}(g^{-},f^{-},f^{+},g^{+})$ receives contributions from three diagrams which are shown in figure \ref{ffgg} and may be evaluated to obtain the expression
\begin{equation}
\left. {\mathcal A}^{\rm t}(g^-,f^-,g^+,f^+)\right|_{z=0} =  \frac{2\langle\l_2,\l_1\rangle^2}{\langle\l_3,\l_1\rangle\langle\l_2,\l_4\rangle}\left(\frac{\langle\l_4,\l_1\rangle}{\langle\l_4,\l_3\rangle}f^{cde}f^{aeb}-\frac{\langle\l_2,\l_1\rangle}{\langle\l_2,\l_3\rangle}f^{ade}f^{ceb}\right).
\end{equation}

The scattering amplitude ${\mathcal A}^{\rm t}(f^-,g^-,g^+,f^+)$ may be obtained from the above expression by interchanging the indices a and b and the labels 1 and 2.
\begin{equation}
\left.{\mathcal A}^{\rm t}(f^-,g^-,g^+,f^+)\right|_{z=0} =
 \frac{2\langle\l_2,\l_1\rangle^2}{\langle\l_3,\l_2\rangle\langle\l_1,\l_4\rangle}\left(\frac{\langle\l_4,\l_2\rangle}{\langle\l_4,\l_3\rangle}f^{cde}f^{bea}-\frac{\langle\l_1,\l_2\rangle}{\langle\l_1,\l_3\rangle}f^{bde}f^{cea}\right).
\end{equation}

The four-fermion scattering amplitude receives contributions from the two Feynman diagrams shown in \ref{fig:4f}.
\FIGURE{
\label{fig:4f}
\epsfig{file=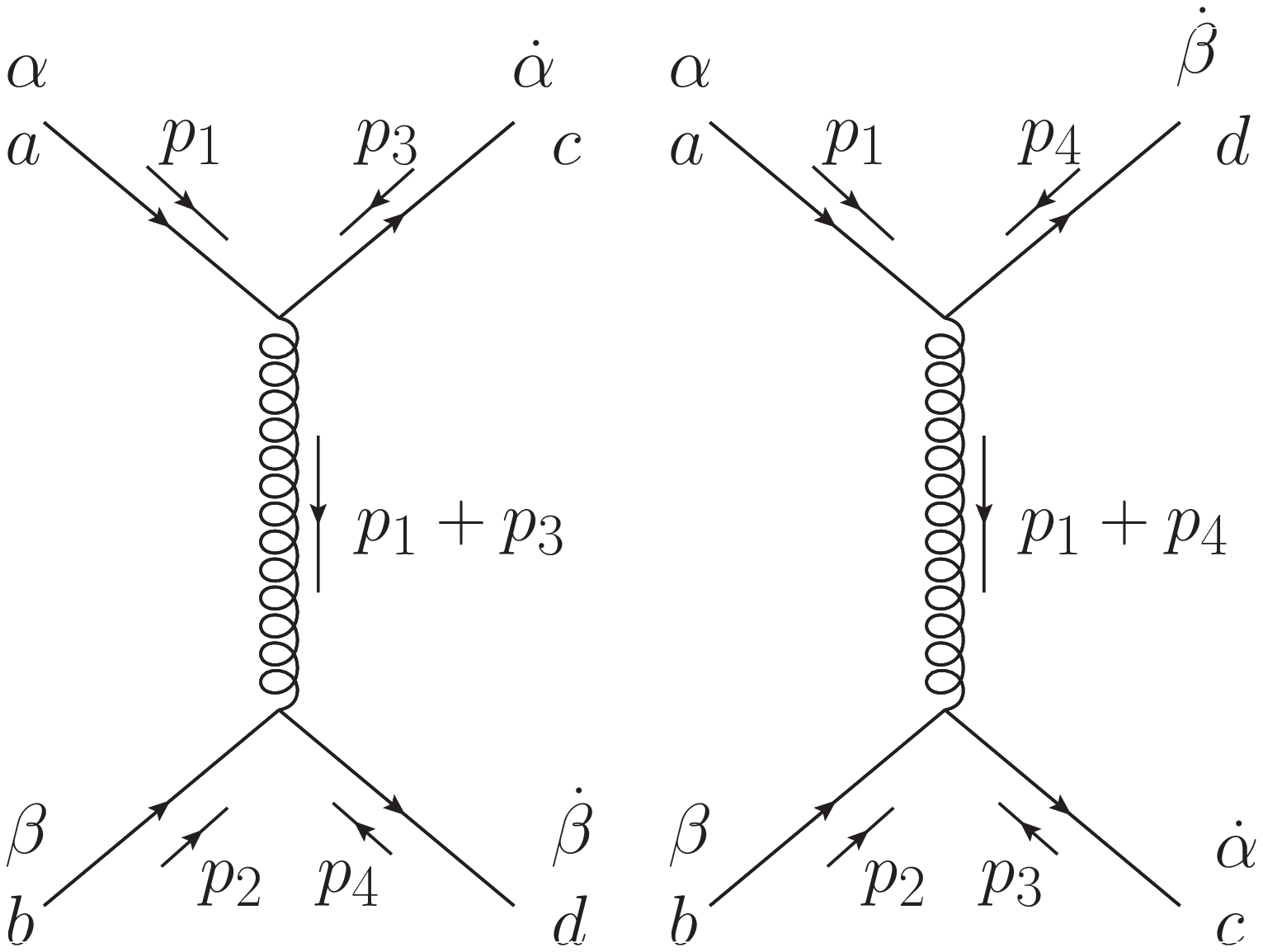,width=0.6\textwidth}
\caption{Feynman diagrams for 4 fermion scattering}
}
These diagrams lead to  
\begin{equation}
\left. {\mathcal A}^{\rm t}(f^-,f^-,f^+,f^+)\right|_{z=0}=
2f^{cae}f^{dbe}\frac{\langle\l_1,\l_2\rangle^2}{\langle\l_1,\l_3\rangle\langle\l_4,\l_2\rangle} - 2f^{dae}f^{cbe}\frac{\langle\l_1,\l_2\rangle^2}{\langle\l_1,\l_4\rangle\langle\l_3,\l_2\rangle}.
\end{equation}

We now BCFW extend these expressions, to obtain, in the limit of $z$ approaching infinity
\begin{equation}
{\mathcal A}^{\rm t}(g^-,f^-,g^+,f^+) \rightarrow
 \frac{2\langle\l_2,\l_1\rangle^2}{z\langle\l_3,\l_2\rangle\langle\l_2,\l_4\rangle}\left(\frac{z\langle\l_4,\l_2\rangle}{\langle\l_4,\l_3\rangle}f^{cde}f^{aeb}-\frac{\langle\l_2,\l_1\rangle}{\langle\l_2,\l_3\rangle}f^{ade}f^{ceb}\right),
\end{equation}
and
\begin{equation}
{\mathcal A}^{\rm t}(f^-,g^-,g^+,f^+) \rightarrow 
 \frac{2\langle\l_2,\l_1\rangle^2}{z\langle\l_3,\l_2\rangle\langle\l_2,\l_4\rangle}\left(\frac{\langle\l_4,\l_2\rangle}{\langle\l_4,\l_3\rangle}f^{cde}f^{bea}-\frac{\langle\l_1,\l_2\rangle}{z\langle\l_2,\l_3\rangle}f^{bde}f^{cea}\right).
\end{equation}

Hence we have, in the limit of $z$ approaching infinity
\begin{equation}
\begin{split}
&\lim_{z \rightarrow \infty} {\mathcal A}^{\rm t}(g^-,f^-,g^+,f^+)+z{\mathcal A}^{\rm t}(f^-,g^-,g^+,f^+)\\
&=\frac{2\langle\l_2,\l_1\rangle^2}{z\langle\l_3,\l_2\rangle\langle\l_2,\l_4\rangle}\left(\frac{\langle\l_4,\l_2\rangle}{\l_4,\l_3}f^{cde}\left(f^{aeb}+f^{bea}\right)\right)\\
&= 0.
\end{split}
\end{equation}

The other nontrivial cancellation, between ${\mathcal A}^{\rm t}(g^{-},f^{-},f^{+},g^{+})$ and ${\mathcal A}^{\rm t}(f^{-},g^{-},f^{+},g^{+})$ can be seen from the above by switching the labels c and d, and 3 and 4. Evidently, the amplitude  ${\mathcal A}^{\rm t}(f^{-},f^{-},f^{+},f^{+})$
also dies off as ${\rm O}\left({1 \over z}\right)$ for large $z$. This proves our result.
\bibliography{references}
\bibliographystyle{JHEP}
\end{document}